\documentclass[showpacs,prc,preprint,nofootinbib,showkeys]{revtex4}
\pdfoutput=1

\usepackage{graphicx}
\usepackage{amsmath}
\usepackage{amssymb}
\usepackage{bm}
\usepackage[usenames]{color}
\usepackage[colorlinks=false,linktocpage=true]{hyperref}

\newcommand{\beq}{\begin{equation}}
\newcommand{\eeq}{\end{equation}}
\newcommand{\bqa}{\begin{eqnarray}}
\newcommand{\eqa}{\end{eqnarray}}

\newcommand{\rs}{{\rm RS}}

\newcommand{\J}{\tilde{{\cal J}}}
\newcommand{\G}{\tilde{{\cal G}}}
\newcommand{\I}{\tilde{{\cal I}}}
\newcommand{\D}{\tilde{{\cal D}}}

\newcommand{\pit}{\tilde{\pi}}
\newcommand{\Pit}{\tilde{\Pi}}
\newcommand{\V}{\tilde{V}}
\newcommand{\alphat}{\alpha_\mathrm{RS}}
\newcommand{\betat}{\beta_\mathrm{RS}}
\newcommand{\dft}{\delta\tilde{f}}
\newcommand{\z}{\zeta}
\newcommand{\F}{\cal F}

\begin{document}


\title{Second-order (2+1)-dimensional anisotropic hydrodynamics}

\author{Dennis Bazow}
\affiliation{Department of Physics, The Ohio State University,
  Columbus, OH 43210 United States}
  
\author{Ulrich Heinz}
\affiliation{Department of Physics, The Ohio State University,
  Columbus, OH 43210 United States}

\author{Michael Strickland}
\affiliation{Department of Physics, Kent State University, 
Kent, OH 44242 United States}

\begin{abstract}
We present a complete formulation of second-order (2+1)-dimensional anisotropic hydrodynamics.  The resulting framework generalizes leading-order anisotropic hydrodynamics by allowing for deviations of the one-particle distribution function from the spheroidal form assumed at leading order.  We derive complete second-order equations of motion for the additional terms in the macroscopic currents generated by these deviations from their kinetic definition using a Grad-Israel-Stewart 14-moment ansatz. The result is a set of coupled partial differential equations for the momentum-space anisotropy parameter, effective temperature, the transverse components of the fluid four-velocity, and the viscous tensor components generated by deviations of the distribution from spheroidal form.  We then perform a quantitative test of our approach by applying it to the case of one-dimensional boost-invariant expansion in the relaxation time approximation (RTA) in which case it is possible to numerically solve the Boltzmann equation exactly.  We demonstrate that the second-order anisotropic hydrodynamics approach provides an excellent approximation to the exact (0+1)-dimensional RTA solution for both small and large values of the shear viscosity.
\end{abstract}


\date{\today }

\maketitle 


\section{Introduction}
\label{sec:introduction}

Fluid dynamics is canonically an effective theory which aims at describing the small frequency and long wavelength dynamics of systems that are close to equilibrium. In the case of a rarefied gas, the Boltzmann transport equation provides a method to obtain the thermodynamical properties along with additional macroscopic parameters of the system. In classical kinetic theory, Grad's method of moments~\cite{Grad:1949} defines these quantities as corresponding moments of the velocity distribution function. In the non-relativistic Grad expansion, the distribution function is obtained by factoring out a local Maxwellian distribution and then expanding the remaining unknown function in terms of Hermite polynomials. In far-from-equilibrium situations the distribution function deviates considerably from Maxwellian form, and consequently the series fails to rapidly converge. The inability to describe far-from-equilibrium situations is due to the form of the leading-order (LO) term in the series. A generalized solution for a phase-space distribution function is \cite{Mintzer}
\begin{equation}
f({\bf x},{\bf p},t)=f_0({\bf x},{\bf p},t)\sum_{\ell,\alpha}a_\alpha({\bf x},t)P^{(\ell)}_\alpha({\bf p},{\bf x},t) \label{eq:f} \; ,
\end{equation}
where $f_0$ is the LO approximation (an arbitrary weight factor), $\ell$ is the degree of the general orthogonal polynomial $P^{(\ell)}_\alpha({\bf p},{\bf x},t)$, $a_\alpha$ are the expansion coefficients, and $\alpha$ is a multi-component index (e.g. the triad $\alpha=i,j,k$).  In order to obtain the most rapid convergence, one wants to choose $f_0$ such that it is as close as possible to the exact solution $f$ of the Boltzmann equation. While $f$ is, of course, unknown, the choice of $f_0$ is guided by general insights into the properties of $f$ for the problem at hand.  

In relativistic kinetic theory, the scalar one-particle phase-space probability density $f(x,p)$ can depend on the four-vectors $p^\mu$ and $x^\mu$ only through scalar combinations. The expansion of $f$ around an isotropic local equilibrium state involves an additional four-vector $u^\mu(x)$ that describes the motion of the local heat bath at point $x$, as well as scalars $T(x)$, $\mu(x)$ characterizing the temperature and chemical potential of the local heat bath. The dependence of the local equilibrium distribution $f_0$ on the energy $E=\sqrt{m^2+|{\bf{p}}|^2}$ of the particles in the local rest frame (which is isotropic in the local rest frame momentum $\bf{p}$) can be written covariantly in terms of the momentum $p^\mu$ and the ``flow velocity'' $u^\mu(x)$ as $f_0\left(\frac{p^\nu u_\nu(x)}{T(x)},\alpha(x)\right)$ where $\alpha(x)\equiv\frac{\mu(x)}{T(x)}$ (J\"uttner distribution). We will often suppress the dependence on $\mu(x)$ and simply write $f_0(p{\cdot}u/T)\equiv f_0(E/T)$ for the local-equilibrium distribution (where $u^\mu$ and $T$ are functions of space-time position $x$).\footnote{$E$, $\bf{p}$ will always be used to denote local rest frame energies 
    and momenta; for four-momentum components in the global frame we will use the 
    notation $p^\mu=(p^0,\bm{p})$.}

An expansion of the general distribution function $f(x,p)$ around local equilibrium takes the form
\begin{equation}
      f(x,p) = f_0\left(\frac{p{\cdot}u(x)}{T(x)},\frac{\mu(x)}{T(x)}\right)+\delta f(x,p)
               \equiv f_0\Bigl(1+(1{-}a f_0)\phi(x,p)\Bigr)
\label{eq:vhexp}
\end{equation}
(with $a=1,-1,0$ for fermions, bosons, or classical distinguishable particles, respectively). The non-equilibrium correction $\phi(x,p)$ above (which gives rise to dissipative currents) is expanded in a suitably chosen set of tensors built from the rest frame energy $E$ and $p^\mu$, with $x$-dependent expansion coefficients. An example of such an expansion is $\phi(x,p)= \epsilon(x) + \epsilon_\mu(x)p^\mu + \epsilon_{\mu\nu}(x) p^\mu p^\nu{\,+\,}\dots$ \cite{Israel:1979wp}. One technical disadvantage of the tensors $1,\,p^\mu,\,p^\mu p^\nu,\,\dots$ is that they are neither orthogonal nor invariant under the Lorentz subgroup that leaves $u^\mu$ invariant. This makes the calculation of the expansion coefficients and a systematic truncation of the expansion problematic \cite{Denicol:2012cn}. A complete set of orthogonal and irreducible tensors built from powers of $p^\mu$ and the rest frame velocity $u^\mu(x)$ was introduced by Anderson \cite{Anderson:1974} and discussed in detail in \cite{deGroot}. They were recently used by Denicol {\it et al.} to define a general moment expansion of $\phi(x,p)$ in terms of momentum moments over $\delta f$ of these tensors multiplied with powers of $E=p{\cdot}u$ \cite{Denicol:2012cn}. The Boltzmann equation could then be rewritten as an infinite set of coupled partial differential equations for these moments. This hierarchy was solved by relating the $\delta f$-moments to the dissipative currents, and truncating the set of these moments by relating moments of higher rank to lower-rank ones using a systematic power-counting in Knudsen and inverse Reynolds numbers.  

Defining as in \cite{Denicol:2012cn} the moment expansion for $\phi(x,p)$ in terms of moments of $\delta f$ with powers of the local energy $E$ multiplied with Anderson polynomials is prejudiced by the assumption, manifest in Eq.~(\ref{eq:vhexp}), that the system is close to local equilibrium $f_0(E/T)$, i.e. in particular close to local momentum isotropy. This assumption breaks down during the very early expansion stage of the systems formed in ultrarelativistic heavy-ion collisions (URHICs) at the Relativistic Heavy Ion Collider at Brookhaven National Laboratory and the Large Hadron Collider at CERN. Due to its initially huge scalar expansion rate, coupled with very large anisotropies between its rapid longitudinal (along the beam direction) but much weaker transverse expansion, the system is initially unable to achieve a state of approximate local thermal equilibrium. It instead features strong anisotropies in the local rest frame momentum distributions when comparing the longitudinal and transverse directions, which results in very different longitudinal and transverse pressures. These are the result of dissipative phenomena related to the finite shear viscosity (relaxation time) of the medium.  The degree of these momentum-space anisotropies grows as one moves out from the center of the system towards the dilute edge of the overlap region.  In the canonical viscous hydrodynamical treatment these large local momentum anisotropies can cause the total (thermal plus viscous) longitudinal pressure to become negative \cite{Martinez:2009mf}. This suggests that the expansion (\ref{eq:vhexp}) around an isotropic one-particle distribution function is breaking down since the shear stress is no longer small compared to the isotropic pressure.

To account for these large early-time deviations from local momentum isotropy non-perturbatively, a framework called ``anisotropic hydrodynamics'' was developed \cite{Martinez:2010sc,Florkowski:2010cf,Ryblewski:2010bs,Martinez:2010sd,Ryblewski:2011aq,Florkowski:2011jg,Martinez:2012tu,Ryblewski:2012rr,Florkowski:2012as,Florkowski:2013uqa}. Anisotropic hydrodynamics extends traditional viscous hydrodynamical treatments to cases in which the local transverse-longitudinal momentum-space anisotropy of the plasma can be large. In order to accomplish this, one expands around an anisotropic background where the momentum-space anisotropies are built into the LO term:
\begin{equation}
   f(x,p) = 
   f_{\rm aniso}\!\left(\frac{\sqrt{p^\mu \Xi_{\mu\nu}(x)p^\nu}}{\Lambda(x)}, \frac{\tilde{\mu}(x)}{\Lambda(x)}\right)
   + \dft(x,p).
\label{eq:ahexp}
\end{equation}
Here $\Xi_{\mu\nu}$ is a second-rank tensor that measures the amount of momentum-space anisotropy and $\Lambda$ is a temperature-like scale which can be identified with the true temperature of the system only in the isotropic equilibrium limit. $\tilde{\mu}(x)$ is the effective chemical potential of the particles. Specifically, LO anisotropic hydrodynamics ({\sc aHydro}) is based on an azimuthally symmetric ansatz for $\Xi_{\mu\nu}(x)$ \cite{Martinez:2010sc} involving a single anisotropy parameter $\xi$ such that $p^\mu \Xi_{\mu\nu}(x)p^\nu$ reduces to ${\bf p}^2 + \xi(x) p_L^2$ in the local rest frame. The leading-order local rest frame distribution thus becomes of Romatschke-Strickland (RS) form \cite{Romatschke:2003ms} which has spheroidal surfaces of constant occupation number. The dynamical equations of {\sc aHydro} were derived from kinetic theory by taking $f(x,p) = f_{\rm aniso}(x,p)$ (i.e. by ignoring the correction $\dft$ in Eq.~(\ref{eq:ahexp})), and using the zeroth and first moments of the Boltzmann equation in the relaxation time approximation \cite{Martinez:2010sc,Martinez:2012tu}. 

To date, the most widely used relativistic viscous hydrodynamic framework has been Israel-Stewart (IS) theory \cite{Israel:1976tn,Israel:1979wp,Muronga:2001zk,Muronga:2003ta,Heinz:2005bw,Chaudhuri:2006jd,Romatschke:2007mq,Song:2007fn,Song:2007ux,Song:2008si,Luzum:2008cw,Romatschke:2009im,Schenke:2010rr,Schenke:2011bn,Gale:2012rq}.  It is based on an expansion of type (\ref{eq:vhexp}) around local momentum isotropy and a perturbative treatment of the dissipative currents generated by $\delta f$. The IS framework and variations upon it have been applied to URHIC phenomenology by many groups (see \cite{Romatschke:2009im,Heinz:2013th,Gale:2013da} and references therein).  Recently, there have been studies to determine the regions of validity of IS theory~\cite{Martinez:2009mf,Huovinen:2008te,Denicol:2010xn}. Initial attempts to improve upon the IS equations (see e.g.~\cite{Muronga:2006zx,Betz:2010cx,Betz:2009zz,Denicol:2010xn}) still employed the Grad 14-moment approximation \cite{Grad:1949} assuming an isotropic background. Intensive theoretical investigation into the application of relativistic fluid dynamics to URHICs have led to methods which attempt to reformulate the method of moments into a more reliable tool as well as methods which abandon the method of moments altogether in favor of a Chapman-Enskog-like expansion \cite{ChapmanEnskog}. These methods include complete second-order treatments~\cite{Denicol:2012cn,Denicol:2012es,Jaiswal:2013npa}, third-order viscous fluid dynamics~\cite{Jaiswal:2013vta},\footnote{There also exists a macroscopic  
       derivation of third-order viscous hydrodynamics starting from the 2nd law of 
       thermodynamics \cite{El:2009vj}.}
and the {\sc aHydro} formulation. {\sc aHydro} differs from these other approaches by making a specific ansatz for $\delta f$ in (\ref{eq:vhexp}), $\delta f{\,=\,}f_\mathrm{aniso}{\,-\,}f_0$, and treating the dissipative currents caused by this specific form of $\delta f$ non-perturbatively. All other mentioned approaches treat $\delta f$ perturbatively.~\footnote{In Ref.~\cite{PeraltaRamos:2012xk} the authors have presented a nonequilibrium effective theory which, for large $\eta/{\cal S}$, has better agreement with exact solution to the Boltzmann equation than Israel-Stewart viscous hydrodynamics; however, at this moment in time, this scheme cannot reproduce the longitudinal free-streaming limit.} 

In Refs.~\cite{Florkowski:2013lza,Florkowski:2013lya} the non-perturbative equations of {\sc aHydro} were compared to various second-order (perturbative) hydrodynamic approaches for (0+1)-dimensional expansion in which case there exists an exact solution to the Boltzmann equation in the relaxation time approximation. In all cases tested, anisotropic hydrodynamics most accurately approximated the exact solution when compared to various second-order viscous hydrodynamical approaches, showing the power of this non-perturbative approach. However, by ignoring the effects from $\dft$ in Eq.~(\ref{eq:ahexp}), {\sc aHydro} is unable to account for viscous effects other than those included in $\delta f = f_\mathrm{aniso}{-}f_0$. In this paper we explore the improvements that can be made by adding to {\sc aHydro} the additional dissipative currents generated by $\dft$ in Eq.~(\ref{eq:ahexp}). This leads to a formalism which we will refer to loosely as ``viscous anisotropic hydrodynamics'' ({\sc vaHydro}). We will continue to use the {\sc aHydro} framework to treat the deviation $f_\mathrm{aniso}{-}f_0$ from local isotropy non-perturbatively while adding the additional effects from $\dft$ in Eq.~(\ref{eq:ahexp}) perturbatively.

The name ``viscous anisotropic hydrodynamics'' for our second-order framework should not be misinterpreted to say that {\sc aHydro} does not include viscous effects. In fact, the energy momentum tensor for {\sc aHydro} (see Eq.~(\ref{eq20}) below) can be written as
\begin{eqnarray}
\label{eq4}
 T^{\mu\nu}_\text{\sc aHydro} &=& 
 [{\cal E} u^\mu u^\nu - {\cal P}_\mathrm{eq}\Delta^{\mu\nu}]
  + [({\cal P}_{\rm eq}{-}{\cal P}_\perp)\Delta^{\mu\nu} 
  + \left({\cal P}_L{-}{\cal P}_\perp\right)z^\mu z^\nu] 
  \nonumber\\
  &\equiv& T_{\rm eq}^{\mu\nu} - \Pi_\text{\sc aHydro}\Delta^{\mu\nu} + \pi^{\mu\nu}_\text{\sc aHydro} \; ,
\end{eqnarray}
where $\Pi_\text{\sc aHydro}$ and $\pi^{\mu\nu}_\text{\sc aHydro}$ describe the bulk and shear viscous pressure components caused by the spheroidal deformation $\delta f = f_\mathrm{aniso}{-}f_0$ of the local momentum distribution. In {\sc vaHydro} we add to $\Pi_\text{\sc aHydro}$ and $\pi^{\mu\nu}_\text{\sc aHydro}$ additional terms $\Pit$ and $\pit^{\mu\nu}$ resulting from deviations from local spheroidal symmetry due the additional term $\delta\tilde f$ in Eq.~(\ref{eq:ahexp}).  We then derive ``perturbative'' transport equations \`a la Israel-Stewart for $\Pit$ and $\pit^{\mu\nu}$, while treating the dynamics of $\Pi_\text{\sc aHydro}$ and $\pi^{\mu\nu}_\text{\sc aHydro}$ non-perturbatively. The relaxation equations for the dissipative (irreversible) currents $\Pit$ and $\tilde\pi^{\mu\nu}$ found in our approach are more complicated in structure than the corresponding Israel-Stewart relaxation equations. This is due to the fact that the minimal tensor basis to describe the underlying isotropic distribution function in Israel-Stewart theory is constructed from the fluid four-velocity and the metric tensor, whereas in the anisotropic formalism a minimal tensor basis involves the full set of Cartesian basis tensors in addition to powers of the fluid four-velocity.

The structure of the paper is as follows. In Sec.~\ref{sec:fluiddyn} we review how to connect hydrodynamics with kinetic theory, and how to derive hydrodynamic forms of the macroscopic currents and energy momentum tensor by expanding the local distribution function around isotropic and anisotropic local momentum distributions. In Sec.~\ref{sec:equations_of_motion} we derive the fundamental dynamical equations of {\sc vaHydro} by taking moments of the kinetic equation. Then, working towards deriving the additional transport equations for $\Pit$ and $\pit^{\mu\nu}$, we show in Sec.~\ref{sec:expansion} how to formulate the expansion of a general local distribution function around the azimuthally symmetric form used in {\sc aHydro}. In Sec.~\ref{sec:residual_moments} and Sec.~\ref{sec:14_moment} a set of equations are derived for the residual moments of the distribution function deviation $\dft$ and then the 14-moment approximation scheme is applied in order to truncate the expansion. Sec.~\ref{sec:aniso_thermo_vars_sec} reviews the quasi-thermodynamic quantities (particle and energy density, longitudinal and transverse pressure) in ``anisotropic equilibrium" as functions of the anisotropy parameter $\xi$ and effective temperature $\Lambda$ and shows that, in the massless particle limit, the $\xi$ dependence can be factored out. Section~\ref{sec:dyneq_dof} contains the main analytic results of this work: For a system of massless degrees of freedom with zero chemical potential, we derive the coupled equations of motion for $\xi$, $\Lambda$, the hydrodynamic flow $u^\mu$, and the new viscous stress tensor $\pit^{\mu\nu}$ arising from the deviation $\dft$ in Eq.~(\ref{eq:ahexp}). In Sec.~\ref{sec:0+1d.} we simplify these equations for the limiting case of (0+1)-dimensional longitudinally boost-invariant expansion and compare their numerical solution to the exact result from the corresponding underlying Boltzmann equation. Our conclusions are presented Sec.~\ref{sec:concl.}. Six appendices contain intermediate steps of the calculations as well as some tabulated results used in the body of the paper. 

Before proceeding, let us define our notation. We use natural units $\hbar=k_B=c=1$. The Minkowski metric tensor is $g^{\mu\nu}={\rm diag}(+,-,-,-)$. Greek indices run from 0 to 3 and Latin indices from 1 to 3. The summation convention for repeated indices (Greek or Latin) is always used. Our tensor basis, in the local rest frame, is $X^\mu_0\equiv u^\mu=(1,0,0,0)$, $X^\mu_1\equiv x^\mu=(0,1,0,0)$, $X^\mu_2\equiv y^\mu=(0,0,1,0)$, and $X^\mu_3\equiv z^\mu=(0,0,0,1)$. The transverse projection operator $\Delta^{\mu\nu}\equiv {-}X^\mu_iX^\nu_i=g^{\mu\nu}{-}u^\mu u^\nu$ is used to project four-vectors and/or tensors into the space orthogonal to $u^\mu$. The notations $A^{(\mu\nu)}\equiv\frac{1}{2}\left(A^{\mu\nu}{+}A^{\nu\mu}\right)$ and $A^{[\mu\nu]}\equiv\frac{1}{2}\left(A^{\mu\nu}{-}A^{\nu\mu}\right)$ denote symmetrization and antisymmetrization, respectively. $A^{\langle \mu \nu\rangle}\equiv\Delta^{\mu\nu}_{\alpha\beta}A^{\alpha\beta}$ where $\Delta^{\mu\nu}_{\alpha\beta}\equiv\Delta^{(\mu}_\alpha\Delta^{\nu)}_\beta-\Delta^{\mu\nu}\Delta_{\alpha\beta}/3$ is the transverse (to $u$) and traceless projector for second-rank tensors. The four-derivative is $\partial_\mu\equiv\partial/\partial x^\mu$, $D\equiv u^\mu\partial_\mu$ is the convective derivative (the time derivative in the comoving frame), $\nabla^\mu\equiv\Delta^{\mu\nu}\partial_\nu$ is the covariant notation for the spatial gradient operator in the local rest frame, and $\theta\equiv\partial_\mu u^\mu=\nabla_\mu u^\mu$ is the scalar expansion rate.

\section{Hydrodynamic tensor decomposition and local momentum (an)isotropy}
\label{sec:fluiddyn}

In this paper we derive macroscopic hydrodynamical equations from an underlying classical kinetic framework. In the kinetic framework a central role is played by the one-particle distribution $f$ which obeys the Boltzmann equation
\begin{equation}
p^\mu\partial_\mu f=C[f] \label{eq:be} \; ,
\end{equation}
where $C[f]$ is the collision kernel. The classical Boltzmann equation is valid for sufficiently dilute and weakly interacting many-particle systems: $\lambda_\mathrm{mfp} \gg \lambda_\mathrm{th}$ where $\lambda_\mathrm{mfp}$ is the particle mean free path and $\lambda_\mathrm{th}$ the thermal wavelength. The validity of the macroscopic hydrodynamic approach is controlled by the Knudsen number $\mathrm{Kn} = \theta\lambda_\mathrm{mfp}\ll 1$ where $\theta$ is the scalar expansion rate, and by the inverse Reynolds numbers $\mathrm{R}_\Pi^{-1}=|\Pi|/{\cal P}_0\ll 1$, $\mathrm{R}_\pi^{-1}=\sqrt{\pi^{\mu\nu}\pi_{\mu\nu}}/{\cal P}_0\ll 1$, and $\mathrm{R}_{\cal N}^{-1}=\sqrt{-V^\mu V_\mu}/{\cal N}_0\ll 1$, where $\Pi$, $\pi^{\mu\nu}$, and $V^\mu$ are the dissipative currents (see below), and ${\cal N}_0$ and ${\cal P}_0$ are the equilibrium particle number density and pressure, respectively \cite{Denicol:2012cn}. By making $\theta$ small enough and preparing the system initially sufficiently close to local equilibrium, we can ensure the simultaneous validity of both approaches. However, the macroscopic hydrodynamic approach (and the equations derived here) remain valid even for strongly coupled systems where Boltzmann transport theory breaks down, as long as  $\mathrm{Kn}\ll 1$ and $\mathrm{R}_i^{-1}\ll 1$. 

We define the average of a momentum-dependent observable $a(p)$ at point $x$ as $\langle a \rangle(x) \equiv \int dP \, a(p) f(x,p)$, with the Lorentz invariant momentum-space measure $dP \equiv d^3p/\left[p^0(2\pi)^3\right]$. In the following we will suppress the $x$ dependence of all momentum moments to simplify the notation. The $n$-th moment of the one-particle distribution function is defined as  $I^{\mu_1 \cdots \mu_n}=\langle p^{\mu_1}\cdots p^{\mu_n}\rangle$. The particle current and energy-momentum tensor are identified as the first and second moments of the one-particle distribution function,
\begin{equation} \label{eq:lab1}
\begin{array}{ll}
j^\mu=\langle p^\mu\rangle \; , & \;\;\;\;\;  T^{\mu\nu}=\langle p^\mu p^\nu\rangle\; .
\end{array}  
\end{equation}
We will always define the velocity field using the Landau prescription where $u^\mu$ is defined by the flow of total momentum, by solving the eigenvalue equation
\begin{equation}
\label{eq7}
T^\mu_{\ \,\nu} u^\nu  ={\cal E} u^\mu .
\end{equation}
Here ${\cal E}$ is the energy density in the local rest frame (Landau frame).

\subsection{Expansion around an isotropic momentum distribution}
\label{subsec:iso_variables}

For later comparison we briefly review the tensor decomposition of the particle current and energy-momentum tensor in the locally isotropic case. We decompose the particle four-momentum $p^\mu$ into parts parallel and orthogonal to $u^\mu$,
\begin{equation} 
p^\mu= E u^\mu+p^{\langle \mu \rangle} , \label{eq:p_dec_iso}
\end{equation}
where $E$ is the local rest frame energy and $p^{\langle \mu \rangle} = \Delta^{\mu\nu}p_\nu$ reduces to the spatial momentum in that frame. Then $j^\mu$ and $T^{\mu\nu}$ can be tensor decomposed as
\begin{equation}\label{eq:iso_jmuTmunu}
\begin{split}
j^\mu&=\langle E \rangle u^\mu+\langle p^{\langle \mu \rangle}\rangle \; , \\
T^{\mu\nu}&=\langle E^2 \rangle u^\mu u^\nu+\frac{1}{3}\Delta^{\mu\nu}\langle \Delta^{\alpha\beta}p_\alpha p_\beta\rangle+\langle p^{\langle \mu}p^{\nu \rangle}\rangle \; ,
\end{split}
\end{equation}
where we have explicitly used the fact that
\begin{equation}\label{eq:pp}
p^{\langle\mu\rangle}p^{\langle\nu\rangle}=p^{\langle\mu}p^{\nu\rangle}+\frac{1}{3}\Delta^{\mu\nu}\Delta^{\alpha\beta}p_{\alpha}p_{\beta} \; .
\end{equation}
The averages in Eq.~(\ref{eq:iso_jmuTmunu}) are related to the macroscopic properties of the system by
\begin{equation} 
\begin{array}{lll}
{\cal N}\equiv\langle E\rangle, & V^\mu\equiv \langle p^{\langle \mu \rangle}\rangle, & {}\\
{\cal E}\equiv \langle E^2 \rangle, & {\cal P}_0+\Pi\equiv -\frac{1}{3}\langle \Delta^{\alpha\beta}p_\alpha\ p_\beta\rangle,  & \pi^{\mu\nu}\equiv\langle p^{\langle \mu}p^{\nu \rangle}\rangle .
\end{array}  
\end{equation}
Here ${\cal N}$ is the particle density and $V^\mu$ is the particle current in the local rest frame, ${\cal E}$ is the energy density in the local rest frame, ${\cal P}_0$ is the thermodynamic pressure, $\Pi$ is the bulk viscous pressure, and $\pi^{\mu\nu}$ is the shear stress tensor.

In local isotropic equilibrium the one-particle distribution function has the form
\begin{equation}
f(x,p)=f_0(x,p)\equiv\left(e^{\beta E-\alpha}+a\right)^{-1} \label{eq:fiso} \;,
\end{equation}
where  $\beta(x)=1/T(x)$ is the inverse local temperature, $\alpha(x)=\mu(x)/T(x)$ is the ratio of the local chemical potential to local temperature, and $a=\pm1,0$ corresponds to Fermi-Dirac, Bose-Einstein, and classical Boltzmann statistics, respectively. For systems which are slightly out of equilibrium, we can linearize $f$ to obtain
\begin{equation} 
\label{eq:iso_exp}
f(x,p)=f_0(x,p)\left(1+\tilde{f}_0(x,p)\phi(x,p)\right) \; , 
\end{equation}
where $\phi(x,p)$ describes the small deviation from local equilibrium. The factor $\tilde{f}_0=1-af_0$ takes into account quantum statistics. The expansion (\ref{eq:iso_exp}) is strictly only useful for systems that are close to equilibrium (i.e. $|\phi|\ll 1$) and can be expected to fail to converge in far-from-equilibrium situations.\footnote{We note in this context that even close to equilibrium the hydrodynamic gradient expansion may not converge either~\cite{Heller:2013fn}.} When $f$ relaxes to $f_0$, the entropy of the system is maximized and entropy production ceases.  Equivalently, irreversible thermodynamic processes are described through $\phi(x,p)$.  We define the equilibrium particle four-current $j^\mu[f_0]\equiv j^\mu_0$ and energy-momentum tensor $T^{\mu\nu}[f_0]\equiv T^{\mu\nu}_0$ by
\begin{equation}\label{tmunu0}
\begin{split}
j^\mu_0 &\equiv {\cal N}_0 u^\mu, \\
T^{\mu\nu}_0 &\equiv \left({\cal E}_0+{\cal P}_0\right)u^\mu u^\nu -{\cal P}_0 g^{\mu\nu}  .
\end{split}
\end{equation}
These quantities can be expressed entirely in terms of the equilibrium thermodynamic properties of the system. By decomposing $j^\mu=j^\mu_0+\delta j^\mu$ and $T^{\mu\nu}=T^{\mu\nu}_0+\delta T^{\mu\nu}$ we see that
\begin{equation}
\begin{split}
\delta j^\mu&=\langle E \rangle_\delta u^\mu+\langle p^{\langle\mu\rangle} \rangle_\delta, \\
\delta T^{\mu\nu} &=\langle E^2 \rangle_\delta u^\mu u^\nu +\frac{1}{3}\Delta^{\mu\nu}\langle \Delta^{\alpha\beta}p_\alpha\ p_\beta\rangle_\delta+\langle p^{\langle\mu}p^{\nu\rangle}\rangle_\delta,
\end{split}
\end{equation}
where $\langle \,\cdots\, \rangle_0\equiv\int dP \,(\,\cdots )f_0$ and $\langle \,\cdots\, \rangle_\delta=\langle \,\cdots\, \rangle-\langle \,\cdots\, \rangle_0$ describe moments taken with the equilibrium distribution $f_0$ and with the deviation from equilibrium $\delta f = f_0 \tilde{f}_0 \phi$, respectively.  

To define $f_0$ in Eq.~(\ref{eq:iso_exp}) we need conditions that fix the local temperature and chemical potential in such a way that $f_0$ optimally approximates the non-equilibrium distribution function $f$ and minimizes $\phi$.  These are the Landau matching conditions: ${\cal N}\equiv {\cal N}_0=\langle E\rangle_0$ and ${\cal E}\equiv {\cal E}_0=\langle E^2\rangle_0$.  These state that, by optimizing $T$ and $\mu$ in $f_0$, $\delta f$ makes no residual contributions to the local energy and particle density ${\cal E}$ and ${\cal N}$:  $\langle E \rangle_\delta = \langle E^2 \rangle_\delta = 0$.
This leads to
\begin{equation}\label{eq:fluid_fields_iso}
\begin{split}
j^\mu &=n_0 u^\mu+V^\mu = j_0^\mu + V^\mu, \\
T^{\mu\nu} &={\cal E}_0 u^\mu u^\nu-\left({\cal P}_0+\Pi\right)\Delta^{\mu\nu}+\pi^{\mu\nu} ,
\end{split}
\end{equation}
where $V^\mu\equiv\langle p^{\langle\mu\rangle}\rangle_\delta$, $\Pi\equiv-\frac{1}{3}\langle\Delta^{\alpha\beta}p_\alpha p_\beta\rangle_\delta$, and $\pi^{\mu\nu}\equiv\langle p^{\langle\mu} p^{\nu\rangle} \rangle_\delta$.

\subsection{Expansion around a spheroidal momentum distribution}
\label{subsec:aniso_variables}

We now repeat the above procedure for an expansion around a ``local anisotropic equilibrium" distribution function $f_\mathrm{aniso}(x,p)$ as in Eq.~(\ref{eq:ahexp}). In this work we assume that in the local rest frame $f_\mathrm{aniso}(x,p)$ is azimuthally symmetric in momentum-space $\left(\langle p_x^2 \rangle = \langle p_y^2 \rangle \equiv \frac{1}{2}\langle p_\perp^2 \rangle\ne \langle p_z^2 \rangle\right)$. We rewrite the decomposition (\ref{eq:p_dec_iso}) in terms of the Cartesian basis vectors in the local rest frame \cite{Martinez:2012tu}:
\begin{equation}
p^\mu = Eu^\mu+p_i X^{\mu}_i \equiv Eu^\mu+p_x x^\mu+p_y y^\mu+p_z z^\mu \; ,
\label{eq:pdecomp} 
\end{equation}
where $(E,p_x,p_y,p_z) $ are the Cartesian components of the four-momentum in the local rest frame. This leads to 
\begin{equation}
\label{eq:jmuTmunu1}
\begin{split}
j^\mu &\equiv \langle p^\mu \rangle=\langle E \rangle u^\mu+\langle p_i \rangle X^{\mu}_i,\\
T^{\mu\nu} &\equiv \langle p^\mu p^\nu \rangle=\langle E^2 \rangle u^\mu u^\nu +\langle p_ip_j\rangle X^\mu_iX^\nu_j + 2\langle Ep_i \rangle  X^{(\mu}_iu^{\nu)}.
\end{split}
\end{equation}
The last term in $T^{\mu\nu}$ vanishes when we adopt the Landau definition for the fluid four-velocity (i.e. we demand that there is no net momentum flow in the local rest frame). Similar to the isotropic case, we use the expansion (\ref{eq:ahexp}) to split Eq.~(\ref{eq:jmuTmunu1}) into its thermodynamical quantities in ``anisotropic equilibrium"\footnote{Since {\sc aHydro} is inherently a dissipative    
    dynamical effective theory we need to define what we mean by ``anisotropic equilibrium''.
    Clearly, entropy production does not vanish when $f=f_\mathrm{aniso}$. The word
    ``equilibrium" in this context serves only to remind the reader of the fact that, in the 
    isotropic limit, $f_\mathrm{aniso}$ reduces to the local equilibrium distribution function 
    $f_0$.} 
and additional irreversible quantities arising from $\dft$:
\begin{equation}
\label{eq19}
\begin{split}
j^\mu &=j^\mu_\mathrm{aniso} +\delta \tilde j^\mu,\\
T^{\mu\nu} &= T^{\mu\nu}_\mathrm{aniso} + \delta\tilde T^{\mu\nu},
\end{split}
\end{equation}
where \cite{Martinez:2009ry}
\begin{equation}
\label{eq20}
\begin{split}
j^\mu_\mathrm{aniso} \equiv j^\mu[f_\mathrm{aniso}] &= \langle E \rangle_\mathrm{aniso} u^\mu\equiv {\cal N}_\mathrm{aniso} u^\mu,\\
T^{\mu\nu}_\mathrm{aniso} \equiv T^{\mu\nu}[f_\mathrm{aniso}] &= 
\langle E^2 \rangle_\mathrm{aniso} u^\mu u^\nu 
+\sum_{i=1}^3\langle p^2_i\rangle_\mathrm{aniso}X^\mu_iX^\nu_i \\
&\equiv {\cal E}_{\rm aniso} u^\mu u^\nu -{\cal P}_\perp\Delta^{\mu \nu} +\left({\cal P}_L-{\cal P}_\perp\right)z^\mu z^\nu ,
\end{split}
\end{equation}
and  
\begin{equation}
\label{eq:jmuTmunu}
\begin{split}
\delta\tilde j^\mu &= \langle E \rangle_{\tilde\delta} u^\mu+\langle p_i \rangle_{\tilde\delta} X^{\mu}_i ,\\
\delta\tilde T^{\mu\nu} &= \langle E^2 \rangle_{\tilde\delta} u^\mu u^\nu +\langle p_ip_j\rangle_{\tilde\delta} X^\mu_iX^\nu_j.
\end{split}
\end{equation}
Here $\langle \,\cdots\, \rangle_\mathrm{aniso}\equiv\int dP \,(\,\cdots )f_\mathrm{aniso}$ and $\langle \,\cdots\, \rangle_{\tilde\delta}\equiv\int dP \,(\,\cdots )\dft$ denote moments taken with $f_\mathrm{aniso}$ and $\dft$, respectively. In Eq.~(\ref{eq20}) we used the fact that $f_\mathrm{aniso}$ is a parity-even function of $p^\mu$ such that $\langle p_i \rangle_\mathrm{aniso}$ and $\langle p_ip_j \rangle_\mathrm{aniso}$ for $i\ne j$ vanish upon symmetric integration. The final term in Eq.~(\ref{eq:jmuTmunu}) can be recast into a form similar to the isotropic $\Pi$ and $\pi^{\mu\nu}$ by using
\begin{equation}
p_ip_jX^\mu_iX^\nu_j=p^{\langle\mu\rangle}p^{\langle\nu\rangle} 
\end{equation}
and Eq.~(\ref{eq:pp}). This yields
\begin{equation}
\langle p_ip_j\rangle_{\tilde\delta} X^\mu_iX^\nu_j=\frac{1}{3}\Delta^{\mu\nu}\langle \Delta^{\alpha\beta}p_\alpha\ p_\beta\rangle_{\tilde\delta}+\langle p^{\langle\mu}p^{\nu\rangle}\rangle_{\tilde\delta}\;.
\end{equation}
As in the case of an expansion of $f(x,p)$ around local equilibrium, we need matching conditions that provide values for the parameters $\xi(x)$ and $\Lambda(x)$ which define the ``optimal anisotropic equilibrium distribution'' $f_\mathrm{aniso}(x,p)$ in Eq.~(\ref{eq:ahexp}). We impose the generalized Landau matching conditions $\langle E\rangle_{\tilde\delta} = \langle E^2\rangle_{\tilde\delta} = 0$, i.e. we demand that the effective temperature $\Lambda$ and the chemical potential $\mu$ in $f_\mathrm{aniso}$ are chosen such that (for given anisotropy parameter $\xi$, see below) $\delta\tilde f$ makes no additional contributions to the energy and particle densities in the local rest frame: $u_\mu\delta \tilde j^\mu=0=u_\mu\delta\tilde T^{\mu\nu} u_\nu$. The relation of the moments ${\cal N}_\mathrm{aniso}(\xi,\Lambda)\equiv\langle E \rangle_\mathrm{aniso}(\xi,\Lambda)$ and ${\cal E}_\mathrm{aniso}(\xi,\Lambda)\equiv\langle E^2 \rangle_\mathrm{aniso}(\xi,\Lambda)$ to the thermal equilibrium particle and energy densities, ${\cal N}_0(T)$ and ${\cal E}_0(T)$, will be discussed in Sec.~\ref{sec:aniso_thermo_vars_sec}.

Putting everything together we obtain the anisotropic decompositions
\begin{equation}
\label{eq24}
\begin{split}
  j^\mu &= {\cal N}_\mathrm{aniso} u^\mu + \tilde V^\mu,\\
  T^{\mu\nu} &= {\cal E}_{\rm aniso} u^\mu u^\nu -\Bigl({\cal P}_\perp+\tilde\Pi\Bigr)
  \Delta^{\mu \nu} +\left({\cal P}_L{-}{\cal P}_\perp\right)z^\mu z^\nu + \tilde\pi^{\mu\nu}
  = T^{\mu\nu}_\mathrm{eq} - \Pi \Delta^{\mu\nu} + \pi^{\mu\nu},
\end{split}
\end{equation}
with 
\begin{subequations} 
\label{eq:fields}
\begin{align}
&{\cal N}_\mathrm{aniso} \equiv u_\mu j^\mu = \langle E \rangle
                                                                        = \langle E \rangle_\mathrm{aniso},\\ 
&\V^\mu \equiv \Delta^{\mu\nu}j_\nu = \langle p_i \rangle X^\mu_i
              = \langle p_i \rangle_{\tilde\delta} X^\mu_i  
              = \left\langle p^{\langle\mu\rangle}\right\rangle_{\tilde\delta},\\
&{\cal E}_\mathrm{aniso} \equiv u_\mu T^{\mu\nu} u_\nu = \langle E^2 \rangle
                                                 = \langle E^2 \rangle_\mathrm{aniso}, \\
&{\cal P}_\perp \equiv \frac{1}{2}\left\langle p^2_x+p^2_y \right\rangle_{\rm aniso}
                          = \langle p^2_\perp \rangle_{\rm aniso}, \\
&{\cal P}_L \equiv \langle p^2_z \rangle_{\rm aniso}, \\
&\Pit \equiv -\frac{1}{3}\langle\Delta^{\alpha\beta}p_\alpha p_\beta\rangle_{\tilde\delta}, \\
&\pit^{\mu\nu} \equiv \langle p^{\langle\mu}p^{\nu\rangle} \rangle_{\tilde\delta} \, .
\end{align}
\end{subequations}
The various pressure components can be obtained from the total $T^{\mu\nu}$ in (\ref{eq24}) by projection as follows: The total isotropic pressure is obtained from
\begin{equation}
-\frac{1}{3} \Delta_{\mu \nu}T^{\mu \nu} \equiv {\cal P}_\mathrm{eq}+\Pi \, .
\end{equation}
It is the sum of the thermodynamic equilibrium pressure ${\cal P}_\mathrm{eq}(T)$, which can be obtained from the energy density through the equation of state (EoS) ${\cal P}_\mathrm{eq}({\cal E}_\mathrm{eq})$,\footnote{In the present work 
     we ignore conserved charges and associated chemical potentials.} 
and the bulk viscous pressure $\Pi$. The appropriate ``equilibrium temperature'' $T(x)=T(\xi(x),\Lambda(x))$ for a given energy-momentum tensor at a given point $x$ is fixed \cite{Martinez:2009ry} such that ${\cal E}_\mathrm{eq}(T)$ matches the ``anisotropic equilibrium energy density'' ${\cal E}_\mathrm{aniso}$ from Eq.~(\ref{eq:fields}c): ${\cal E}_\mathrm{eq}(T) = {\cal E}_\mathrm{aniso}(\xi,\Lambda)$ (Landau matching). The total bulk viscous pressure $\Pi$ is thus calculable as
\begin{equation}
\label{eq27}
\Pi = -\frac{1}{3} \Delta_{\mu \nu}T^{\mu \nu} - {\cal P}_\mathrm{eq}({\cal E}),
\end{equation}
where ${\cal E}=u_\mu T^{\mu\nu} u_\nu$, with $u^\mu$ being the time-like normalized eigenvector of $T^{\mu\nu}$ (see Eq.~(\ref{eq7})). Applying this projection to the decomposition (\ref{eq24}) we see that the bulk viscous pressure $\Pi$ can be written as
\begin{equation}
\label{eq28}
\Pi = \frac{2{\cal P}_\perp+{\cal P}_L}{3}-{\cal P}_\mathrm{eq}+\Pit
    \equiv \Pi_\text{\sc aHydro} +\Pit .
\end{equation}
The total shear stress tensor is obtained from
\begin{equation}
\label{eq29}
\pi^{\mu\nu} = T^{\langle\mu\nu\rangle} \equiv \Delta^{\mu\nu}_{\alpha\beta}T^{\alpha\beta} \, ,
\end{equation}
which, when applied to the decomposition (\ref{eq24}), yields
\begin{eqnarray}
\label{eq29a}
\pi^{\mu\nu} &=&  \left({\cal P}_L{-}{\cal P}_\perp\right)
                            \left(\frac{\Delta^{\mu\nu}}{3}+z^\mu z^\nu\right) + \pit^{\mu\nu} 
= \left({\cal P}_\perp{-}{\cal P}_L\right)\,\frac{x^\mu x^\nu{+}y^\mu y^\nu{-}2 z^\mu z^\nu}{3}
       + \pit^{\mu\nu}
\nonumber\\
&\equiv&  \pi^{\mu\nu}_\text{\sc aHydro}   + \pit^{\mu\nu}. 
\end{eqnarray}
Equations~(\ref{eq28}) and (\ref{eq29a}) split the bulk and shear viscous pressures $\Pi$ and $\pi^{\mu\nu}$ into terms associated with the underlying phase-space distributions $f_\mathrm{aniso}{-}f_0$ and $\dft$ through Eqs.~(\ref{eq:fields}d-g). This separation is ambiguous until we specify the anisotropy parameter $\xi(x)$ in $f_\mathrm{aniso}$. This requires an additional matching condition. We use our general expressions (\ref{eq24}) and (\ref{eq:fields}) and compute 
\begin{eqnarray}
\label{eq30}
\left(\frac{1}{2}\left( x_\mu x_\nu{+}y_\mu y_\nu\right) -z_\mu z_\nu\right)T^{\mu\nu} 
&=& {\cal P}_\perp - {\cal P}_L + \frac{x_\mu x_\nu{+}y_\mu y_\nu{-}2z_\mu 
z_\nu}{2}\,\pit^{\mu\nu}
\nonumber\\
&=& {\cal P}_\perp - {\cal P}_L + \frac{x_\mu x_\nu{+}y_\mu y_\nu{-}2z_\mu z_\nu}{2}
   \,\langle p^{\langle\mu}p^{\nu\rangle} \rangle_{\tilde\delta}. 
\end{eqnarray}
The matching condition for $\xi$ stipulates that this parameter should be chosen such that $f_\mathrm{aniso}$ captures all of the pressure anisotropy ${\cal P}_L{-}{\cal P}_\perp$,\footnote{This is not always possible, though: no choice of $\xi$ in $f_\mathrm{aniso}$ can result in 
    a negative longitudinal pressure ${\cal P}_L<0$ as it occurs e.g. in a color flux tube. In 
    such a situation one would let $\xi\to\infty$, thereby obtaining ${\cal P}_L\to0$ from 
    $f_\mathrm{aniso}$, and still be left with a remaining nonzero contribution to the pressure 
    anisotropy ${\cal P}_L{-}{\cal P}_\perp$ from $\pit^{\mu\nu}$.}   
i.e. that it receives no contribution from $\dft$ and the last term in (\ref{eq30}) is zero: 
\begin{equation}
\label{eq31}
\frac{x_\mu x_\nu{+}y_\mu y_\nu{-}2z_\mu z_\nu}{2}\,\pit^{\mu\nu}=\frac{x_\mu x_\nu{+}y_\mu y_\nu{-}2z_\mu z_\nu}{2}\, \langle p^{\langle\mu}p^{\nu\rangle} \rangle_{\tilde\delta}=0.
\end{equation}
As a consequence, the pressure anisotropy can be extracted directly from the following projection of a general $T^{\mu\nu}$ tensor:
\begin{equation}
\label{eq32}
 {\cal P}_\perp - {\cal P}_L=\left(\frac{1}{2}\left( x_\mu x_\nu{+}y_\mu y_\nu\right) -z_\mu z_\nu\right)T^{\mu\nu}. 
\end{equation}
For the decomposition (\ref{eq28}) we need additionally the relation between $(2{\cal P}_\perp{+}{\cal P}_L)/3$ and ${\cal P}_\mathrm{eq}$. Their difference defines the bulk viscous pressure in anisotropic hydrodynamics. For fixed parameters $\Lambda$, $\mu$, and $\xi$ in $f_\mathrm{aniso}$, this difference still depends on the particle mass and their interactions. It thus can be thought of as an additional (non-equilibrium) equation of state for a system in ``anisotropic equilibrium'' (``anisotropic EoS'') that embodies the reaction of its pressure to a spheroidal deformation of its local momentum distribution, and which depends on the assumed model for the many-body system, i.e. on the specific functional form of $f_\mathrm{aniso}$. For a gas of non-interacting massless particles (assumed in all applications presented here) the difference $(2{\cal P}_\perp{+}{\cal P}_L)/3{\,-\,}{\cal P}_\mathrm{eq}$ \cite{Martinez:2009ry} and $\Pit$ both vanish (see Sec.~\ref{sec:aniso_thermo_vars_sec}), and so does the total bulk viscous pressure. 

Let us summarize: Given an initial particle current $j^\mu(x)$ and energy-momentum tensor $T^{\mu\nu}(x)$ (obtained, for example, as output from some theoretical description of the pre-equilibrium stage of a heavy-ion collision), we determine the local flow velocity $u^\mu(x)$ from Eq.~(\ref{eq7}) by finding its normalized time-like eigenvector, compute the local energy density ${\cal E}$ from Eq.~(\ref{eq:fields}c), the associated equilibrium pressure from the EoS ${\cal P}_\mathrm{eq}={\cal P}({\cal E})$, the pressure anisotropy from (\ref{eq32}), the difference between $(2{\cal P}_\perp{+}{\cal P}_L)/3$ and the thermal equilibrium pressure ${\cal P}_\mathrm{eq}$ from our model for $f_\mathrm{aniso}$ (or, more generally, from some ``anisotropic EoS''), and then solve Eqs.~(\ref{eq27})-(\ref{eq29a}) for $\Pit$ and $\pit^{\mu\nu}$. With ${\cal N}_\mathrm{aniso}$ and $\tilde V^\mu$ from Eqs.~(\ref{eq:fields}a,b) the anisotropic hydrodynamic decomposition (\ref{eq24}) of $j^\mu$ and $T^{\mu\nu}$ is complete, and the further evolution of the system can be determined by solving the {\sc vaHydro} equations of motion. These will be derived in the following sections.

For notational convenience from here on we drop the subscript ``aniso'' on ${\cal N}$ and ${\cal E}$; ${\cal N}_0$ and ${\cal E}_0$ continue to denote the local equilibrium values of these quantities.

\section{Viscous anisotropic hydrodynamic equations of motion}
\label{sec:equations_of_motion}

In this section we derive the hydrodynamic equations of motion by taking the zeroth and first moments of the Boltzmann equation. Taking moments implies multiplying (\ref{eq:be}) by integer powers of the four-momentum and integrating over momentum-space. This process results in the following $n$-th ($n\ge 0$) moment equation:
\begin{equation}
\partial_{\mu_1} \langle p^{\mu_1}\cdots p^{\mu_{n+1}} \rangle={\cal C}^{\mu_1\cdots \mu_{n}}. 
\end{equation}
Here ${\cal C}^{\mu_1 \cdots \mu_n}\equiv{\cal C}_0^{\mu_1 \cdots \mu_n}$ where we have defined the $n$-th rank collisional tensors ${\cal C}^{\mu_1 \cdots \mu_n}_r\equiv \left\langle E^rp^{\mu_1}\cdots p^{\mu_n} C[f] \right\rangle$. The case $n=r=0$ corresponds to the scalar ${\cal C}\equiv\langle C[f]\rangle$. It should be noted that a vanishing ${\cal C}^{\mu_1\cdots\mu_n}$ corresponds to a conservation law for the $(n+1)$-st moment of the one-particle distribution function, e.g. $\partial_\mu\langle p^\mu\rangle=0$ (conservation of particles) and $\partial_\mu\langle p^\mu p^\nu\rangle=0$ (conservation of energy-momentum). The equations of spheroidal anisotropic hydrodynamics are derived in the following subsections in a general (2+1)-dimensional space-time expansion subject to the constraint of longitudinal boost-invariance.

General parametrizations of the fluid four-velocity and the four-vector $z^\mu$ are obtained by a sequence of Lorentz transformations applied to the local rest frame basis vectors~\cite{Ryblewski:2011aq,Martinez:2012tu}:
\begin{eqnarray}
u^\mu &=& \left(u_0 {\rm cosh}\vartheta,u_x,u_y,u_0 {\rm sinh}\vartheta\right) \, , \\
z^\mu &=& \left({\rm sinh}\vartheta,0,0,{\rm cosh}\vartheta\right) .
\end{eqnarray}
Here $u_x$ and $u_y$ are the two transverse components of the fluid four-velocity in the longitudinal rest frame, and $\vartheta$ is the longitudinal fluid rapidity. The normalization condition $u^\mu u_\mu=1$ implies that
\begin{equation}
u^2_0=1+{\bf u}^2_\perp \; .
\end{equation}
We denoted the two-component vector in the transverse plane as ${\bf u}_\perp\equiv\left(u_x,u_y\right)$. In heavy-ion collision phenomenology it is convenient to transform to Milne ($\tau$-$\varsigma$) coordinates where $\tau=\sqrt{t^2-z^2}$ is the longitudinal proper time and $\varsigma=\tanh^{-1}\left(z/t\right)$ is the space-time rapidity. This coordinate system is particularly well-adapted to longitudinally boost-invariant systems where $\vartheta=\varsigma$; the necessary differential operators then reduce to \cite{Florkowski:2010cf,Martinez:2012tu}
\begin{equation}\label{diff_operators}
\begin{split}
D&=u^\mu \partial_\mu=u_0\partial_\tau+{\bf u}_\perp\cdot {\bf \nabla}_\perp \; , \\
\theta &=\partial_\mu u^\mu=\partial_\tau u_0+{\bf \nabla}_\perp\cdot {\bf u}_\perp
+\frac{u_0}{\tau} \equiv \theta_\perp + \frac{u_0}{\tau} \;, \\
D_{\rm L}&=z^\mu\partial_\mu=\frac{\partial_\varsigma}{\tau} \;,\\
\theta_{\rm L}&=\partial_\mu z^\mu=0 \;,\\
u_\nu D_{\rm L}z^\nu&=u_\nu z^\mu\partial_\mu z^\nu=\frac{u_0}{\tau} \;.
\end{split}
\end{equation}

\subsection{Zeroth moment of the Boltzmann equation}
\label{subsec:bem0}

The zeroth moment of the Boltzmann equation gives
\begin{equation}
\partial_\mu j^\mu = D{\cal N}+{\cal N}\theta+\partial_\mu \V^\mu={\cal C} \label{eq:be0_tmp} ,
\end{equation}
where we used the relationship (\ref{eq24}). For theories where only elastic processes are important we have particle number conservation (${\cal C}=0$). However, in non-equilibrium quantum field theories inelastic processes become important~\cite{Biro:1993qt,Baier:2000sb,Xu:2004mz,Xu:2007aa}, which means that there is particle production/annihilation; in general, there must then be a non-vanishing source term, ${\cal C}\ne 0$. It should be pointed out that net baryon number is conserved, but there is gluon production so if $f$ is the gluon distribution there is no conservation of gluon number. Denoting the action of the local time derivative $D$ by a dot, Eq.~(\ref{eq:be0_tmp}) can be written as an equation of motion for the rest frame particle density ${\cal N}$:
\begin{equation}
\label{eq:be0}
\dot{\cal N}=-{\cal N}\theta-\partial_\mu \V^\mu+{\cal C}.
\end{equation}
%

\subsection{First moment of the Boltzmann equation}
\label{subsec:bem1}

The first moment of the Boltzmann equation is equivalent to the requirement of energy-momentum conservation: $\partial_\mu T^{\mu\nu}=0$. With the viscous anisotropic hydrodynamic decomposition of $T^{\mu\nu}$ given in (\ref{eq24}) this conservation law yields
\begin{equation}
\begin{split}
\partial_\mu T^{\mu\nu}&=u^\nu D({\cal E}{+}{\cal P}_\perp{+}\Pit)
  + u^\nu ({\cal E}{+}{\cal P}_\perp{+}\Pit)\theta
  + ({\cal E}{+}{\cal P}_\perp{+}\Pit) D u^\nu
   - \partial^\nu({\cal P}_\perp{+}\Pit) \\
&+ z^\nu D_L({\cal P}_{\rm L}{-}{\cal P}_\perp)
  + z^\nu({\cal P}_{\rm L}{-}{\cal P}_\perp)\theta_{\rm L} 
  + ({\cal P}_{\rm L}{-}{\cal P}_\perp)D_L z^\nu + \partial_\mu\pit^{\mu\nu}=0.
\end{split}
\end{equation}
Projecting these four equations on the fluid four-velocity yields an equation of motion for the rest frame energy density ${\cal E}$: 
\begin{equation}
u_\nu\partial_\mu T^{\mu\nu}=\dot{{\cal E}} + ({\cal E}{+}{\cal P}_\perp{+}\Pit)\theta
+ ({\cal P}_{\rm L}{-}{\cal P}_\perp)u_\nu D_L  z^\nu+u_\nu\partial_\mu\pit^{\mu\nu}=0.
\label{eq:par}
\end{equation}
The projections $\Delta^\alpha_{\ \nu} \partial_\mu T^{\mu\nu}$ transverse to $u^\mu$ yield equations of motion for the fluid four-velocity $u^\mu$:
\begin{eqnarray}
\label{eq:perp} 
&&\Delta^\alpha_{\ \nu} \partial_\mu T^{\mu\nu}
= ({\cal E}{+}{\cal P}_\perp{+}\Pit)\dot{u}^\alpha 
 - \nabla^\alpha ({\cal P}_\perp{+}\Pit)
+ \Delta^\alpha_{\ \nu}\partial_\mu\pit^{\mu\nu}
\\\nonumber
&&\qquad\qquad + z^\alpha D_L ({\cal P}_{\rm L}{-}{\cal P}_\perp)
+ z^\alpha ({\cal P}_{\rm L}{-}{\cal P}_\perp)\theta_{\rm L} 
+  ({\cal P}_{\rm L}{-}{\cal P}_\perp)D_L z^\alpha 
-({\cal P}_{\rm L}{-}{\cal P}_\perp)u^\alpha u_\nu D_L z^\nu=0.
\end{eqnarray}
In addition to the spatial pressure gradients in the second and third term on the right hand side, which represent the standard hydrodynamic force including viscous corrections from $\dft$, we note the appearance in the second line of additional driving terms proportional to the pressure anisotropy ${\cal P}_{\rm L}{-}{\cal P}_\perp$ introduced by the momentum-space deformation in $f_\mathrm{aniso}$.

Equations (\ref{eq:par}) and (\ref{eq:perp}) are the fundamental equations of relativistic viscous anisotropic hydrodynamics. Due to the normalization condition $u^\mu u_\mu=1$, one of the four equations in (\ref{eq:perp}) is redundant. We can thus ignore the $\alpha{\,=\,}0$ equation in (\ref{eq:perp}). For longitudinally boost-invariant systems (see (\ref{diff_operators})) Eqs.~(\ref{eq:par}) and (\ref{eq:perp}) simplify to
\begin{equation}
\label{eq:macroBEm1}
\begin{split}
\dot{\cal E}+({\cal E}{+}{\cal P}_\perp{+}\Pit)\theta+({\cal P}_{\rm L}{-}{\cal P}_\perp)\frac{u_0}{\tau}+u_\nu\partial_\mu\pit^{\mu\nu}&=0, \\
 ({\cal E}{+}{\cal P}_\perp{+}\Pit)\dot{u}_x + \partial_x ({\cal P}_\perp{+}\Pit)
 +u_x (\dot{\cal P}_\perp{+}\dot{\Pit}) + ({\cal P}_\perp{-}{\cal P}_{\rm L})\frac{u_0 u_x}{\tau}
 -\Delta^{1\nu}\partial^\mu\pit_{\mu\nu} &=0, \\
 ({\cal E}{+}{\cal P}_\perp{+}\Pit)\dot{u}_y + \partial_y ({\cal P}_\perp{+}\Pit)
 +u_y (\dot{\cal P}_\perp{+}\dot{\Pit}) + ({\cal P}_\perp{-}{\cal P}_{\rm L})\frac{u_0 u_y}{\tau}
 -\Delta^{2\nu}\partial^\mu\pit_{\mu\nu} &=0,
\end{split}
\end{equation}
where dots denote the operation $D=u_0\partial_\tau + {\bf u}_\perp\cdot {\bf \nabla}_\perp$, and we substituted $\nabla_\mu=\partial_\mu - u_\mu D$ in the last two equations. These {\em viscous anisotropic} hydrodynamic ({\sc vaHydro}) equations differ from the {\em anisotropic} hydrodynamic ({\sc aHydro}) equations given in Eq.~(2.47) of Ref.~\cite{Martinez:2012tu} only by the terms involving the additional bulk and shear viscous pressure components $\Pit$ and $\pit^{\mu\nu}$. Some additional rearrangements, using the definition of the velocity shear tensor $\sigma^{\mu\nu}\equiv\nabla^{\langle\mu}u^{\nu\rangle}$, yield the following three equations of motion:
\begin{eqnarray}
\label{eq:aniso_eqns1}
\!\!\!\!&&\dot{\cal E} = -({\cal E}{+}{\cal P}_\perp)\theta_\perp
     - ({\cal E}{+}{\cal P}_{\rm L})\frac{u_0}{\tau}-\Pit\theta+\pit^{\mu\nu}\sigma_{\mu\nu},
\nonumber\\
\!\!\!\!&&({\cal E}{+}{\cal P}_\perp{+}\Pit)\dot{u}_\perp = -\partial_\perp({\cal P}_\perp{+}\Pit)
    - u_\perp(\dot{\cal P}_\perp{+}\dot{\Pit}) 
    - u_\perp({\cal P}_\perp{-}{\cal P}_{\rm L})\frac{u_0}{\tau} 
    + \left(\frac{u_x\Delta^1_{\ \nu}{+}u_y\Delta^2_{\ \nu}}{u_\perp}\right)
    \partial_\mu\pit^{\mu\nu},
\nonumber\\
\!\!\!\!&&({\cal E}{+}{\cal P}_\perp{+}\Pit)u_\perp\dot\phi_u 
= - D_\perp({\cal P}_\perp{+}\Pit) 
   - \frac{u_y\partial_\mu\pit^{\mu 1}{-}u_x\partial_\mu\pit^{\mu 2}}{u_\perp}.
\end{eqnarray}
Here $u_\perp$ is the magnitude of the transverse flow vector ${\bf u}_\perp$, $\phi_u=\tan^{-1}(u_y/u_x)$ its azimuthal direction, and we used the following shorthand notation: 
\begin{eqnarray}
\label{eq46}
&& {\bf \nabla}_\perp\equiv (\partial_x,\partial_y),\quad 
     \partial_\perp = \frac{{\bf u}_\perp\cdot {\bf \nabla}_\perp}{u_\perp},\quad
     D_\perp = \hat{{\bf z}}\cdot(\hat{{\bf u}}_\perp \times \nabla_\perp)
     = \frac{u_x\partial_y-u_y\partial_x}{u_\perp}.
\end{eqnarray}    
The first equation (\ref{eq:aniso_eqns1}) expresses energy conservation; we rewrote the last term using the transversality of $\pit^{\mu\nu}$, $\Delta^\mu_{\ \alpha}\pit^{\alpha\nu}{\,=\,}\pit^{\mu\nu}{\ \leftrightarrow\ }u_\mu\pit^{\mu\nu}{\,=\,}0$. The second and third equation in (\ref{eq:aniso_eqns1}) are obtained from the linear combinations $(u_1\Delta^1_{\ \nu}{+}u_2\Delta^2_{\ \nu}) \partial_\mu T^{\mu\nu}$ and $(u_2\Delta^1_{\ \nu}{-}u_1\Delta^2_{\ \nu}) \partial_\mu T^{\mu\nu}$, respectively, and describe the transverse dynamics. 

The system of equations (\ref{eq:be0}) and (\ref{eq:aniso_eqns1}) is not closed. To close it we need the equations of state that give (i) ${\cal P}_\mathrm{eq}$ in terms of the energy density ${\cal E}$ and (ii) ${\cal P}_{\perp,L}$ in terms of the equilibrium pressure ${\cal P}_\mathrm{eq}$, as well as additional evolution or ``transport'' equations for the dissipative flows $\V^\mu$, $\Pit$, and $\pit^{\mu\nu}$. To derive the latter we begin in Sec.~\ref{sec:expansion} with a general expansion of the distribution function around ``anisotropic equilibrium", followed in Sec.~\ref{sec:residual_moments} by a derivation of the evolution equations for the dissipative flows. We then close the set of equations for the desired dissipative flows by performing the 14-moment approximation scheme in Sec.~\ref{sec:14_moment}.

\section{Expansion of the one-particle distribution function around an anisotropic background}
\label{sec:expansion}

\subsection{Leading-order ansatz}
\label{subsec:LO_dist}

In this paper we consider systems that are, to leading order, spheroidal in momentum-space in the local rest frame. This is accomplished by introducing an ansatz for the leading-order one-particle distribution function in the local rest frame (LRF) with two different effective temperatures in longitudinal and transverse directions:
\begin{equation}
f=f_0\left(\sqrt{\frac{m^2+p^2_\perp}{\Lambda^2_\perp}+\frac{p^2_z}{\Lambda^2_\parallel}}\right) \; . \label{eq:lo_df}
\end{equation}
We now define dimensionless parameters $\lambda_i\in (0,1]$ ($i=\perp,\parallel$) such that $\Lambda_i\equiv\lambda_i\Lambda$ where $\Lambda$ is the effective temperature of the partons. A family of concentric spheroids is defined by surfaces of constant
\begin{equation}
\lambda^2=\frac{p^2_x+p^2_y}{\lambda^2_\perp}+\frac{p^2_z}{\lambda^2_\parallel} \, ,
\end{equation}
where $p_x$, $p_y$, and $p_z$ are the Cartesian LRF momentum components. The semi-axes of the spheroid have lengths of $\lambda_\perp\lambda$, $\lambda_\perp\lambda$, and $\lambda_\parallel\lambda$. Without loss of generality we let $\lambda_\perp=1$. The quadric surface 
\begin{equation}\label{quadric_surface}
\frac{p^2_x+p^2_y}{\lambda^2}+\frac{p^2_z}{\lambda^2_\parallel\lambda^2}=1 \, ,
\end{equation}
can be parameterized in terms of spherical coordinates by: $p_x=\lambda{\rm sin}\theta{\rm cos}\phi$, $p_y=\lambda{\rm sin}\theta{\rm sin}\phi$, and $p_z=\lambda_\parallel\lambda{\rm cos}\theta$. (For fixed $\lambda_\parallel$, $\lambda$ generates a confocal family of ellipsoids of revolution.) 
The Jacobian for this coordinate system is
\begin{equation}\label{jocabian}
J(\lambda,\theta,\phi;\lambda_\parallel)=\lambda_\parallel\lambda^2{\rm sin}\theta \; .
\end{equation}
For massless particles, the momentum-space integration measure becomes
\begin{equation}
\label{eq51}
 dP \equiv \frac{1}{(2\pi)^3} \frac{d^3p}{E_p} = \frac{1}{(2\pi)^3}\, \lambda_\parallel
 \frac{\lambda^2 d\lambda\, d(\cos\theta)\,d\phi}
        {\sqrt{\lambda^2\sin^2\theta{+}\lambda^2_\parallel\lambda^2\cos^2\theta}}
 =   \frac{\lambda_\parallel}{(2\pi)^3} \lambda d\lambda\, d\phi\, 
      \frac{d(\cos\theta)}{\sqrt{\sin^2\theta{+}\lambda^2_\parallel\cos^2\theta}}.
\end{equation}

The momentum dependence of a spheroidal distribution function can thus be parametrized by two degrees of freedom, $\Lambda$ and $\lambda_\parallel$. Writing $\lambda^2_\parallel\equiv (1+\xi)^{-1}$ one obtains the Romatschke-Strickland (RS) distribution function \cite{Romatschke:2003ms} 
\begin{equation}
\label{eq52}
f_\rs(x,p;\xi,\Lambda,\mu)=f_0\left(\frac{\sqrt{{\bf p}^2_\perp{+}(1{+}\xi)p^2_z{+}m^2}}{\Lambda};\frac{\tilde{\mu}}{\Lambda}\right)\;,
\end{equation}
where $f_0$ has the same functional dependence as in local thermal equilibrium\footnote{Later 
    in this paper we will set the particle mass $m$ to zero. Then the specific functional 
    dependence of $f_0$ on the LRF energy is mostly irrelevant for our theory, since it 
    enters only through the thermodynamic integrals (\ref{eq:thermo_funcs}) below whose 
    actual values factors out in the present work.}
and $\xi$, $\Lambda$, and $\tilde{\mu}$ are all functions of $x$.\footnote{Since we propose to incorporate bulk viscous corrections perturbatively, 
it is not necessary to incorporate them in the leading-order distribution function as originally proposed in \cite{Martinez:2012tu}.
As a result, we can take $\Phi$ in Sec.~2I of \cite{Martinez:2012tu} to be zero.}

This can be written covariantly as
\begin{equation}
\label{eq53}
f_\rs(x,p;\xi,\betat,\alphat)=f_0\!\left(\betat(x)\sqrt{p^\mu \Xi_{\mu\nu}(x;\xi) p^\nu};\,\alphat(x)\right)
\equiv f_0\!\left(\betat E_\mathrm{RS}{-}\alphat\right) ,
\end{equation}
where $\betat\equiv 1/\Lambda$, $\alphat\equiv\tilde{\mu}/\Lambda$, $E_\mathrm{RS}\equiv\sqrt{p^\mu \Xi_{\mu\nu}p^\nu}$, and $\Xi_{\mu\nu}$ is an $x$-dependent symmetric tensor given by
\begin{equation}
\label{eq:Xi}
\Xi_{\mu\nu}(x;\xi)=u_\mu(x) u_\nu(x) + \xi(x) z_\mu(x) z_\nu(x)\,.
\end{equation}
For later convenience, we introduce $\z_0\equiv\betat\sqrt{p^\mu \Xi_{\mu\nu}p^\nu}{-}\alphat$ and then consider small deviations $\delta\z=\z{-}\z_0$. We can then formally expand $f$ in a power series
\begin{equation}
\label{eq55}
f(\z)=f_0(\z_0)+\sum_{n=1}^\infty \frac{1}{n!}f^{(n)}_0(\z_0)\left(\delta\z\right)^n\,;
\end{equation}
the first term is the ``anisotropic equilibrium'' RS function while the second term is a series expansion for the full $\delta\tilde{f}$ which determines the dissipative currents $\V^\mu$, $\Pit$, and $\pit^{\mu\nu}$ in Eq.~(\ref{eq24}). 

We can also expand around local thermal equilibrium by expanding in deviations from $y_0\equiv\beta u_\mu p^\mu-\alpha$:
\begin{equation}
\label{eq56}
f(y)=f_0(y_0)+\sum_{n=1}^\infty \frac{1}{n!}f^{(n)}_0(y_0)\left(\delta y\right)^n.
\end{equation}
Now the first term is the local equilibrium distribution, and the sum is a series expansion for the full $\delta f$ which determines the dissipative currents in Eq.~(\ref{eq:fluid_fields_iso}). 

Equating $T^{\mu\nu}$ from Eq.~(\ref{eq:fluid_fields_iso}) and Eq.~(\ref{eq24}) we find 
\begin{eqnarray}
\label{eq:pi_pit}
&&\pi^{\mu\nu} - \pit^{\mu\nu} = \left({\cal P}_\perp{-}{\cal P}_L\right)\,\frac{x^\mu x^\nu{+}y^\mu y^\nu{-}2 z^\mu z^\nu}{3}\,,
\\\nonumber
&&\Pi -\Pit = \frac{2{\cal P}_\perp+{\cal P}_L}{3}-{\cal P}_\mathrm{eq}\,.
\end{eqnarray}
These relate the series expansions (\ref{eq55}) and (\ref{eq56}). For massless particles both terms on the left hand side of the last equation are zero, and so is the right hand side; it thus yields no information. The first equation in (\ref{eq:pi_pit}), however, will prove useful in deriving the evolution equation for $\pit^{\mu\nu}$.

\section{Equations of motion for the residual dissipative flows}
\label{sec:residual_moments}

We now use the Boltzmann equation for $\dft$ to derive exact evolution equations for the residual currents $\Pit$, $\V^\mu$ and $\pit^{\mu\nu}$~\cite{Denicol:2012cn,Denicol:2010xn}. Starting from their kinetic definitions one finds
\begin{eqnarray}
\dot{\Pit}&=&-\frac{m^2}{3}D\int dP\,\dft\;,\label{eq:evolve_res_moments}\\
\dot{\V}^{\langle\mu\rangle} &=&\Delta^\mu_\nu D\int dP\,p^{\langle\nu\rangle}\dft\;,\\
\dot{\pit}^{\langle\mu\nu\rangle}&=&\Delta^{\mu\nu}_{\alpha\beta}D\int dP\, p^{\langle\alpha}p^{\beta\rangle}\dft\;.\label{eq:evolve_res_moments2}
\end{eqnarray}
Moving the convective derivative $D=u\cdot\partial$ on the right hand side under the integral, we see that it acts nontrivially on $E=p\cdot u(x)$, the projectors $\Delta^{\nu_1\cdots\nu_\ell}_{\alpha_1\cdots\alpha_\ell}$ implicit in the definition of $p^{\langle\nu_1}\cdots p^{\nu_\ell\rangle}$ (which also depend on $u(x)$),\footnote{Since derivatives of $u$ are orthogonal to $u$, derivatives of 
    $\Delta^{\nu_1\cdots\nu_\ell}_{\alpha_1\cdots\alpha_\ell}$ (which is a purely 
    space-like projector) always contain factors pointing parallel to $u$; therefore all 
    these terms get annihilated by the space-like projector 
    $\Delta^{\mu_1\cdots\mu_\ell}_{\nu_1\cdots\nu_\ell}$ in front of the integral 
    in (\ref{eq:evolve_res_moments}).}
and the residual distribution function $\dft(x,p)$. This last term, $D(\dft)\equiv\delta\dot{\tilde{f}}$,  is obtained from the Boltzmann equation Eq.~(\ref{eq:be}) which can be written in the form
\begin{equation} 
\label{eq:evolve_be}
\delta\dot{\tilde{f}}=-\dot{f}_\rs - \frac{1}{E}\Bigl(p{\cdot}\nabla (f_\rs{+}\dft)-C[f]\Bigr) .
\end{equation}
Substituting this together into the expressions Eq.~(\ref{eq:evolve_res_moments})-(\ref{eq:evolve_res_moments2}) one obtains the following equations of motion: 
\begin{align}
-\frac{3}{m^2}\dot{\Pit}-{\cal C}_{-1}& =\dot{\beta}_\mathrm{RS}\J_{0,0,1}
+\frac{\betat}{2}\dot{\xi}\J^{zz}_{0,0,-1}-\dot{\alpha}_\mathrm{RS}\J_{0,0,0}
+\left(\I_{0,1,0}-\I_{0,0,0}\right)\theta
\notag \\
&-\left(\I^{zz}_{-2,0,0}-\I^{xx}_{-2,0,0}\right)z^\mu z^\nu\sigma_{\mu\nu}
-\Pit\theta-\nabla_\mu\left\langle E^{-1}p^{\langle\mu\rangle}\right\rangle_{\tilde\delta}-\left\langle E^{-2}p^{\mu}p^\nu\right\rangle_{\tilde\delta}\nabla_\mu u_\nu ,  \label{Scalar_n} 
\end{align}
\begin{align}
\dot{\V}^{\left\langle \mu \right\rangle }-{\cal C}^{\left\langle \mu
\right\rangle }_{-1}& =\betat\J^{ij}_{1,0,-1}X^\mu_iX^\nu_j\dot{u}_\nu 
+\J^{ij}_{-1,0,1}X^\mu_iX^\nu_j\nabla_\nu\beta_\rs
+\frac{\betat}{2}\J^{ijzz}_{-1,0,-1}X^\mu_iX^\nu_j\nabla_\nu\xi
\notag \\
&-\betat\xi\J^{ijkz}_{-1,0,-1}X^\mu_iX^\nu_jX^\lambda_k\nabla_\nu z_\lambda
-\J^{ij}_{-1,0,0}X^\mu_iX^\nu_j\nabla_\nu\alphat
\notag \\
&-\V^\lambda\nabla_\lambda u^\mu-\V^\mu\theta
-\Delta^\mu_\alpha\nabla_\beta\left\langle E^{-1}p^{\langle\alpha\rangle}p^{\langle\beta\rangle}\right\rangle_{\tilde\delta}
-\left\langle E^{-2}p^{\langle\mu\rangle}p^{\alpha}p^{\beta}\right\rangle_{\tilde\delta}\nabla_\alpha u_\beta,  \label{Vector_n}
\end{align}
\begin{align}
\label{Tensor_n}
\dot{\pit}^{\langle\mu\nu\rangle }-{\cal C}^{\langle\mu\nu\rangle}_{-1}
& =\dot{\beta}_\rs\left(\J^{ij}_{0,0,1} X^{(\mu}_iX^{\nu )}_j+\Delta^{\mu\nu}\J_{2,1,1}\right)
-\dot{\alpha}_\rs\left(\J^{ij}_{0,0,0}X^{(\mu}_iX^{\nu )}_j+\Delta^{\mu\nu}\J_{2,1,0}\right)
\notag \\
&+\frac{\betat}{2}\dot{\xi}\left(\J^{ijzz}_{0,0,-1}X^{(\mu}_iX^{\nu )}_j+\Delta^{\mu\nu}
\J^{zz}_{2,1,-1}\right) 
\notag \\
&-\betat\xi\left(\J^{ijkz}_{0,0,-1}X^{(\mu}_iX^{\nu )}_jX^\lambda_k+\Delta^{\mu\nu}\J^{zz}_{2,1,-1}z^\lambda\right)(\dot{z}_\lambda+u^\alpha\nabla_\lambda z_\alpha)
\nonumber \\
&+\betat\left(\J^{ijk\ell}_{0,0,-1}X^{(\mu}_iX^{\nu )}_jX^\lambda_kX^\alpha_\ell 
  +\Delta^{\mu\nu}\J^{k\ell}_{2,1,-1}X^\lambda_k X^\alpha_\ell\right)
   \nabla_\lambda u_\alpha
\notag \\
&-\frac{5}{3}\pit^{\mu\nu}\theta
-2\pit^{\langle\mu}_\lambda\sigma^{\nu\rangle\lambda} 
+2\pit^{\langle\mu}_\lambda\omega^{\nu\rangle\lambda}
+2\Pit\sigma^{\mu\nu}
-\Delta^{\mu\nu}_{\alpha\beta}\nabla_\lambda\left\langle
E^{-1}p^{\langle\alpha\rangle}p^{\langle\beta\rangle}p^{\langle\lambda\rangle}\right\rangle_{\tilde\delta}
\nonumber \\
&-\left\langle E^{-2}p^{\langle\mu}p^{\nu\rangle}p^{\langle\alpha\rangle}p^{\langle\beta\rangle}\right\rangle_{\tilde\delta}\nabla_\alpha u_\beta \,.  
\end{align}%
These equations involve the generalized collision terms \cite{Denicol:2012cn}
\begin{equation}
\label{General_Col_term}
{\cal C}_{r}^{\left\langle \mu _{1}\cdots \mu _{\ell }\right\rangle }=\int dP E^r 
p^{\left\langle \mu_1\right. }\cdots p^{\left. \mu_\ell\right\rangle} C[f] \,,
\end{equation}
with $r=-1$. Here $\sigma^{\mu\nu}$ is the velocity shear tensor defined earlier, $\omega^{\mu\nu}\equiv\nabla^{[\mu}u^{\nu]}$ is the vorticity tensor, and we introduced the following auxiliary thermodynamic functions (with $E_\mathrm{RS}$ defined in Eq.~(\ref{eq53})):
\begin{equation}
\label{eq:thermo_funcs}
\begin{split}
\I^{i_1\cdots i_\ell}_{nqr}&\equiv\frac{(-1)^q}{(2q+1)!!}\int dP E^{n-2q}E^r_\rs \left(\Delta^{\alpha\beta}p_\alpha p_\beta\right)^q 
p_{i_1}\cdots p_{i_\ell} f_\rs \;,\\
\J^{i_1\cdots i_\ell}_{nqr}&\equiv\frac{(-1)^q}{(2q+1)!!}\int dP E^{n-2q}E^r_\rs \left(\Delta^{\alpha\beta}p_\alpha p_\beta\right)^q 
p_{i_1}\cdots p_{i_\ell} f_\rs\tilde{f}_\rs \;.
\end{split}
\end{equation}
Closing this system of equations requires an approximation for $\dft$ in order to evaluate the terms $\langle \,\cdots\, \rangle_{\tilde\delta}$ on the r.h.s. of Eqs.~(\ref{Scalar_n})-(\ref{Tensor_n}). This step will be described in the next section. We point out that the standard evolution equations for the irreversible currents do not involve time derivatives (in the comoving frame) of the hydrodynamic fields. We can ensure this here, too, by replacing the time derivatives $\dot{\alpha}_\rs$, $\dot{\beta}_\rs$, and $\dot{u}^\mu$ appearing on the r.h.s. of Eqs.~(\ref{Scalar_n})-(\ref{Tensor_n}) by spatial gradients, using the equations of anisotropic hydrodynamics (\ref{eq:be0}), (\ref{eq:par}), and (\ref{eq:perp}).
The only difference between the derivation of the equations of motion for the irreducible moments $\Pit$, $\V^\mu$, and $\pit^{\mu\nu}$ defined here and that presented in \cite{Denicol:2010xn} for the corresponding irreducible currents $\Pi$, $V^\mu$, and $\pi^{\mu\nu}$ that arise in the expansion (\ref{eq:iso_exp}) around local equilibrium is the occurrence of the spheroidally deformed distribution $f_\rs$ instead of the equilibrium distribution $f_0$ on the right hand side of the Boltzmann equation (\ref{eq:evolve_be}). This modification is responsible for the occurrence of the factors $E_\mathrm{RS}$ in the integrands of Eq.~(\ref{eq:thermo_funcs}), and of the factors $\betat$ and $\alphat$ in the equations of motion. Note that for massless particles the spheroidal deformation effects in $f_\rs$ can be factored out from the integrals (\ref{eq:thermo_funcs}), the remainder being given by the standard equilibrium thermodynamic integrals (see Eqs.~(\ref{eq:alpha2-2}) and (\ref{eq:alpha2-2a})). For details of the derivation of Eqs.~(\ref{Scalar_n})-(\ref{Tensor_n}) we refer to Appendix~\ref{appa}.

\section{The 14-moment approximation}
\label{sec:14_moment}
\subsection{Truncation procedure}
\label{truncation}

\noindent 
We here use the 14-moment approximation of Grad \cite{Grad:1949} and Israel and Stewart \cite{Israel:1979wp}:
\begin{equation} \label{q_expanded}
\begin{split}
\frac{\dft}{f_\rs\tilde{f}_\rs} &\equiv \alpha - \beta_{\mu} p^{\mu} 
+ w_{\mu \nu} p^{\mu} p^{\nu} \\
&= \alpha - \beta E + E^2 w - \frac{w}{3} \Delta^{\mu \nu} p_{\mu} p_{\nu}
+ \left(2 E w_{\langle\mu\rangle} - v_{\langle\mu\rangle}\right)p^{\langle\mu\rangle}
+ w_{\langle \mu \nu \rangle} p^{\langle\mu} p^{\nu\rangle}\, ,
\end{split}
\end{equation}
where $\beta\equiv\beta^{\mu} u_{\mu}$ and $w \equiv w^{\mu \nu} u_{\mu} u_{\nu}= -w^{\mu \nu} \Delta_{\mu \nu}$ are scalars, $v^{\langle\mu\rangle} \equiv \beta^{\langle \mu \rangle}$ and $w^{\langle\mu\rangle} \equiv w^{\langle \mu \rangle \beta} u_{\beta}$ are four-vectors orthogonal to $u^{\mu}$, and $ w^{\langle \mu \nu \rangle}$ is the traceless and locally spacelike part of $w^{\mu \nu}$. Inserting Eq.~(\ref{q_expanded}) into the definition (\ref{eq:jmuTmunu}) of the residual dissipative flows we find
\begin{eqnarray}
\label{inv1}
0&=&\alpha\J_{1,0}-\beta\J_{2,0}+w\left(\J_{3,0}+\J_{3,1}\right)+w_{\langle\mu\nu\rangle}\rho^{\mu\nu}_{10}\;,
\\\label{69} 
0&=&\alpha\J_{2,0}-\beta\J_{3,0}+w\left(\J_{4,0}+\J_{4,1}\right)+w_{\langle\mu\nu\rangle}\rho^{\mu\nu}_{20}\;, 
\\\label{70} 
\Pit&=&\alpha\J_{2,1}-\beta\J_{3,1}+w\left(\J_{4,1}+\frac{5}{3}\J_{4,2}\right)+w_{\langle\mu\nu\rangle}\rho^{\mu\nu}_{30}\;, 
\\\label{71}
\V^\mu &=&2w_{\langle\nu\rangle}\J^{ij}_{1,0}X^\mu_iX^\nu_j 
-v_{\langle\nu\rangle}\J^{ij}_{0,0}X^\mu_iX^\nu_j \;, 
\\\label{72}
0 &=&2w_{\langle\nu\rangle}\J^{ij}_{2,0}X^\mu_iX^\nu_j 
-v_{\langle\nu\rangle}\J^{ij}_{1,0}X^\mu_iX^\nu_j \;, 
\\\label{inv2}
\pit^{\alpha\beta} &=&\alpha\varphi^{\alpha\beta}_{21}-\beta\varphi^{\alpha\beta}_{31}+w\left(\varphi^{\alpha\beta}_{41}+\varphi^{\alpha\beta}_{42}\right)+\lambda^{\alpha\beta\mu\nu}w_{\mu\nu}\;,
\end{eqnarray}
where the first two equations are the constraints from employing the Landau matching conditions $\langle E\rangle_{\tilde\delta} = \langle E^2\rangle_{\tilde\delta} = 0$ and the l.h.s of Eq.~(\ref{72}) is zero due to the definition of the fluid velocity in the Landau frame which states that $\langle E p^{\langle\mu\rangle}\rangle=0$. The auxiliary tensors $\rho^{\mu\nu}_{nq}$ and $\varphi^{\alpha\beta}_{nq}$ are defined as
\begin{eqnarray}
\label{74}
\rho^{\mu\nu}_{nq} &\equiv& \left(\J^{zz}_{nq}-\J^{xx}_{nq}\right)z^\mu z^\nu\;, 
\\\label{75}
\varphi^{\alpha\beta}_{nq}&\equiv&
\left(\frac{(2q+1)!!}{3}\J_{nq}-\J^{xx}_{n-2,q-1}\right)\Delta^{\alpha\beta}
+\left(\J^{zz}_{n-2,q-1}-\J^{xx}_{n-2,q-1}\right)z^\alpha z^\beta\;.
\end{eqnarray}
The ``anisotropic thermodynamic integrals" with two subscripts are defined by $\J^{i_1\cdots i_\ell}_{nq}\equiv\J^{i_1\cdots i_\ell}_{n,q,0}$. We can express the parameters occurring in the second line of Eq.~(\ref{q_expanded}) through the dissipative flows by inverting the relations (\ref{inv1})-(\ref{inv2}). To decouple equations (\ref{71}) and (\ref{72}) is straightforward:
\begin{eqnarray}
v^{\mu} = \mathcal{B}_v^{\mu\nu}\V_\nu, \quad
w^{\mu} = \mathcal{B}_w^{\mu\nu}\V_\nu.
\label{pi_w}
\end{eqnarray}
Here, using the shorthand notation $\D^{i_1\cdots i_\ell}_{nq}\equiv\J^{i_1\cdots i_\ell}_{n+1,q}\J^{i_1\cdots i_\ell}_{n-1,q}-\bigr(\J^{i_1\cdots i_\ell}_{nq}\bigr)^2$,
\begin{eqnarray}
\mathcal{B}_v^{\mu\nu} &\equiv& 
\frac{\J^{xx}_{0,0}}{\D^{xx}_{1,0}}\Delta^{\mu\nu}-\left(\frac{\J^{zz}_{2,0}}{\D^{zz}_{1,0}}
- \frac{\J^{xx}_{2,0}}{\D^{xx}_{1,0}}\right)z^\mu z^\nu \,,
\\
\mathcal{B}_w^{\mu\nu} &\equiv& 
\frac{\J^{xx}_{1,0}}{\D^{xx}_{1,0}}\Delta^{\mu\nu}-\left(\frac{\J^{zz}_{1,0}}{\D^{zz}_{1,0}}
- \frac{\J^{xx}_{1,0}}{\D^{xx}_{1,0}}\right)z^\mu z^\nu\,.
\end{eqnarray}
Equations (\ref{inv1})-(\ref{70}) and (\ref{inv2}) are written in matrix form ${\cal A} \bm{b}=\bm{c}$, where
\begin{eqnarray}
\label{79}
&&{\cal A}\equiv 
 \begin{pmatrix}
\J_{1,0} & -\J_{2,0} & \J_{3,0}+\J_{3,1} &0&0&0&0& 0 &\rho^{zz}_{1,0} 
\\
\J_{2,0} & -\J_{3,0} & \J_{4,0}+\J_{4,1} &0&0&0&0& 0 &\rho^{zz}_{2,0}
\\
\J_{2,1} & -\J_{3,1} & \J_{4,1}+\frac{5}{3}\J_{4,2} &0&0&0&0& 0 &\rho^{zz}_{2,1}
\\  
\varphi^{xx}_{21} & -\varphi^{xx}_{31} & \varphi^{xx}_{41}+\varphi^{xx}_{42}&  \lambda^{1111} & 0 & 0 & \lambda^{1122} & 0 & \lambda^{1133} 
\\
0&0&0&0 & 2\lambda^{1212} & 0 & 0 & 0 & 0 
\\
0&0&0& 0  & 0  & 2\lambda^{1313} & 0 & 0 & 0  
\\
\varphi^{xx}_{21} & -\varphi^{xx}_{31} & \varphi^{xx}_{41}+\varphi^{xx}_{42}&  \lambda^{1122} & 0  & 0 & \lambda^{1111} & 0 & \lambda^{1133} 
\\
0&0&0&  0  & 0  & 0  & 0  & 2\lambda^{1313} & 0 
\\
\varphi^{zz}_{21} & -\varphi^{zz}_{31} & \varphi^{zz}_{41}+\varphi^{zz}_{42}&  \lambda^{1133} & 0  & 0 & \lambda^{1133} & 0 & \lambda^{3333}
 \end{pmatrix} ,
\\ \label{80}
&&\bm{b}\equiv 
 \begin{pmatrix}
  \alpha & \beta & w & w_{11} & w_{12} & w_{13} & w_{22} & w_{23} & w_{33} 
 \end{pmatrix}^T \,,
\\ \label{81}
&&\bm{c}\equiv 
 \begin{pmatrix}
  0& 0 & \Pit & \pit_{11} & \pit_{12} & \pit_{13} & \pit_{22} & \pit_{23} & \pit_{33} 
 \end{pmatrix}^T \,.
\end{eqnarray}
%
Matrix diagonalization yields
\begin{eqnarray} 
\label{abw}
\alpha &=& \mathcal{A}_{\Pi \alpha} \Pit +\mathcal{A}^{\mu\nu}_{\pi\alpha}\pit_{\mu\nu}\, ,\qquad
\beta = \mathcal{A}_{\Pi \beta} \Pit+\mathcal{A}^{\mu\nu}_{\pi\beta}\pit_{\mu\nu}\, ,
\\ \label{VW}
w &=& \mathcal{A}_{\Pi w} \Pit+\mathcal{A}^{\mu\nu}_{\pi w}\pit_{\mu\nu} \, , \qquad
w^{\langle \mu \nu \rangle}  = \mathcal{C}_{\Pi w}^{\mu\nu}\Pit
+ \left(\mathcal{C}_{\pi w}\right)^{\alpha\beta\mu\nu}\pit_{\alpha\beta} \, ;
\end{eqnarray}
the coefficients ${\cal A}_{\Pi\alpha}$, ${\cal A}^{\mu\nu}_{\Pi\alpha}$, etc. are listed in Appendix~\ref{sec:14_coefficients}.
%
%
Defining further
\begin{eqnarray}
\label{82}
\lambda_\Pi &\equiv&\mathcal{A}_{\Pi \alpha} -\mathcal{A}_{\Pi \beta}E+\frac{4}{3}\mathcal{A}_{\Pi w}E^2-\frac{1}{3}\mathcal{A}_{\Pi w}m^2\;,
\\\label{83}
\lambda^{\mu\nu}_\pi &\equiv&\mathcal{A}^{\mu\nu}_{\pi \alpha} -\mathcal{A}^{\mu\nu}_{\pi \beta}E+\frac{4}{3}\mathcal{A}^{\mu\nu}_{\pi w}E^2-\frac{1}{3}\mathcal{A}^{\mu\nu}_{\pi w}m^2\;,
\\\label{84}
\lambda^{\mu\nu}_V &\equiv& 2E\mathcal{B}^{\mu\nu}_w -\mathcal{B}^{\mu\nu}_v\;,\quad
\lambda^{\mu\nu}_\Pi \equiv \mathcal{C}^{\mu\nu}_{\Pi w}\;,\quad
\lambda^{\alpha\beta\mu\nu}_\pi \equiv \mathcal{C}^{\alpha\beta\mu\nu}_{\pi w}\;,
\end{eqnarray}
we write the distribution function expanded around an anisotropic background in the 14-moment approximation as
\begin{equation}
\label{eq:f14moment}
f=f_\rs+\left[\lambda_\Pi\Pit+\lambda^{\mu\nu}_\pi\pit_{\mu\nu}+\lambda_V^{\mu\nu}\V_\nu p_{\langle\mu\rangle}+\left(\lambda_\Pi^{\mu\nu}\Pit+\lambda_\pi^{\mu\nu\alpha\beta}\pit_{\alpha\beta}\right)p_{\langle\mu}p_{\nu\rangle}\right]f_\rs\tilde{f}_\rs\;,
\end{equation}
where $\lambda_\Pi$, $\lambda^{\mu\nu}_\pi$, $\lambda^{\mu\nu}_V$, $\lambda^{\mu\nu}_\Pi$, and $\lambda^{\alpha\beta\mu\nu}_\pi$ are all functions of $u{\cdot}p$, $\alpha_\rs$, $\beta_\rs$, and $\xi$. For a locally isotropic medium, the tensor structures of $\lambda_V^{\mu\nu}$ and $\lambda_\pi^{\mu\nu\alpha\beta}$ have the form of a scalar multiplied by the (isotropic) basis tensors $\Delta^{\mu\nu}$ and $\Delta^{\mu\nu\alpha\beta}$ while $\lambda^{\mu\nu}_\pi$ and $\lambda^{\mu\nu}_\Pi$ vanish identically. The coupling to $\V^\mu$ and $\pit^{\mu\nu}$ is more complicated here than in the isotropic formalism due to the local breaking of rotational invariance at leading-order. This symmetry breaking also manifests itself in an additional scalar contribution to $\dft$ involving the shear stress tensor $\pit^{\mu\nu}$ and a rank-two tensor contribution involving the bulk viscous pressure $\Pit$ (``bulk-shear couplings'').

\subsection{Equations of motion}
\label{subsec:eom}

The evolution equations for the dissipative flows $\Pit$,  $\V^\mu$, and $\pit^{\mu\nu}$ can now be obtained by inserting the closed form of the single-particle distribution function (\ref{eq:f14moment}) into the expectation values $\langle \,\cdots\, \rangle_{\tilde\delta}$ on the r.h.s. of the equations of motion (\ref{Scalar_n})-(\ref{Tensor_n}). After some algebra the relaxation equation (\ref{Scalar_n}) for the bulk viscous pressure takes the form
\begin{equation}
\label{pi_tmp}
\begin{split}
-\frac{3}{m^2}\dot{\Pit}&={\cal C}_{-1}
+{\cal W}
+\beta_{\Pi\perp}\theta+\beta_{\Pi L}z^\mu z^\nu\sigma_{\mu\nu}
-\Pit\theta-\lambda_{\Pi V}^{\mu\nu}\nabla_\mu\V_\nu-\tau_{\Pi V}^{\mu}\V_\mu \\
& \hspace{5mm}
-\delta_{\Pi\Pi}^{\mu\nu}\Pit\nabla_\mu u_\nu-\pit_{\alpha\beta}\delta_{\Pi\pi}^{\mu\nu\alpha\beta}\nabla_\mu u_\nu \; .
\end{split}
\end{equation}
Similarly, we obtain from Eqs.~(\ref{Vector_n}) and (\ref{Tensor_n}) 
\begin{eqnarray}
\label{V_tmp}
\dot{\V}^{\langle\mu\rangle}
&=& {\cal C}_{-1}^{\langle\mu\rangle}+{\cal Z}^{\mu}
-\V^\lambda\nabla_\lambda u^\mu-\V^\mu\theta
-\ell^{\mu\nu}_{V\Pi}\nabla_\nu\Pit-\tau^\mu_{V\Pi}\Pit-\delta^{\mu\nu\alpha\beta}_{VV}\V_\nu\nabla_\alpha u_\beta
\nonumber\\
&& \!\!\!\!\!\!\!\!
+\ell^{\mu\mu\alpha\beta}_{V\pi}\nabla_\nu\pit_{\alpha\beta}
+\tau^{\mu\alpha\beta}_{V\pi}\pit_{\alpha\beta}
\,  ,
\\
\label{pimunu_tmp}
\dot{\pit}^{\left\langle \mu \nu \right\rangle }
&=& {\cal C}_{-1}^{\left\langle\mu \nu \right\rangle }+{\cal K}^{\mu\nu}+{\cal L}^{\mu\nu}
+{\cal H}^{\mu\nu\lambda}\left(\dot{z}_\lambda+u^\alpha\nabla_\lambda z_\alpha\right)
+{\cal Q}^{\mu\nu\lambda\alpha}\nabla_\lambda u_\alpha 
\nonumber\\
&& \!\!\!\!\!\!\!\!
-\frac{5}{3}\pit^{\mu\nu}\theta
-2\pit^{\langle\mu}_\lambda\sigma^{\nu\rangle\lambda} 
+2\pit^{\langle\mu}_\lambda\omega^{\nu\rangle\lambda}
+2\Pit\sigma^{\mu\nu}  
\nonumber\\
&& \!\!\!\!\!\!\!\!
-\ell_{\pi V}^{\mu\nu\alpha\beta}\nabla_\alpha\V_\beta
-\tau_{\pi V}^{\mu\nu\lambda}\V_\lambda
-\Pit\delta_{\pi\Pi}^{\mu\nu\alpha\beta}\nabla_\alpha u_\beta
-\delta_{\pi\pi}^{\mu\nu\alpha\beta\sigma\lambda}\pit_{\sigma\lambda}\nabla_\alpha u_\beta
. \quad
\end{eqnarray}
The dissipative forces ${\cal W}$, ${\cal Z}^{\mu}$ etc. and transport coefficients $\lambda^{\mu\nu}_{\Pi V}$, $\tau^\mu_{\Pi V}$ etc. appearing in Eqs.~(\ref{pi_tmp}), (\ref{V_tmp}), and (\ref{pimunu_tmp}) are tabulated in Appendix~\ref{sec:transport_coefficients}. We note that the tensor coefficients ${\cal Q}^{\mu\nu\lambda\alpha}$ and ${\cal H}^{\mu\nu\lambda}$ are related to the shear viscosity since they couple to derivatives of the fluid four-velocity.\footnote{The four-vector $z^\mu$ 
    is an implicit function of $u^\mu$ since in frames other than the local rest frame it is 
    obtained by a Lorentz boost by $u^\mu$.} 
In an anisotropic plasma the coefficients multiplying the longitudinal and transverse gradients of the fluid four-velocity can be different, implying that there could be two different shear viscosities. This has been pointed out in other contexts, see e.g. the discussion in \cite{Rebhan:2011vd}. It is relatively straightforward to show that in the isotropic limit, $\xi \rightarrow 0$, the transverse and longitudinal shear viscosities are the same. In fact, all of the transport coefficients controlling the evolution of the residual dissipative flows arising from $\dft$ have a tensorial structure that can be decomposed into a transverse and longitudinal part. The former differ from those in~\cite{Denicol:2012cn} since they are expressed in terms of the ``anisotropic thermodynamic integrals"~(\ref{eq:thermo_funcs}) involving $f_\rs$ instead of the equilibrium distribution $f_0$, while the latter vanish in the isotropic limit.

\section{Thermodynamical quantities in ``anisotropic equilibrium"}
\label{sec:aniso_thermo_vars_sec}

We have already mentioned that we will consider macroscopic properties arising from moments of $f_\rs$ as ``thermodynamic'' in nature, although they do not describe an equilibrium state. They gain a standard thermodynamic interpretation in the $\xi\to0$ limit. For completeness, we restate the particle current and energy-momentum tensor:
\begin{equation}\label{tmunu_rs}
\begin{split}
j^\mu_\rs&={\cal N}u^\mu\;,\\
T^{\mu\nu}_\rs&=({\cal E}+{\cal P}_\perp)u^\mu u^\nu-{\cal P}_\perp g^{\mu\nu}+({\cal P}_{\rm L}-{\cal P}_\perp)z^\mu z^\nu \; .
\end{split}
\end{equation}
This is the energy-momentum tensor for LO azimuthally-symmetric anisotropic hydrodynamics \cite{Martinez:2012tu,Ryblewski:2011aq}; it contains the limit of ideal hydrodynamics for $\xi\to 0$. Connecting the ``anisotropic equilibrium" quantities with moments of $f_\rs$ and assuming a gas of massless particles one finds that these quantities can be factored \cite{Martinez:2010sc}:
\begin{equation} \label{eq:thermo_vars}
\begin{split}
{\cal N}(\Lambda,\xi)&=\langle E \rangle_\rs={\cal R}_0(\xi)\,{\cal N}_0(\Lambda), \\
{\cal E}(\Lambda,\xi)&=\langle E^2 \rangle_\rs={\cal R}(\xi)\,{\cal E}_0(\Lambda), \\
{\cal P}_\perp(\Lambda,\xi)&=\langle p^2_\perp \rangle_\rs={\cal R}_\perp(\xi)\,{\cal P}_0(\Lambda),  \\
{\cal P}_{\rm L}(\Lambda,\xi)&=\langle p^2_{\rm L} \rangle_\rs={\cal R}_{\rm L}(\xi)\,{\cal P}_0(\Lambda).
\end{split}
\end{equation}
We note that masslessness is the only assumption needed for this factorization. All anisotropy factors ${\cal R}_i$ are normalized such that they approach unity in the isotropic limit  $\xi\to0$. The second factors ${\cal N}_0$, ${\cal E}_0$, and ${\cal P}_0$ are the isotropic thermodynamic properties of the system which only depend on the functional form of $f_0$ in Eq.~(\ref{eq52}). Matching the energy densities ${\cal E}_0(T)={\cal E}(\xi,\Lambda)$ as described above Eq.~(\ref{eq27}) leads to the ``dynamical Landau matching'' condition $T={\cal R}^{1/4}(\xi)\Lambda$ for the corresponding local equilibrium temperature \cite{Martinez:2010sc}. 

The function ${\cal R}_0$ simply arises from the differences between the spheroidal and spherical Jacobian factors in momentum-space:
\begin{equation}
{\cal R}_0(\xi)=\frac{1}{\sqrt{1+\xi}} \; .
\end{equation}
The ${\cal R}$ function is given by
\begin{equation}
\label{eq:h_solidAngle}
{\cal R}(\xi)={\cal R}_0(\xi)\int \frac{d\Omega}{4\pi} \sqrt{\mathrm{sin}^2\theta+\frac{\mathrm{cos}^2\theta}{1{+}\xi}} 
\end{equation}
and has a simple geometric meaning: It is the normalized surface area of a unit ellipsoid rotated around its minor axis, with semi-axes $1$ and $(1+\xi)^{-1}$. All of the ${\cal R}$ functions in Eq.~(\ref{eq:thermo_vars}) can be computed analytically~\cite{Rebhan:2008uj,Martinez:2009ry}:
\begin{subequations}
\begin{align}
{\cal R}(\xi) &= \frac{1}{2}\left(\frac{1}{1+\xi} +\frac{\arctan\sqrt{\xi}}{\sqrt{\xi}} \right) \, , \\
{\cal R}_\perp(\xi)  &= \frac{3}{2 \xi} \left( \frac{1+(\xi^2{-}1){\cal R}(\xi)}{\xi + 1}\right) \, , \\
{\cal R}_L(\xi) &= \frac{3}{\xi} \left( \frac{(\xi{+}1){\cal R}(\xi)-1}{\xi+1}\right) \, .
\end{align}
\label{eq:rfuncs}
\end{subequations}
Since it was possible in anisotropic equilibrium to factor out the deformation effects from the local equilibrium properties ${\cal E}_0$ and ${\cal P}_0$, it is easy to implement the equilibrium equation of state (EoS) ${\cal P}_0({\cal E}_0)$. In this paper we consider a conformal massless gas for which an ideal EoS ${\cal E}_0(\Lambda)=3{\cal P}_0(\Lambda)$ is the appropriate choice.

\section{Dynamical equations of motion for the anisotropic degrees of freedom}
\label{sec:dyneq_dof}

With the RS form as the underlying LO distribution function, it is convenient to evolve the system in terms of the kinematical parameters $\xi$ and $\Lambda$, rather than ${\cal P}_\perp$ and ${\cal P}_{\rm L}$. In the remainder of this paper we deal with a gas of massless particles, such that the factorization of Eq.~(\ref{eq:thermo_vars}) is valid. In this situation the bulk pressure vanishes ($\Pi=\Pi_\rs=\Pit=0$). For simplicity, we consider the case of zero chemical potential\footnote{The case of finite net baryon number, which requires a nonzero 
    baryon chemical potential, has been considered in the context of leading-order 
    (0+1)-dimensional {\sc aHydro} in Ref.~\cite{Florkowski:2012as}.}
and set $V^\nu\sim \nabla (\mu/T)=0$.

In the following we will use the relaxation time approximation (RTA) for the scattering kernel,
\begin{equation}
\label{rta}
C[f]=-\Gamma\ p\cdot u\, \bigl[f({\bf p};\Lambda,\xi){-}f_0(|{\bf p}|;T)\bigr] ,
\end{equation}
where $\Gamma$ is the relaxation rate, which is assumed to be momentum-independent. With this collision kernel, we now derive the explicit form of the equations of motion in RTA for (2+1)-dimensional boost-invariant second-order anisotropic hydrodynamics.

\subsection{Zeroth moment of the Boltzmann equation}
\label{subsec:bem0-2}
%
The scalar collisional moment that is needed on the right hand side of Eq.~(\ref{eq:be0}) can be written as ${\cal C}= \Gamma \left(\langle E\rangle_\rs-\langle E\rangle_0\right)$. Using the factorized expression (\ref{eq:thermo_vars}) for the particle number density ${\cal N}$ in Eq.~(\ref{eq:be0}), the relation ${\cal N}(\Lambda)\sim\Lambda^3$ for massless particles, and the Landau matching relation $T={\cal R}^{1/4}\Lambda$, one can rewrite Eq.~(\ref{eq:be0}) in terms of the RS parameters $\Lambda$ and $\xi$ as \cite{Martinez:2010sc}
\begin{equation}
\label{eq:de1}
\frac{\dot{\xi}}{1+\xi}-6\frac{\dot{\Lambda}}{\Lambda}-2\theta=
2\Gamma\left(1-\sqrt{1{+}\xi}\,{\cal R}^{3/4}(\xi)\right).
\end{equation}
Since we set $\V^\mu=0$, this agrees with the corresponding equation in {\sc aHydro} \cite{Martinez:2010sc}.
\vspace*{-3mm}
\subsection{First moment of the Boltzmann equation}
\label{subsec:bem1-2}
\vspace*{-2mm}
We now use Eqs.~(\ref{eq:thermo_vars}) to rewrite the energy-momentum conservation equations (\ref{eq:aniso_eqns1}) in terms of evolution equations for $\xi$, $\Lambda$, and the flow velocity:
\begin{subequations}
\label{eq:de2}
\begin{eqnarray}
\label{eq:de2a}
&&{\cal R}' \dot\xi + 4 {\cal R}\frac{\dot\Lambda}{\Lambda} = 
- \left({\cal R}{+}\frac{1}{3} {\cal R}_\perp\right) \theta_\perp
- \left({\cal R}{+}\frac{1}{3} {\cal R}_L\right) \frac{u_0}{\tau} 
+\frac{\pit^{\mu\nu}\sigma_{\mu\nu}}{{\cal E}_0(\Lambda)} ,
\\
\label{eq:de2b}
&&\left[3{\cal R}{+}{\cal R}_\perp\right]\dot{u}_\perp = 
-{\cal R}_\perp' \partial_\perp \xi 
- 4  {\cal R}_\perp \frac{\partial_\perp\Lambda}{\Lambda}
-u_\perp\Bigl({\cal R}_\perp' \dot\xi{+}4{\cal R}_\perp \frac{\dot\Lambda}{\Lambda}\Bigr)
\nonumber\\
&&\hspace*{2.8cm}
-u_\perp({\cal R}_\perp{-}{\cal R}_L)\frac{u_0}{\tau}  
+ \frac{3}{{\cal E}_0(\Lambda)}
\left(\frac{u_x\Delta^1_{\ \nu}+u_y\Delta^2_{\ \nu}}{u_\perp}\right) \partial_\mu\pit^{\mu\nu} ,
\\
\label{eq:de2c}
&&\left[3{\cal R}{+}{\cal R}_\perp\right] u_\perp\dot{\phi}_u = 
-{\cal R}_\perp' D_\perp\xi 
- 4 {\cal R}_\perp \frac{D_\perp\Lambda}{\Lambda}
-\frac{3}{{\cal E}_0(\Lambda)}\left(\frac{u_y\partial_\mu\pit^{\mu 1}-u_x\partial_\mu\pit^{\mu 2}}{u_\perp}\right).\qquad
\end{eqnarray}
\label{96}
\end{subequations}
Here all ${\cal R}$ functions depend on $\xi$, and primes denote derivatives with respect to $\xi$.

%
\vspace*{-3mm}
\subsection{Evolution equation for $\pit^{\mu\nu}$}
\label{subsec:eveq}
\vspace*{-3mm}

Equation~(\ref{pimunu_tmp}) will close the system of {\sc vaHydro} equations. The relaxation time approximation (\ref{rta}) for the collision kernel gives
\begin{equation}
{\cal C}^{\langle\mu\nu\rangle}_{-1}
=-\Gamma\Delta^{\mu\nu}_{\alpha\beta}\int dP \,p^\alpha p^\beta (f{-}f_0)
=-\Gamma\Delta^{\mu\nu}_{\alpha\beta}\int dP \,p^\alpha p^\beta \, \delta f 
=-\Gamma\pi^{\mu\nu},
\end{equation}
where $f{-}f_0=\delta f$ gives rise to $\pi^{\mu\nu}$ rather than the anisotropic  shear tensor $\pit^{\mu\nu}$. Equations~(\ref{eq:pi_pit}) and (\ref{tmunu_rs}) give
\begin{equation}
\pi^{\mu\nu}=\left(T^{\mu\nu}_\rs-T^{\mu\nu}_0\right)+\tilde{\pi}^{\mu\nu}\;,
\end{equation}
which lets us replace $\pi^{\mu\nu}$ by $\pit^{\mu\nu}$. Using the matching relation $T={\cal R}^{1/4}(\xi)\Lambda$ and the EoS ${\cal E}_0(\Lambda)=3{\cal P}_0(\Lambda)$ to express $T^{\mu\nu}_0$ in terms of ${\cal P}_0(\Lambda)$ we find
\begin{equation}
T^{\mu\nu}_\rs-T^{\mu\nu}_0=\left[\left({\cal R}(\xi)-{\cal R}_\perp(\xi)
\right)\Delta^{\mu\nu}+\left({\cal R}_{\rm L}(\xi)-{\cal R}_\perp(\xi)\right)
z^\mu z^\nu\right]{\cal P}_0(\Lambda).
\end{equation}
Using these ingredients, as well as the identity \cite{Baier:2006um}
\begin{equation}
\dot{\pit}^{\left\langle \mu \nu \right\rangle}\equiv\Delta^{\mu\nu}_{\alpha\beta}\dot\pit^{\alpha\beta} = D\left(\Delta^{\mu\nu}_{\alpha\beta}\pit^{\alpha\beta}\right) - \pi_{\alpha\beta} \dot{\Delta}^{\mu\nu}_{\alpha\beta} = \dot\pit^{\mu\nu} + 2\dot{u}_\alpha \pit^{\alpha(\mu} u^{\nu)}
\end{equation}
to rewrite the left hand side of Eq.~(\ref{pimunu_tmp}), the evolution equation for $\pit^{\mu\nu}$ becomes
\bqa
\label{eq:pimunu}
\dot{\pit}^{\mu \nu} &=& - 2\dot{u}_\alpha \pit^{\alpha(\mu} u^{\nu)}
-\Gamma\Bigl[
\bigl({\cal P}(\Lambda,\xi){-}{\cal P}_\perp(\Lambda,\xi)\bigr)\Delta^{\mu\nu}
+\bigl({\cal P}_{\rm L}(\Lambda,\xi){-}{\cal P}_\perp(\Lambda,\xi)\bigr) z^\mu z^\nu
+\pit^{\mu\nu}
\Bigr]\qquad
\nonumber\\
&& + {\cal K}^{\mu\nu}+{\cal L}^{\mu\nu} + \, {\cal H}^{\mu\nu\lambda}\left(\dot{z}_\lambda 
     +u^\alpha\nabla_\lambda z_\alpha\right)
     +{\cal Q}^{\mu\nu\lambda\alpha}\nabla_\lambda u_\alpha 
\\
&& -\frac{5}{3}\pit^{\mu\nu}\theta
     -2\pit^{\langle\mu}_\lambda\sigma^{\nu\rangle\lambda}   
     +2\pit^{\langle\mu}_\lambda\omega^{\nu\rangle\lambda}
     -\delta_{\pi\pi}^{\mu\nu\alpha\beta\sigma\lambda}\pit_{\sigma\lambda}\nabla_\alpha u_\beta,
\nonumber
\eqa
where, as announced above, we neglected terms coupling the shear stress to $\Pit$ and $\V^\mu$. 

Equations (\ref{eq:de1}), (\ref{96}), and (\ref{eq:pimunu}) are the main analytic results of this paper and define the {\sc vaHydro} framework for massless systems with longitudinal boost-invariance.

\section{Application: (0+1)-dimensional expansion}
\label{sec:0+1d.}

In this section we present and solve the boost-invariant {\sc vaHydro} equations for a simplified situation without transverse expansion. For transversely homogeneous systems undergoing boost-invariant longitudinal expansion, the Boltzmann equation (\ref{eq:be}) with an RTA collision kernel (\ref{rta}) can be solved exactly \cite{Florkowski:2013lya,Florkowski:2013lza,Baym:1984np}, and this can be used to determine the efficacy of various approximation schemes. The transport coefficients to be used in the different hydrodynamic approximations can be related to the RTA relaxation rate $\Gamma$ by matching to the exact solution at asymptotically late times when the system approaches local momentum isotropy ($\xi\to 0$). Since in this application we assume that the system consists of distinguishable massless particles (Boltzmann statistics), the thermodynamic $\I$ and $\J$ integrals in Eq.~(\ref{eq:thermo_funcs}) are identical.

\subsection{Reduced (0+1)-dimensional {\sc vaHydro} equations} 

In the situation just described there are no transverse derivatives, the comoving time derivative $\dot{A}=DA$ simply becomes $dA/d\tau$, and the shear stress tensor $\tilde\pi^{\mu\nu}$ is fully defined by a single non-vanishing component $\pit\equiv\pit^{z}_{z}=-\pit^{zz}$: at $z{\,=\,}0$, $\pit^{\mu}_{\ \nu}=\mathrm{diag}(0,-\pit/2,-\pit/2,\pit)$. Eqs.~(\ref{eq:de1}) and (\ref{eq:de2a}) simplify to
\begin{equation}
\label{xi_p_0+1d}
\begin{split}
\frac{\dot\xi}{1{+}\xi}-6\frac{\dot\Lambda}{\Lambda}&=
\frac{2}{\tau}+2\Gamma\left(1-\sqrt{1{+}\xi}\,{\cal R}^{3/4}(\xi)\right)\;,
\\
{\cal R}'(\xi) \dot\xi + 4 {\cal R}(\xi) \frac{\dot\Lambda}{\Lambda} &=
- \left({\cal R}(\xi) + \frac{1}{3} {\cal R}_L(\xi)\right) \frac{1}{\tau} 
+\frac{\pit}{{\cal E}_0(\Lambda)\tau},
\end{split}
\end{equation}
In this case, Eqs.~(\ref{eq:de2b}) and (\ref{eq:de2c}) for the transverse flow velocity become redundant, and the evolution equation (\ref{eq:pimunu}) for $\pit$ becomes (after some algebra)
\begin{eqnarray}
\label{pi_0+1d}
\dot{\pit}&=&
-\Gamma\Bigl[\bigl({\cal R}(\xi){\,-\,}{\cal R}_{\rm L}(\xi)\bigr){\cal P}_0(\Lambda)+\pit\Bigr] 
-\lambda (\xi)\frac{\pit}{\tau}
\nonumber\\
&&+4\Bigl[
   \frac{\dot{\Lambda}}{\Lambda}\bigl({\cal R}_{\rm L}(\xi){\,-\,}{\cal R}(\xi)\bigr)
  +\Bigl(\frac{1{+}\xi}{\tau}-\frac{\dot{\xi}}{2}\Bigr)
  \bigl(3{\cal R}^{zzzz}_{-1}(\xi){\,-\,}{\cal R}^{zz}_{1}(\xi)\bigr)
  \Bigr]  {\cal P}_0(\Lambda),
\end{eqnarray}
where 
${\cal R}^{zzzz}_{-1}$ and ${\cal R}^{zz}_{1}$ are given in Eq.~(\ref{C6}) and $\lambda(\xi)$ is defined in Eq.~\ref{lambda}. We now proceed to solve the three coupled equations (\ref{xi_p_0+1d}) and (\ref{pi_0+1d}). 

In Eqs.~(\ref{xi_p_0+1d}) and (\ref{pi_0+1d}) all dissipative transport effects are controlled by a single para\-meter, the relaxation rate $\Gamma$. It entered these equations through the collision terms ${\cal C}$ and ${\cal C}_{-1}^{\langle\mu\nu\rangle}$ in Eqs.~(\ref{eq:be0}) and (\ref{pimunu_tmp}). The specific way in which $\Gamma$ influences the viscous anisotropic hydrodynamic evolution is thus a consequence of the relaxation time time approximation for the collision kernel. We will compare the results from the {\sc vaHydro} equations (\ref{xi_p_0+1d}) and (\ref{pi_0+1d}) with the exact solution of the Boltzmann equation for the same system, which is described in Appendix~\ref{sec:exact_solution}. Comparison of the collision kernel assumed in Eq.~(\ref{eq:col-term}) with the one used to derive the {\sc vaHydro} equations (\ref{rta}), leads to the identification $\tau_\mathrm{eq}=1/\Gamma$.  This is consistent with the asymptotic behavior of the local energy density ${\cal E}={\cal R}(\xi){\cal E}_0(\Lambda)$ at very late times when the system becomes locally isotropic in momentum space, $\xi\to0$. Expanding the {\sc vaHydro} equations (\ref{xi_p_0+1d}) and (\ref{pi_0+1d}) around $\xi=0$ for fixed $\Gamma$, using the asymptotic time-dependence of $\xi$ to eliminate $\xi$ in terms of $\tau$ \cite{Martinez:2010sc}, one finds
\begin{equation}
\lim_{\tau \rightarrow \infty} {\cal E}(\tau) = 
\frac{D}{\tau^{4/3}} \left( 1 - \frac{16}{45} \frac{1}{\Gamma \tau} + {\cal O}\left(\tau^{-2}\right) \right) ,
\label{epsAH}
\end{equation}
which for $\Gamma=1/\tau_\mathrm{eq}$ agrees with the exact result (\ref{epsKIN}) from the Boltzmann equation.

In other viscous hydrodynamic approaches the dissipative effects are typically characterized by a different transport parameter, the specific shear viscosity $\bar\eta=\eta/{\cal S}$, where ${\cal S}$ is the entropy density and $\eta$ the shear viscosity.\footnote{Remember that we 
     ignore heat flow $\V^\mu$ and consider massless particles, hence the heat conductivity 
     and bulk viscosity are zero.} 
To compare our new {\sc vaHydro} approach to other approaches in the literature, we translate $\Gamma$ into $\bar\eta$ using the relationship 
\begin{equation}
\eta=\frac{4}{5} \tau_\mathrm{eq}{\cal P}_0
\end{equation}
obtained in \cite{Denicol:2010xn} for a massless Boltzmann gas, by taking moments of the Boltzmann equation expanded around an isotropic local equilibrium state (see also the discussion in Sec.~VII of Ref.~\cite{Florkowski:2013lya} where this result is obtained without moment expansion). All approaches will be compared at the same value of $\bar\eta$, using
\begin{equation}
  \Gamma = \frac{1}{\tau_\mathrm{eq}} = \frac{T}{5\bar{\eta}}
                 = \frac{{\cal R}^{1/4}(\xi)\Lambda}{5\bar{\eta}}.
\end{equation}

Note that this second-order matching no longer contains the factor of two encountered in leading-order {\sc aHydro} where one found $\Gamma_\text{\sc aHydro} = 2/\tau_{\rm eq}$ \cite{Martinez:2010sc}. As expected for a conformal (massless system), for fixed specific shear viscosity the relaxation rate $\Gamma$ is proportional to the thermal equilibrium temperature of the system.

\begin{figure}[t!]
\begin{center}
\includegraphics[width=1\linewidth]{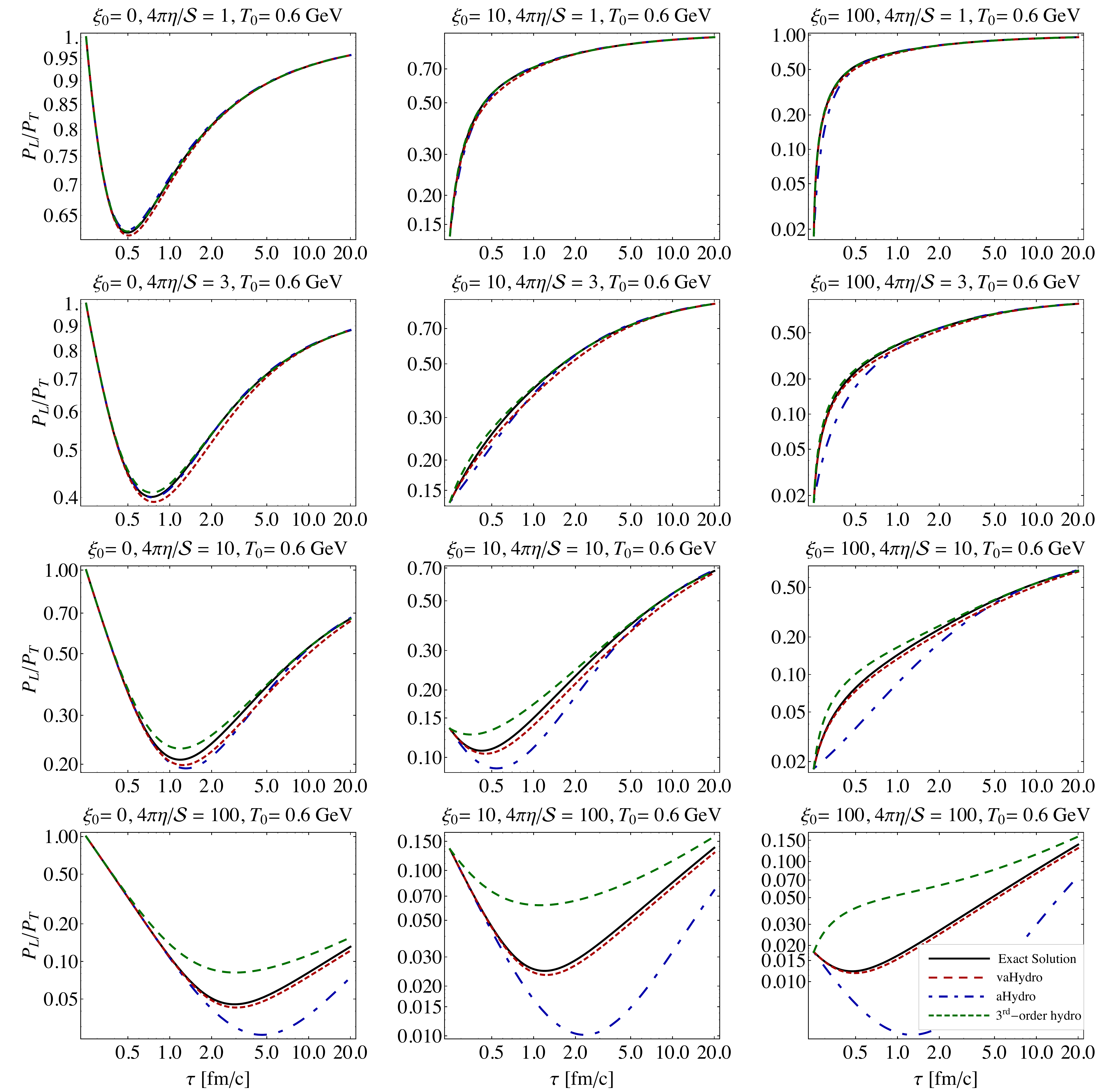}
\end{center}
\vspace{-7mm}
\caption{Ratio of the longitudinal to transverse pressure for $4\pi\bar{\eta}\in\{1,3,10,100\}$ (rows) and $\xi_0\in\{0,10,100\}$ (columns). The black solid, red short-dashed, blue dashed-dotted, and green long-dashed lines are the results obtained from the exact solution of the Boltzmann equation, NLO anisotropic hydrodynamics ({\sc vaHydro}), LO anisotropic hydrodynamics ({\sc aHydro}), and third-order viscous hydrodynamics, respectively. The initial conditions in this figure are $T_0=600$\,MeV, $\pit_0=0$, and $\tau_0=0.25$\,fm/$c$.}
\label{fig:pratio600}
\end{figure}

\begin{figure}[t!]
\begin{center}
\includegraphics[width=\linewidth]{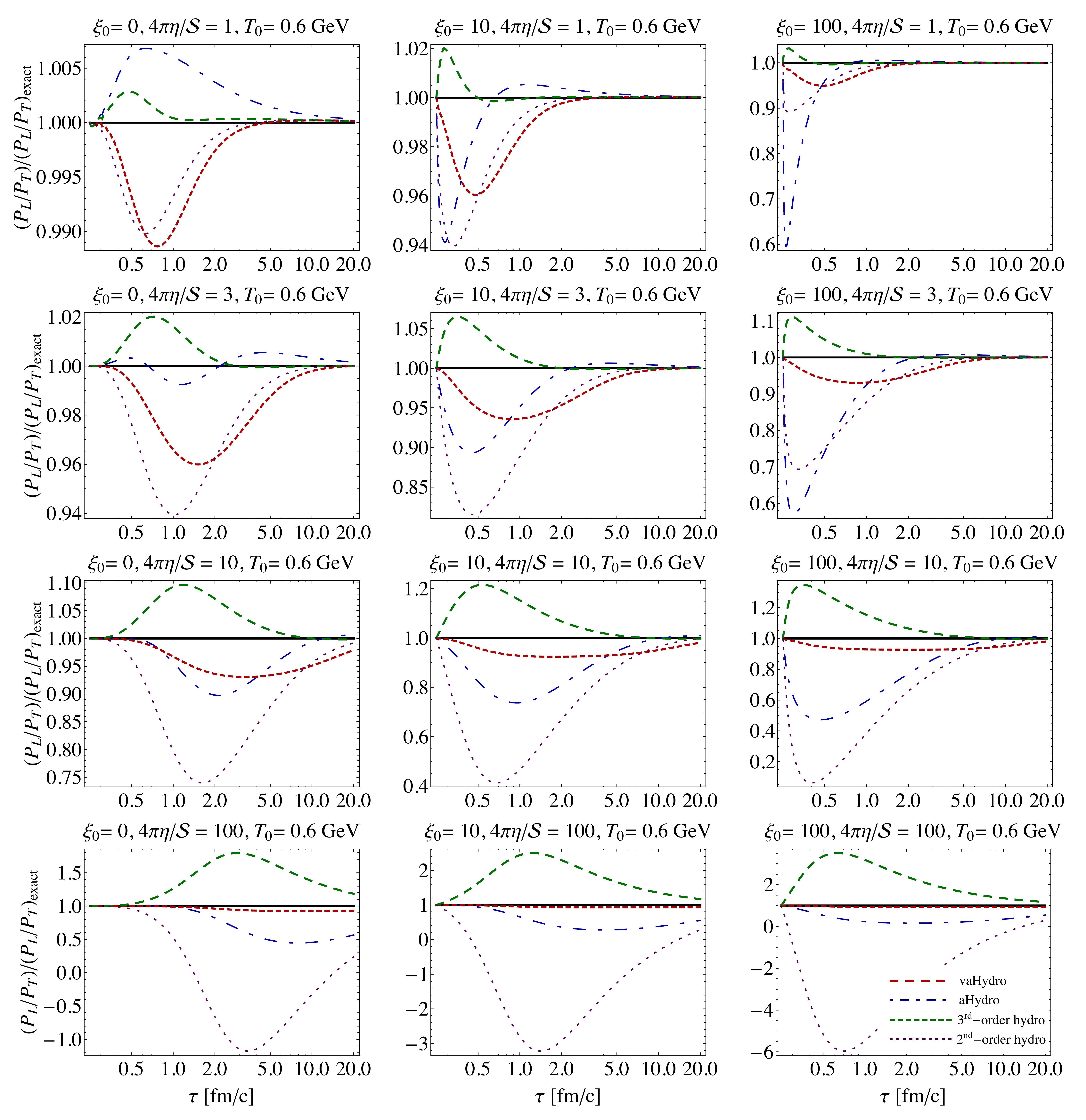}
\end{center}
\vspace{-7mm}
\caption{The same results as in Fig.~\ref{fig:pratio600}, but now plotted as ratios of the approximations to the exact (Boltzmann equation) result. An additional set of purple dotted curves shows results from second-order viscous hydrodynamics \cite{Denicol:2010xn,Denicol:2012cn,Denicol:2012es,Jaiswal:2013npa}.}
\label{fig:pratio600err}
\end{figure}

\subsection{Pressure anisotropy}

We initialize the system with a Romatschke-Strickland distribution function having initial conditions $T_0=600$\,MeV and $\pit_0=0$ at $\tau_0=0.25$\,fm/$c$.  At this initial time we take different values $\xi_0$ for the initial momentum-space anisotropy parameter $\xi_0$. In Fig.~\ref{fig:pratio600} we plot the pressure ratio, ${\cal P}_{\rm L}/{\cal P}_\perp$, for four different values of the shear viscosity to entropy ratio $4\pi\bar{\eta}\in\{1,3,10,100\}$ (rows) and three different initial momentum anisotropies corresponding to $\xi_0\in\{0,10,100\}$ (columns). The black solid, red short-dashed, blue dashed-dotted, and green long-dashed lines are the results obtained from the exact solution of the Boltzmann equation, viscous anisotropic hydrodynamics ({\sc vaHydro}), LO anisotropic hydrodynamics ({\sc aHydro}), and third-order viscous hydrodynamics \cite{Jaiswal:2013vta}, respectively. One sees that in all cases shown {\sc vaHydro} is very close to the exact solution. It is closer to the exact solution than the leading-order {\sc aHydro}. For $4 \pi \eta/{\cal S} \lesssim 10$ the third-order viscous hydrodynamical equations of Jaiswal \cite{Jaiswal:2013vta} reach similar accuracy as {\sc vaHydro}, but for the extreme case of $4 \pi \eta/{\cal S} = 100$ third-order hydrodynamics begins to break down whereas {\sc vaHydro} continues to perform well.

In order to more accurately assess the relative precision of the different approximations, in Fig.~\ref{fig:pratio600err} we plot the ratio of the various approximate results to the exact result for the pressure anisotropy for the same cases shown in Fig.~\ref{fig:pratio600}.  We additionally include the corresponding approximate result obtained by using the second-order viscous hydrodynamic equations of Denicol {\it et al.} \cite{Denicol:2010xn,Denicol:2012cn,Denicol:2012es} as a dotted purple line. The black line in all panels is a visual guide for the reader, indicating the exact solution of the Boltzmann equation. Once again {\sc vaHydro} is seen to yield the best overall approximation in all situations, with third-order hydrodynamics a close second for sufficiently small specific shear viscosities. We also point out that, among the approximations explored here, the second-order viscous hydrodynamic equations
of Denicol {\it et al.} which were shown in \cite{Denicol:2010xn,Florkowski:2013lya,Florkowski:2013lza} to work better than Israel-Stewart theory, provide the poorest approximation to the exact solution, in all cases studied. 

\begin{figure}[t!]
\begin{center}
\includegraphics[width=\linewidth]{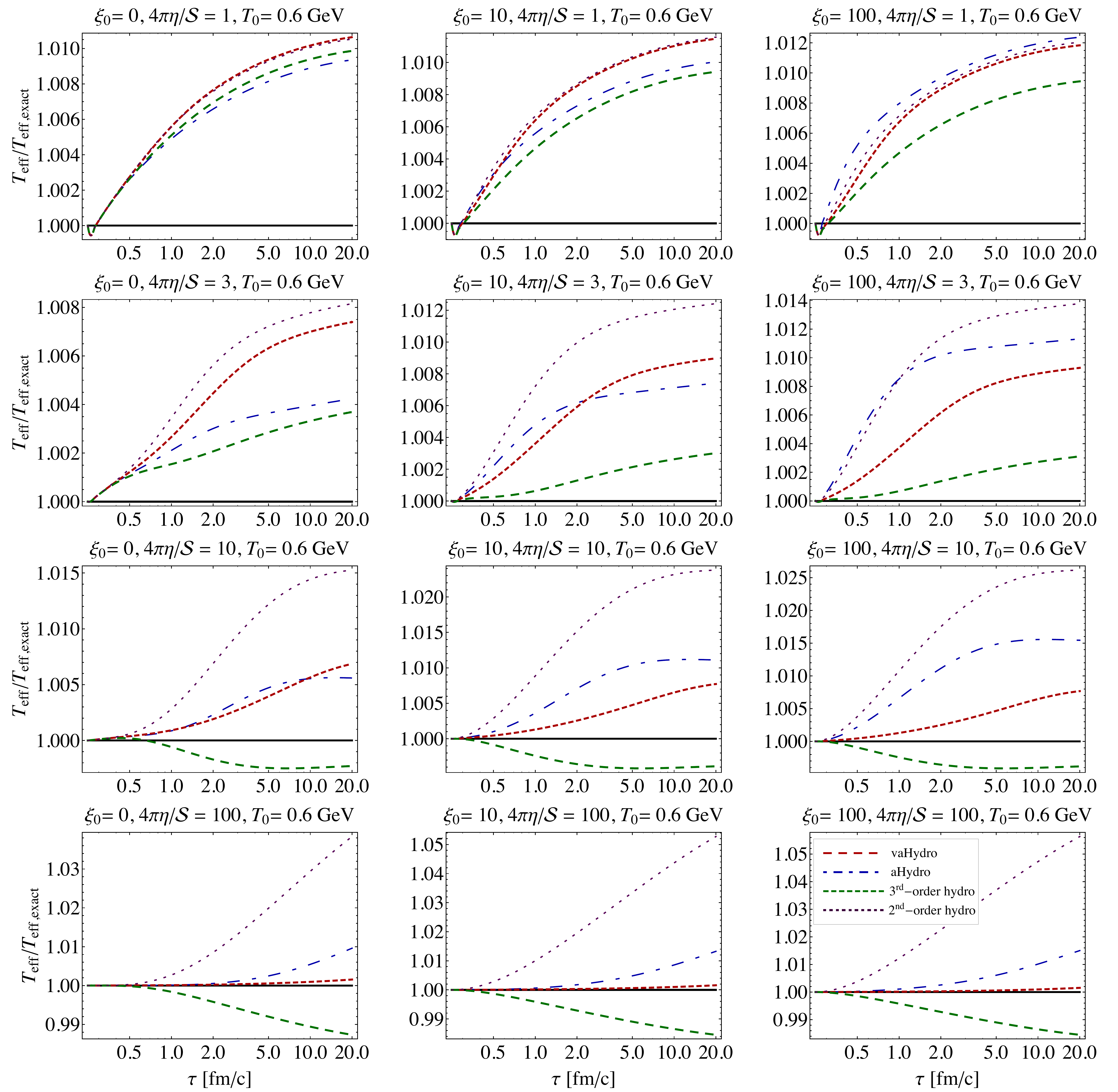}
\end{center}
\vspace{-7mm}
\caption{(Color online) The effective temperature scaled by the value obtained from the exact solution of the Boltzmann equation for the scenarios and approximations shown in Fig.~\ref{fig:pratio600err}. 
}
\label{fig:teff600err}
\end{figure}

\subsection{Effective temperature}

As another measure of accuracy of the various approximations, in Fig.~\ref{fig:teff600err} we plot the effective temperature scaled by the value obtained from the exact solution of the Boltzmann equation, for the same parameter sets as in Figs.~\ref{fig:pratio600} and \ref{fig:pratio600err}. One sees that all approaches shown give quite accurate approximations to the effective temperature, with errors not exceeding $\sim 5.5\%$ over the entire parameter range shown. Once again, however, {\sc vaHydro} outperforms all other approaches, especially for large initial anisotropies $\xi_0$, while second-order viscous hydrodynamics (DMNR \cite{Denicol:2012es}) provides the poorest approximation among those shown. 

\begin{figure}[t!]
\begin{center}
\includegraphics[width=0.55\linewidth]{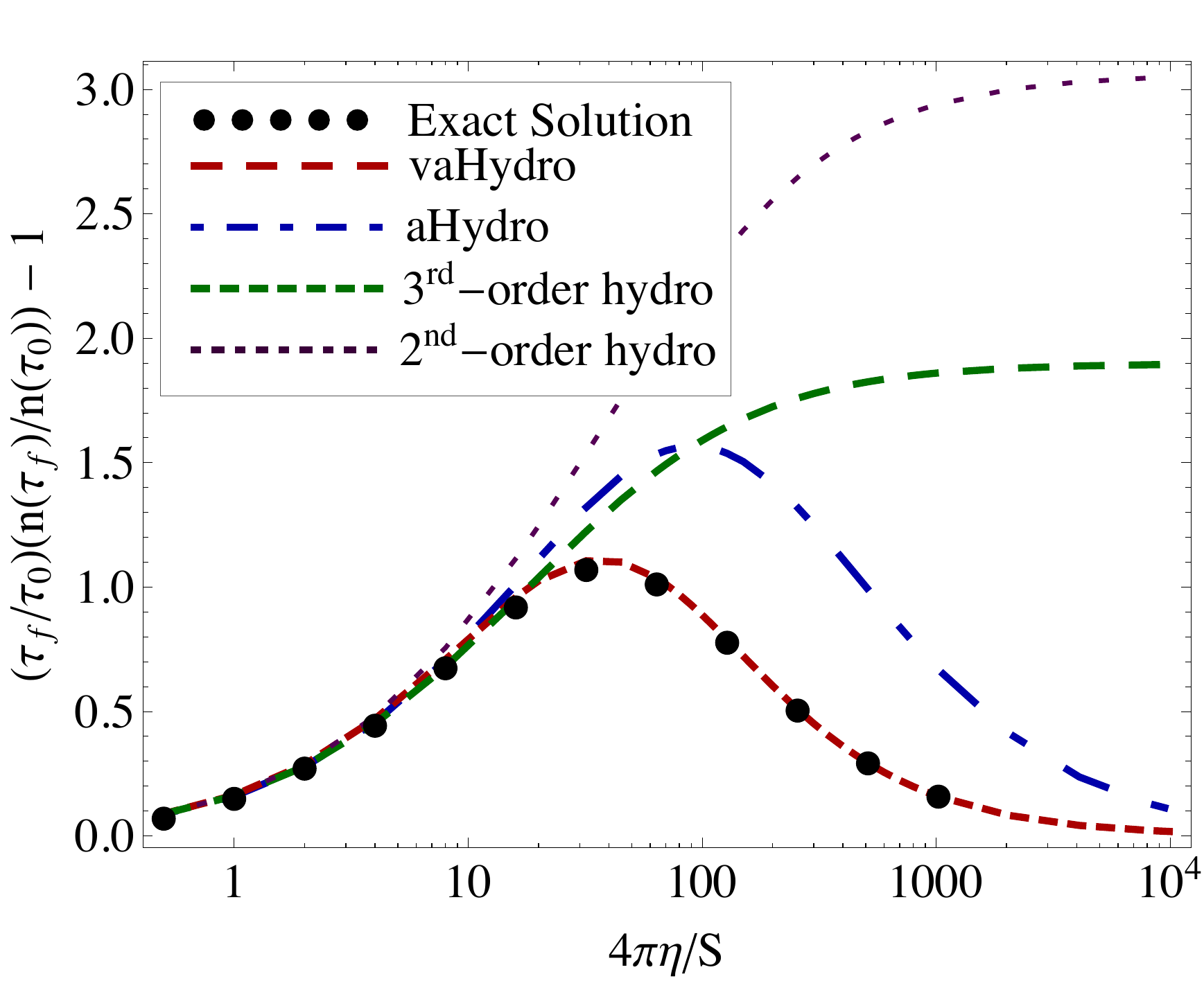}
\end{center}
\vspace{-4mm}
\caption{(Color online) Particle production measure $\Delta_n = (\tau_f n(\tau_f))/(\tau_0 n(\tau_0)) -1$ as a function of $4 \pi \eta/{\cal S}$.  The black points, red dashed line, blue dashed-dotted line, green dashed line, and purple dotted line correspond to the exact solution of the Boltzmann equation, viscous anisotropic hydrodynamics, LO anisotropic hydrodynamics, third-order viscous hydrodynamics, and second-order viscous hydrodynamics, respectively.  The initial conditions in this figure are ${\rm T}_0=600\, \rm{MeV}$, $\xi_0=0$, $\tilde\pi_0=0$, and $\tau_0=0.25\, {\rm fm/c}$.  The freeze-out temperature was taken to be $T_f=150$ MeV.
}
\label{fig:entropy-gen}
\end{figure}

\subsection{Particle production}

In Fig.~\ref{fig:entropy-gen} we plot $\Delta_n \equiv (\tau_f n(\tau_f))/(\tau_0 n(\tau_0)) -1$ which measures particle production by viscous heating. Here $\tau_f$ is the freeze-out time, defined by when the effective temperature, $T_{\rm eff} \propto {\cal E}^{1/4}$, drops below $T_f = 150$\,MeV.   For this figure, we used an isotropic initial condition with $\xi_0 = 0$ and $T_0 = 600$\,MeV at $\tau_0 = 0.25$\,fm/$c$. Physically, the particle production $\Delta_n$ should to go to zero in the limit of ideal hydrodynamics ($\eta/{\cal S}\to 0$) due to entropy conservation, and in the free-streaming limit ($\eta/{\cal S}\to\infty$) due to lack of interactions (this can be seen by the fact that in both the free-streaming and ideal fluid limits, the particle density drops like $1/\tau$ in (0+1)-dimesional expansion). This figure shows that the anisotropic hydrodynamic framework reproduces the correct asymptotic behavior for this quantity as $\eta/{\cal S} \rightarrow \infty$, whereas both the second- and third-order viscous hydrodynamic approaches produce large amounts of additional particles in this limit. While LO {\sc aHydro} describes the $\bar\eta\to\infty$ trend qualitatively correctly, it falls short quantitatively. The {\sc vaHydro} approach, on the other hand, is seen to reproduce the exact result with impressive precision for all values of the specific shear viscosity.

\section{Conclusions}
\label{sec:concl.}

In this paper we derived the dynamical equations for viscous second-order anisotropic hydrodynamics ({\sc vaHydro}) by considering a general expansion of the one-particle phase-space distribution around a locally momentum-anisotropic background. The leading-order term in this reorganized approach was assumed to be of spheroidal form, and deviations from this form were expanded perturbatively in terms of the residual moments. The evolution equations for the dissipative flows were derived from their kinetic definition. This set of equations was then truncated by using the Grad-Israel-Stewart 14-moment approximation scheme which allows one to express the distribution function entirely in terms of the macroscopic fluid-dynamical fields. By taking moments of the underlying microscopic kinetic theory provided by the Boltzmann equation, the fundamental equations for (2+1)-dimensional anisotropic hydrodynamics were obtained in terms of the thermodynamical and macroscopic quantities of the system. For a boost-invariant system of massless degrees of freedom, these equations were then be reformulated in terms of the kinematical parameters of the system: the anisotropy parameter $\xi$, the effective transverse temperature $\Lambda$, and the transverse components of the fluid four-velocity. They were supplemented by evolution equations for the viscous stress tensor $\pit^{\mu\nu}$, which were derived from the dynamical evolution equations for the moments of the residual distribution function $\dft$. For massless systems, the bulk pressure vanishes, and we further restricted our attention to systems with constant chemical potential such that we could ignore heat flow. The resulting dynamical equations define the (2+1)-dimensional {\sc vaHydro} framework and are given in Eqs.~(\ref{eq:de1}) and (\ref{eq:de2}).

In order to evaluate the efficacy of the {\sc vaHydro} approach we then considered the case of transversally homogeneous, longitudinally boost-invariant (0+1)-dimensional expansion for which there exists an exact solution of the Boltzmann equation in the relaxation time approximation. We compared numerical results obtained from {\sc vaHydro}, {\sc aHydro}, 2nd-order viscous hydrodynamics, and 3rd-order viscous hydrodynamics approximations to the exact (0+1)-dimensional RTA solution. We found that generally {\sc vaHydro} agrees with the exact solution better than the three other approaches, particularly in the limit of large shear viscosity to entropy ratio $\bar\eta$. For $\bar\eta$ values below ten times the lower bound of $1/4\pi$, 3rd-order hydrodynamics as formulated by Jaiswal was found to produce results with similar accuracy as {\sc vaHydro} for the initial temperature considered herein. 

Viscous hydrodynamics breaks down when the forces driving the system out of local equilibrium (i.e. the expansion rate and the velocity shear tensor) become too large. {\sc vaHydro} presents no exception from that general rule. In heavy-ion collisions, the largest such dissipative force results from the strong difference between the longitudinal and transverse expansion rates at early times. The viscous hydrodynamic response to this force is handled non-perturbatively in {\sc aHydro}, improving the efficiency of the macroscopic theory compared to treatments that rely on an expansion around isotropic local equilibrium. This improvement is particularly impressive in the case of (0+1)-dimensional expansion studied in Sec.~\ref{sec:0+1d.} where the difference between longitudinal and transverse expansion rates is maximal. In (3+1)-dimensional expansion, there will be additional dissipative force components resulting from anisotropic expansion rates in the transverse plane, caused by strongly inhomogeneous initial pressure profiles. We expect them to be similar in {\sc vaHydro} and Israel-Stewart theory, resulting in similar limitations of both approaches as far as transverse flow anisotropies are concerned. As time proceeds, the flow anisotropies decrease as a result of viscous damping, resulting in smaller deviations from ideal fluid behavior (i.e. smaller values for both $\delta f$ and $\dft$) and thus increasingly better performance of both approaches.

We note that the assumption of massless degrees of freedom made in the present work, which allowed us to factor out the local momentum anisotropy effects from the thermodynamic quantities and use this to convert the viscous anisotropic hydrodynamic equations (\ref{eq:be0}) and (\ref{eq:aniso_eqns1}) into the evolution equations (\ref{eq:de1}) and (\ref{eq:de2}) for the kinematic variables $\xi$ and $\Lambda$, is not necessary and can be relaxed in future work. Instead of the procedure followed here one would then directly solve the coupled set of equations (\ref{eq:aniso_eqns1}) and (\ref{eq:pimunu}), supplemented by an evolution equation for $\Pit$ derived from Eq.~(\ref{pi_tmp}) and an ``anisotropic EoS'' that relates $(2{\cal P}_\perp+{\cal P}_L)/3$ to the equilibrium pressure ${\cal P}_0$ as described in Sec.~\ref{subsec:aniso_variables}. For massive particle systems the transport coefficients will be given by more complicated thermodynamic and collisional integrals that will require numerical evaluation or replacement by phenomenological values. The structure of the equations, however, will not change (except for the addition of an evolution equation for the viscous bulk pressure).  

The {\sc vaHydro} equations derived in this work can describe inhomogeneous systems that undergo anisotropic transverse expansion while remaining boost-invariant along the beam direction. 
Since the expansion around a locally anisotropic momentum distribution results in smaller deviations $\dft$ of the distribution function from the leading-order ansatz, {\sc vaHydro} has a smaller shear inverse Reynolds number $\tilde{\mathrm{R}}_\pi^{-1}=\sqrt{\pit^{\mu\nu}\pit_{\mu\nu}}/{\cal P}_0$ than is the case for standard viscous hydrodynamics.  As a result, the {\sc vaHydro} framework should yield results that are quantitatively more reliable, particularly when it comes to the early stages of QGP hydrodynamical evolution and near the transverse edges of the overlap region where the system is approximately free streaming. 
Numerical solution of the {\sc vaHydro} equations will be explored in future work.

\acknowledgments

We thank G. Denicol and D. Rischke for valuable comments on the first version of this manuscript. This work was supported by the U.S. Department of Energy, Office of Science, Office of Nuclear Physics under Awards No. \rm{DE-SC0004286} and (within the framework of the JET Collaboration) DE-SC0004104. 
M.S. also thanks the organizers of the Yukawa Institute for Theoretical Physics, Kyoto University ``New Frontiers in QCD'' workshop (YITP-T-13-05) where the final stages of this work were completed.

\appendix

\section{Derivation of the general equations of motion}
\label{appa}

To work out the right hand side of the Boltzmann equation (\ref{eq:evolve_be}) we need the four-derivative of $f_\rs$:
\begin{equation}
\label{eq:a1}
\partial_\lambda f_\rs = f_\rs\tilde{f}_\rs \left[ \partial_\lambda\alphat 
- E_\rs \partial_\lambda\beta_\rs -\betat  \partial_\lambda E_\rs \right] ,
\end{equation}
where $E_\rs(x)\equiv\sqrt{p^\mu p^\nu\Xi_{\mu\nu}(x)}$. With the definition (\ref{eq:Xi}) we find
\begin{equation}
\label{eq:a2}
\partial_\lambda \Xi_{\mu\nu}=2u_{(\mu}\partial_\lambda u_{\nu )}
+\left(\partial_\lambda \xi\right)z_\mu z_\nu
+2\xi z_{(\mu}\partial_\lambda z_{\nu )} ,
\end{equation}
and thus for the last term in (\ref{eq:a1})
\bqa
\label{eq:dfrs}
\partial_\lambda E_\rs = \frac{\betat}{E_\rs}\left(Ep^\mu(\partial_\lambda u_\mu)
-\xi p_z p^\mu(\partial_\lambda z_\mu) + \frac{p_z^2}{2} (\partial_\lambda \xi)\right),
\eqa
where $p_z\equiv-p^\lambda z_\lambda$ is the $z$-component of the momentum in the local rest frame.

To obtain the equation of motion for the shear-stress tensor, we apply the convective derivative to its kinetic definition which gives
\begin{equation} 
\label{eq:s1}
\begin{split}
\dot{\pit}^{\langle\mu\nu\rangle}&=
\Delta^{\mu\nu}_{\alpha\beta} D \int dP \,p^{\langle \alpha}p^{\beta\rangle}\dft \\
&={\cal C}^{\langle\mu\nu\rangle}_{-1}- \!\! \int dP \,p^{\langle\mu}p^{\nu\rangle}\dot{f}_{\rs}
- \!\! \int dP \, E^{-1}p^{\langle\mu}p^{\nu\rangle}p^\lambda\nabla_\lambda f_\rs- \!\! \int dP \, E^{-1}p^{\langle\mu}p^{\nu\rangle}p^\lambda\nabla_\lambda \dft \, .
\end{split}
\end{equation}
The first integral in Eq.~(\ref{eq:s1}) can be simplified by using $\dot{f}_\rs=u^\lambda\partial_\lambda f_\rs$ and noticing that the resulting term proportional to $\dot{u}_\lambda$ in Eq.~(\ref{eq:dfrs}) is parity odd. As a result,
\begin{equation} 
\begin{split}
-\int dP \, p^{\langle\mu}p^{\nu\rangle}\dot{f}_{\rs}&=
\dot{\beta}_\rs\int dP \, E_\rs p^{\langle\mu}p^{\nu\rangle}
f_\rs\tilde{f}_\rs
+\frac{\beta_\rs}{2}\dot{\xi}\int dP \,E^{-1}_\rs p^{\langle\mu}p^{\nu\rangle}p^2_z
f_\rs\tilde{f}_\rs \\
& \hspace{5mm}
-\beta_\rs\xi\dot{z}_\lambda
\int dP \,E^{-1}_\rs p^{\langle\mu}p^{\nu\rangle}p^{\langle\lambda\rangle} p_z
f_\rs\tilde{f}_\rs 
-\dot{\alpha}_\rs\int dP \, p^{\langle\mu}p^{\nu\rangle}
f_\rs\tilde{f}_\rs\;.
\end{split}
\end{equation}
By expressing the above tensors in terms of their irreducible forms (into a part symmetric, orthogonal to $u^\mu$, and a traceless part) we arrive at
\begin{eqnarray}
-\int dP \, p^{\langle\mu}p^{\nu\rangle}\dot{f}_{\rs}&=&
\dot{\beta}_\rs\left(\J^{ij}_{0,0,1}X^{(\mu}_iX^{\nu )}_j+\Delta^{\mu\nu}\J_{2,1,1}\right)
-\dot{\alpha}_\rs\left(\J^{ij}_{0,0,0}X^{(\mu}_iX^{\nu )}_j+\Delta^{\mu\nu}\J_{2,1,0}\right)
\nonumber\\ && \hspace{1cm}
+\frac{\beta_\rs}{2}\dot{\xi}
\left(\J^{ijzz}_{0,0,-1}X^{(\mu}_iX^{\nu )}_j+\Delta^{\mu\nu}\J^{zz}_{2,1,-1}\right) 
\nonumber\\ && \hspace{2cm}
-\beta_\rs\xi\left(\J^{ijkz}_{0,0,-1}X^{(\mu}_iX^{\nu )}_jX^\lambda_k+\Delta^{\mu\nu}\J^{zz}_{2,1,-1}z^\lambda\right)\dot{z}_\lambda \;,
\end{eqnarray}
where we have used Eq.~(\ref{eq:thermo_funcs}). Once again, using parity arguments, the second integral in Eq.~(\ref{eq:s1}) can be simplified by using $\nabla_\lambda f_\rs=\Delta^\alpha_\lambda\partial_\lambda f_\rs$,
\begin{equation}\label{eq:pimunu_grad}
\begin{split}
-\int dP \, E^{r-1}p^{\langle\mu}p^{\nu\rangle}p^\lambda\nabla_\lambda f_{\rs}&=
\beta_\rs\left(\nabla_\lambda u_\alpha\right)\int dP \, \frac{E^r}{E_\rs}p^{\langle\mu}p^{\nu\rangle}p^\lambda p^\alpha f_{\rs}\tilde{f}_\rs
\\ & \hspace{1cm}
+\frac{\beta_\rs}{2}\left(\nabla_\lambda\xi\right)
\int dP \, \frac{E^{r-1}}{E_\rs}p^{\langle\mu}p^{\nu\rangle}p^\lambda p^2_z f_{\rs}\tilde{f}_\rs 
\\ & \hspace{2cm}
-\beta_\rs\xi\left(\nabla_\lambda z_\alpha\right)
\int dP \, \frac{E^{r-1}}{E_\rs}p^{\langle\mu}p^{\nu\rangle}p^\lambda p^\alpha f_{\rs}\tilde{f}_\rs\;.
\end{split}
\end{equation}
Decomposing $p^\lambda$ and $p^\alpha$ into parts parallel and orthogonal to the fluid four-velocity, and then using the definition of $\Delta^{\mu\nu\alpha\beta}$, Eq.~(\ref{eq:pimunu_grad}) can be written as 
\begin{equation}
\begin{split}
-\int dP \, E^{r-1}p^{\langle\mu}p^{\nu\rangle}p^\lambda\nabla_\lambda f_{\rs}&=\beta_\rs\left(\J^{ijk\ell}_{r,0,-1}X^{(\mu}_iX^{\nu )}_jX^\lambda_kX^\alpha_\ell+\Delta^{\mu\nu}\J^{k\ell}_{r+2,1,-1}X^\lambda_kX^\alpha_\ell\right)
\nabla_\lambda u_\alpha \\
& \hspace{5mm}
-\beta_\rs\xi
\left(\J^{ijkz}_{r,0,-1}X^{(\mu}_iX^{\nu )}_jX^\lambda_k+\Delta^{\mu\nu}\J^{zz}_{r+2,1,-1}z^\lambda\right)u^\alpha\nabla_\lambda 
z_\alpha \;.
\end{split}
\end{equation}

The remaining terms in Eq.~(\ref{Tensor_n}) arise from moments of the residual distribution function $\dft$. It is relatively straight forward to show, using partial differentiation, that the third integral in Eq.~(\ref{eq:s1}) can be written as 
\begin{align}
\int dP \, E^{-1}p^{\langle\mu}p^{\nu\rangle}p^\lambda\nabla_\lambda \dft &=-\frac{5}{3}\pit^{\mu\nu}\theta
-2\pit^{\langle\mu}_\lambda\sigma^{\nu\rangle\lambda} 
+2\pit^{\langle\mu}_\lambda\omega^{\nu\rangle\lambda}
+2\Pit\sigma^{\mu\nu}
\nonumber \\
&-\Delta^{\mu\nu}_{\alpha\beta}\nabla_\lambda\left\langle
E^{-1}p^{\langle\alpha\rangle}p^{\langle\beta\rangle}p^{\langle\lambda\rangle}\right\rangle_{\tilde\delta}
-\left\langle E^{-2}p^{\langle\mu}p^{\nu\rangle}p^{\langle\alpha\rangle}p^{\langle\beta\rangle}\right\rangle_{\tilde\delta}\nabla_\alpha u_\beta \;,
\end{align}
where we made extensive use of the relativistic Cauchy-Stokes decomposition
\begin{equation}
\partial_\mu u_\nu=u_\mu\dot{u}_\nu+\frac{1}{3}\Delta_{\mu\nu}\theta+\sigma_{\mu\nu}+\omega_{\mu\nu}\;.
\end{equation}

\newpage
\section{14-moment coefficients}
\label{sec:14_coefficients}
In this appendix we list all of the parameters necessary to describe the residual non-equilibrium distribution function in the 14-moment ansatz (\ref{eq:f14moment}) for second-order anisotropic hydrodynamics. We introduce the shorthand notations
\begin{equation}
\G_{n,m}\equiv\J_{n,0}\J_{m,0}-\J_{n-1,0}\J_{m+1,0},
\end{equation}
and with the help of the auxiliary functions
\begin{eqnarray}
{\cal A}_{\pi\alpha} &\equiv& \rho^{zz}_{21}\D_{3,0}{+}(\rho^{zz}_{20}{-}\rho^{zz}_{21})\J_{3,0}\J_{3,1}{+}\rho^{zz}_{20}\J^2_{3,1}{-}\rho^{zz}_{10}\J_{3,1}\J_{4,0}
{-}(\rho^{zz}_{20}{-}\rho^{zz}_{21})\J_{2,0}\J_{4,1}
\nonumber\\
&&\hspace*{1.5cm}  
+\rho^{zz}_{10}(\J_{3,0}{-}\J_{3,1})\J_{4,1}{-}\frac{5}{3}(\rho^{zz}_{20}\J_{2,0}{-}\rho^{zz}_{10}\J_{3,0})\J_{4,2}\;,
\end{eqnarray}
\begin{eqnarray}
{\cal A}_{\pi\beta} &\equiv& \rho^{zz}_{21}\G_{2,3}{+}(\rho^{zz}_{21}\J_{2,0}{-}\rho^{zz}_{20}\J_{2,1})\J_{3,1}{+}\rho^{zz}_{10}\J_{2,1}\J_{4,0}{+}(\rho^{zz}_{20}-\rho^{zz}_{21})\J_{1,0}\J_{4,1}
\nonumber\\
&&\hspace*{1.5cm}  
-\rho^{zz}_{10}(\J_{2,0}-\J_{2,1})\J_{4,1}{-}\rho^{zz}_{20}\J_{2,1}\J_{3,0}{+}\frac{5}{3}(\rho^{zz}_{20}\J_{1,0}{-}\rho^{zz}_{10}\J_{2,0})\J_{4,2}\;,
\end{eqnarray}
\begin{eqnarray}
{\cal A}_{\pi w} &\equiv& -\rho^{zz}_{21}\D_{2,0}{-}(\rho^{zz}_{2,0}\J_{2,0}{-}\rho^{zz}_{10}\J_{3,0})\J_{2,1}
{-}(\rho^{zz}_{10}\J_{2,0}{-}\rho^{zz}_{20}\J_{1,0})\J_{3,1}\;, 
\end{eqnarray}
\begin{eqnarray}
{\cal C}_{\pi w} &\equiv& \D_{3,0} \J_{2,1}{+}\J_{3,1} (\G_{2,3} {-} \J_{2,1} \J_{3,0} {+} \J_{2,0} \J_{3,1}) 
\nonumber\\
&&\ 
{+}(\D_{2,0} + \J_{2,0} \J_{2,1} {-} \J_{1,0} \J_{3,1}) \J_{4,1}{+} \frac{5}{3} \D_{2,0} \J_{4,2})\;,
\end{eqnarray}
we define
\begin{eqnarray}
\F \equiv
&&
2{\cal C}_{\pi w}(\lambda^{1133})^2
+(\lambda^{1111}+\lambda^{1122})\bigl(
{\cal A}_{\pi\alpha}\varphi^{zz}_{21}+{\cal A}_{\pi\beta}\varphi^{zz}_{31}-{\cal A}_{\pi w}(\varphi^{zz}_{41}+\varphi^{zz}_{42})
-{\cal C}_{\pi w}\lambda^{3333}\bigr)
\nonumber\\
&&\hspace{1.5cm} 
-2\lambda^{1133}\bigl(
{\cal A}_{\pi\alpha}\varphi^{xx}_{21}+{\cal A}_{\pi\beta}\varphi^{xx}_{31}-{\cal A}_{\pi w}(\varphi^{xx}_{41}+\varphi^{xx}_{42})\bigr)
\;.
\end{eqnarray}
The functions $\rho^{\alpha\beta}_{nq}$ and $\varphi^{\alpha\beta}_{nq}$ are defined in (\ref{74}) and (\ref{75}). The coefficients contributing in scalar combinations to $\dft$ are:
\begin{eqnarray}
{\cal A}_{\Pi\alpha} &\equiv&
-\frac{1}{\F} \Bigl[
                 2\lambda^{1133} \Bigl(
     (\varphi^{xx}_{41}{+}\varphi^{xx}_{42}) (\rho^{zz}_{2,0} \J_{2,0}{-}\rho^{zz}_{1,0}\J_{3,0})
    - \varphi^{xx}_{31} \bigl[\rho^{zz}_{2,0}(\J_{3,0}{+}\J_{3,1}){-}\rho^{zz}_{1,0} (\J_{4,0}{+}\J_{4,1})\bigr]
    \Bigr)
\nonumber\\
&&\ 
   +\Bigl(\lambda^{3333}(\lambda^{1111}{+}\lambda^{1122}){-}2 (\lambda^{1133})^2\Bigr)
     \Bigl(\D_{3,0} - \J_{3, 0} \J_{3, 1} + \J_{2, 0} \J_{4, 1}\Bigr) 
\\
&&\ 
    - (\lambda^{1111}{+}\lambda^{1122}) 
    \Bigl(
    (\varphi^{zz}_{41}{+}\varphi^{zz}_{42}) (\rho^{zz}_{2,0} \J_{2,0}{-}\rho^{zz}_{1,0}\J_{3, 0})
    -\varphi^{zz}_{31} \bigl[\rho^{zz}_{2,0} (\J_{3, 0}{+}\J_{3, 1}){-}\rho^{zz}_{1,0} 
                                                                  (\J_{4, 0}{+}\J_{4, 1})\bigr]
      \Bigr)
      \Bigr]\;,
\nonumber
\end{eqnarray}
\begin{eqnarray}
{\cal A}_{\Pi\beta} &\equiv&
-\frac{1}{\F} \Bigl[
                 2\lambda^{1133} \Bigl(
     (\varphi^{xx}_{41}{+}\varphi^{xx}_{42}) (\rho^{zz}_{2,0} \J_{1,0}{-}\rho^{zz}_{1,0}\J_{2,0})
    - \varphi^{xx}_{21} \bigl[\rho^{zz}_{2,0}(\J_{3,0}{+}\J_{3,1}){-}\rho^{zz}_{1,0} (\J_{4,0}{+}\J_{4,1})\bigr]
    \Bigr)
\nonumber\\
&&\ 
   -\Bigl(\lambda^{3333}(\lambda^{1111}{+}\lambda^{1122}){-}2 (\lambda^{1133})^2\Bigr)
     \Bigl(\G_{2,3} - \J_{1, 0} \J_{4, 1} + \J_{2, 0} \J_{3, 1}\Bigr) 
\nonumber\\
&&\ 
    - (\lambda^{1111}{+}\lambda^{1122}) 
    \Bigl(
    (\varphi^{zz}_{41}{+}\varphi^{zz}_{42}) (\rho^{zz}_{2,0} \J_{1,0}{-}\rho^{zz}_{1,0}\J_{2, 0})
\\
&&\ 
    -\varphi^{zz}_{21} \bigl[\rho^{zz}_{2,0} (\J_{3, 0}{+}\J_{3, 1}){-}\rho^{zz}_{1,0} 
                                                                  (\J_{4, 0}{+}\J_{4, 1})\bigr]
      \Bigr)
      \Bigr]\;,
\nonumber
\end{eqnarray}
\begin{eqnarray}
{\cal A}_{\Pi w} &\equiv&
\frac{1}{\F} \Bigl[
                 \lambda^{1133} \Bigl(
     \varphi^\perp_{31}(\rho^{zz}_{2,0} \J_{1,0}{-}\rho^{zz}_{1,0}\J_{2,0})
     +\varphi^\perp_{21}(\rho^{zz}_{1,0} \J_{2,0}{-}\rho^{zz}_{2,0}\J_{2,0})
    \Bigr)
\nonumber\\
&&\ 
   +\Bigl(\lambda^{3333}(\lambda^{1111}{+}\lambda^{1122}){-}2 (\lambda^{1133})^2\Bigr)
     \D_{2,0} 
\\
&&\ 
    + (\lambda^{1111}{+}\lambda^{1122}) 
    \Bigl(
    \varphi^{zz}_{31}(\rho^{zz}_{2,0} \J_{1,0}{-}\rho^{zz}_{1,0}\J_{2, 0})
    - \varphi^{zz}_{21}(\rho^{zz}_{2,0} \J_{3,0}{-}\rho^{zz}_{2,0}\J_{2, 0})
      \Bigr)
      \Bigr]\;,
\nonumber
\end{eqnarray}
and
\begin{equation}
\begin{split}
{\cal A}^{\mu\nu}_{\pi\alpha} &\equiv -{\cal A}^{xx}_{\pi\alpha}\Delta^{\mu\nu}+\left({\cal A}^{zz}_{\pi\alpha}-{\cal A}^{xx}_{\pi\alpha}\right)z^\mu z^\nu\;,
\end{split}
\end{equation}
\begin{equation}
\begin{split}
{\cal A}^{\mu\nu}_{\pi\beta} &\equiv -{\cal A}^{xx}_{\pi\beta}\Delta^{\mu\nu}+\left({\cal A}^{zz}_{\pi\beta}-{\cal A}^{xx}_{\pi\beta}\right)z^\mu z^\nu\;,
\end{split}
\end{equation}
\begin{equation}
\begin{split}
{\cal A}^{\mu\nu}_{\pi w} &\equiv -{\cal A}^{xx}_{\pi w}\Delta^{\mu\nu}+\left({\cal A}^{zz}_{\pi w}-{\cal A}^{xx}_{\pi w}\right)z^\mu z^\nu\;,
\end{split}
\end{equation}
where
\begin{eqnarray}
{\cal A}^{xx}_{\pi\alpha} &\equiv&
-\frac{\lambda^{1133}}{\F} {\cal A}_{\pi\alpha}\;,
\end{eqnarray}
\begin{eqnarray}
{\cal A}^{zz}_{\pi\alpha} &\equiv&
\frac{\lambda^{1111}+\lambda^{1122}}{\F} {\cal A}_{\pi\alpha}\;,
\end{eqnarray}
\begin{equation}
{\cal A}^{xx}_{\pi\beta} \equiv
\frac{\lambda^{1133}}{\F} {\cal A}_{\pi\beta}\;,
\end{equation}
\begin{equation}
{\cal A}^{zz}_{\pi\beta} \equiv
-\frac{\lambda^{1111}+\lambda^{1122}}{\F} {\cal A}_{\pi\beta}\;,
\end{equation}
\begin{equation}
{\cal A}^{xx}_{\pi w} \equiv\frac{\lambda^{1133}}{\F} {\cal A}_{\pi w}\;,
\end{equation}
\begin{equation}
{\cal A}^{zz}_{\pi w} \equiv
-\frac{\lambda^{1111}+\lambda^{1122}}{\F} {\cal A}_{\pi w}\;,
\end{equation}
The parameters which contribute to $\dft$ at second-order in the momenta four-vector are:
\begin{equation}
\begin{split}
({\cal C}_{\Pi w})_{\mu\nu} &\equiv -({\cal C}_{\Pi w})_{xx}\Delta_{\mu\nu}+\left(({\cal C}_{\Pi w})_{zz}-({\cal C}_{\Pi w})_{xx}\right)z_\mu z_\nu\;,
\end{split}
\end{equation}
\begin{eqnarray}
{\cal C}^{xx}_{\Pi w} &\equiv&
\frac{1}{\F} \Bigl[
                 (\varphi^{xx}_{41}+\varphi^{xx}_{42}) \Bigl(
     \varphi^{zz}_{21}(\rho^{zz}_{10}\J_{3,0}{-}\rho^{zz}_{20}\J_{2,0}){+}(\rho^{zz}_{20}\J_{1,0}{-}\rho^{zz}_{10}\J_{2,0})
    \Bigr)
\nonumber\\
&&\ 
   +\varphi^{xx}_{21}\Bigl((\varphi^{zz}_{41}{+}\varphi^{zz}_{42})(\rho^{zz}_{20}\J_{2,0}{-}\rho^{zz}_{10}\J_{3,0}){+}\rho^{zz}_{31}(\rho^{zz}_{20}(\J_{2,0}{+}\J_{3,1}){-}\rho^{zz}_{10}(\J_{4,0}{+}\J_{4,1}))
   \Bigr) 
\nonumber\\
&&\ 
   +\varphi^{xx}_{31}\Bigl((\varphi^{zz}_{41}{+}\varphi^{zz}_{42})(\rho^{zz}_{10}\J_{2,0}{-}\rho^{zz}_{20}\J_{1,0}){+}\rho^{zz}_{21}(\rho^{zz}_{20}(\J_{2,0}{+}\J_{3,1}){-}\rho^{zz}_{10}(\J_{4,0}{+}\J_{4,1}))
   \Bigr) 
\\
&&\ 
    - \lambda^{3333}
    \Bigl(
    (\varphi^{xx}_{41}{+}\varphi^{xx}_{42})\D_{2,0}{+}\varphi^{xx}_{31}(\G_{2,3}+\J_{2,0}\J_{3,1}-\J_{1,0}\J_{4,1})+\varphi^{xx}_{21}(\D_{3,0}-\J_{3,0}\J_{3,1}+\J_{2,0}\J_{4,1})
      \Bigr)
      \nonumber\\
&&\ 
    - \lambda^{1133}
    \Bigl(
    (\varphi^{zz}_{41}{+}\varphi^{zz}_{42})\D_{2,0}{+}\varphi^{zz}_{31}(\G_{2,3}+\J_{2,0}\J_{3,1}-\J_{1,0}\J_{4,1})+\varphi^{zz}_{21}(\D_{3,0}-\J_{3,0}\J_{3,1}+\J_{2,0}\J_{4,1})
      \Bigr)
      \Bigr]\;,
\nonumber
\end{eqnarray}
\begin{eqnarray}
{\cal C}^{zz}_{\Pi w} &\equiv&
-\frac{1}{\F} \Bigl[
                 -2\lambda^{1133} \Bigl(
     (\varphi^{xx}_{41}{+}\varphi^{xx}_{42})\D_{2,0}{+}\varphi^{xx}_{31}(\G_{2,3}{+}\J_{2,0}\J_{3,1}{-}\J_{1,0}\J_{4,1}){+}\varphi^{xx}_{21}(\D_{3,0}{-}\J_{3,0}\J_{3,1}{+}\J_{2,0}\J_{4,1}
    \Bigr)
\nonumber\\
&&\hspace*{1cm} 
  +(\lambda^{1111}{+}\lambda^{1122}) \Bigl(
     (\varphi^{zz}_{41}{+}\varphi^{zz}_{42})\D_{2,0}{+}\varphi^{zz}_{31}(\G_{2,3}{+}\J_{2,0}\J_{3,1}{-}\J_{1,0}\J_{4,1})
     \nonumber\\
&&\hspace*{3.8cm}
 {+}\varphi^{zz}_{21}(\D_{3,0}{-}\J_{3,0}\J_{3,1}{+}\J_{2,0}\J_{4,1})
    \Bigr)
      \Bigr]\;,
\end{eqnarray}
and the rank-four tensor $({\cal C}_{\pi w})^{\mu\nu\alpha\beta}$. The number of nonzero components that we need to compute are reduced by using the fact that there is azimuthal symmetry in momentum-space and $({\cal C}_{\pi w})^{\mu\nu\alpha\beta}$ is symmetric in the pairs of indices $\alpha$, $\beta$ and $\mu$, $\nu$, i.e. 
\begin{equation}
\begin{split}
({\cal C}_{\pi w})_{2222} &= ({\cal C}_{\pi w})_{1111}\;,\\
({\cal C}_{\pi w})_{2211} &= ({\cal C}_{\pi w})_{1122}\;,\\
({\cal C}_{\pi w})_{2233} &= ({\cal C}_{\pi w})_{1133}\;,\\
({\cal C}_{\pi w})_{3322} &= ({\cal C}_{\pi w})_{3311}\;.
\end{split}
\end{equation}
The nonzero components of $({\cal C}_{\pi w})_{\mu\nu\alpha\beta}$ are:
\begin{eqnarray}
({\cal C}_{\pi w})_{1111} &\equiv&
\frac{1}{3\F}\frac{1}{\lambda^{1111}-\lambda^{1122}}
\Bigl[
{\cal C}_{\pi w}\bigl(\lambda^{1111}-(\lambda^{1133})^2\bigr)
\nonumber\\
&&\ 
-{\cal A}_{\pi\alpha}\Bigl(\lambda^{1133}\varphi^{xx}_{21}{-}\lambda^{1111}\varphi^{zz}_{21}\Bigr)
{-}{\cal A}_{\pi\beta}\Bigl(\lambda^{1133}\varphi^{xx}_{31}{-}\lambda^{1111}\varphi^{zz}_{31}\Bigr)
\nonumber\\
&&\ 
+{\cal A}_{\pi w}\Bigl(\lambda^{1133}(\varphi^{xx}_{41}{+}\varphi^{xx}_{42}){-}\lambda^{1111}(\varphi^{zz}_{41}{+}\varphi^{zz}_{42})
\Bigr)
\Bigr]
\end{eqnarray}
\begin{eqnarray}
({\cal C}_{\pi w})_{1122} &\equiv&
-\frac{1}{3\F}\frac{1}{\lambda^{1111}-\lambda^{1122}}
\Bigl[
{\cal C}_{\pi w}\bigl((\lambda^{1133})^2-\lambda^{1122}\lambda^{3333}\bigr)
\nonumber\\
&&\ 
+{\cal A}_{\pi\alpha}\Bigl(\lambda^{1122}\varphi^{zz}_{21}{-}\lambda^{1133}\varphi^{xx}_{21}\Bigr)
{+}{\cal A}_{\pi\beta}\Bigl(\lambda^{1122}\varphi^{zz}_{31}{-}\lambda^{1133}\varphi^{xx}_{31}\Bigr)
\nonumber\\
&&\ 
+{\cal A}_{\pi w}\Bigl(\lambda^{1122}(\varphi^{zz}_{41}{+}\varphi^{zz}_{42}){-}\lambda^{1133}(\varphi^{xx}_{41}{+}\varphi^{xx}_{42})
\Bigr)
\Bigr]
\end{eqnarray}
\begin{equation}
({\cal C}_{\pi w})_{1133} \equiv
\frac{1}{3\F}\Bigl({\cal C}_{\pi w}\lambda^{1133}{-}{\cal A}_{\pi\alpha}\varphi^{xx}_{21}{-}{\cal A}_{\pi\beta}\varphi^{xx}_{31}{+}{\cal A}_{\pi w}(\varphi^{xx}_{41}{+}\varphi^{xx}_{42})\Bigr)\;,
\end{equation}
\begin{equation}
({\cal C}_{\pi w})_{1212} \equiv \frac{1}{2\lambda^{1212}}\;,
\end{equation}
\begin{equation}
({\cal C}_{\pi w})_{1313} \equiv \frac{1}{2\lambda^{1313}}\;,
\end{equation}
\begin{equation}
({\cal C}_{\pi w})_{2323} \equiv \frac{1}{2\lambda^{1313}}\;,
\end{equation}
\begin{eqnarray}
({\cal C}_{\pi w})_{3311} &\equiv& 
\frac{1}{3\F}{\cal C}_{\pi w}\lambda^{1133}\;,
\end{eqnarray}
\begin{eqnarray}
({\cal C}_{\pi w})_{3333} &\equiv& 
-\frac{1}{3\F}{\cal C}_{\pi w}(\lambda^{1111}+\lambda^{1122})\;.
\end{eqnarray}
%

\section{Dissipative forces and transport coefficients}
\label{sec:transport_coefficients}

In this appendix we list all of the dissipative driving terms and transport coefficients derived in this paper for second-order anisotropic hydrodynamics. For the bulk viscous pressure they are
\begin{eqnarray}
{\cal W}&\equiv&\left(\dot{\beta}_\rs\J_{0,0,-1}+\frac{\betat}{2}\dot{\xi}\J^{zz}_{0,0,-1}-\dot{\alpha}_\rs\J_{0,0,0}\right)\;,\\
\beta_{\Pi\perp}&\equiv& \I_{0,1}-\I_{0,0}\;,\\ 
\beta_{\Pi L}&\equiv& \I^{zz}_{-2,0}-\I^{xx}_{-2,0}\;,\\
\delta_{\Pi\Pi}^{\mu\nu}&\equiv& \left[\mathcal{A}_{\Pi \alpha}\J^{ij}_{-2,0} -\mathcal{A}_{\Pi \beta}\J^{ij}_{-1,0}+\frac{4}{3}\mathcal{A}_{\Pi w}\J^{ij}_{0,0}-\frac{m^2}{3}\mathcal{A}_{\Pi w}\J^{ij}_{-2,0}\right.\nonumber \\
&&\hspace{1.2cm}\left.+(\lambda_\Pi)_{\sigma\rho}\left(\J^{ijk\ell}_{-2,0}X_k^\sigma X_\ell^\rho+\Delta^{\sigma\rho}\J^{ij}_{0,1}\right)\right]X_i^\mu X_j^\mu\;,\\
\delta_{\Pi\pi}^{\mu\nu\alpha\beta}&\equiv& \left[\mathcal{A}^{\alpha\beta}_{\pi \alpha}\J^{ij}_{-2,0} -\mathcal{A}^{\alpha\beta}_{\pi \beta}\J^{ij}_{-1,0}+\frac{4}{3}\mathcal{A}^{\alpha\beta}_{\pi w}\J^{ij}_{0,0}-\frac{m^2}{3}\mathcal{A}^{\alpha\beta}_{\pi w}\J^{ij}_{-2,0}\right. \nonumber\\
&&\hspace{1.2cm}\left.+\left(\lambda_\pi\right)^{\alpha\beta}_{\sigma\rho}\J^{ijk\ell}_{-2,0}X_k^\sigma X_\ell^\rho\right]X_i^\mu X_j^\nu\;,\\
\lambda_{\Pi V}^{\mu\nu}&\equiv& \left(2(\mathcal{B}_w)^\nu_\alpha\J^{ij}_{0,0}-(\mathcal{B}_v)^\nu_\alpha\J^{ij}_{-1,0}\right)X_i^\mu X_j^\alpha\;,\\
\tau_{\Pi V}^\mu &\equiv& \nabla_\nu\left(2(\mathcal{B}_w)^\mu_\alpha\J^{ij}_{0,0}X_i^\nu X_j^\alpha-(\mathcal{B}_v)^\mu_\alpha\J^{ij}_{-1,0}X_i^\nu X_j^\alpha\right)\;.
\end{eqnarray}
The dissipative forces and transport coefficients for the particle-diffusion current are
\begin{eqnarray}
{\cal Z}^{\mu} &\equiv& \betat\J^{ii}_{1,0,-1}X^\mu_iX^\nu_i\dot{u}_\nu 
+\J^{ii}_{-1,0,1}X^\mu_iX^\nu_i\nabla_\nu\beta_\rs
+\frac{\betat}{2}\J^{iizz}_{-1,0,-1}X^\mu_iX^\nu_i\nabla_\nu\xi 
\nonumber\\
&& -\J^{ii}_{-1,0,0}X^\mu_iX^\nu_i\nabla_\nu\alphat
      -\betat\xi\J^{iikz}_{-1,0,-1}X^\mu_iX^\nu_iX^\lambda_k\nabla_\nu z_\lambda\;,
\\
\ell^{\mu\nu}_{V\Pi}&\equiv& \Delta^\mu_\alpha\Bigl[\mathcal{A}_{\Pi \alpha}\J^{ij}_{-1,0} 
-\mathcal{A}_{\Pi \beta}\J^{ij}_{0,0}+\frac{4}{3}\mathcal{A}_{\Pi w}\J^{ij}_{1,0}
-\frac{m^2}{3}\mathcal{A}_{\Pi w}\J^{ij}_{-1,0}
\nonumber\\
&&\qquad
+\lambda_{\Pi,\sigma\lambda}\left(\J^{ijk\ell}_{-1,0}X_k^\sigma X_\ell^\lambda
+\Delta^{\sigma\lambda}\J^{ij}_{1,1}\right)\Bigr]X_i^\alpha X_j^\nu\;,
\\
\tau^\mu_{V\Pi}&\equiv&\Delta^\mu_\alpha\nabla_\nu\Bigl[\Bigl(\mathcal{A}_{\Pi \alpha}\J^{ij}_{-1,0} 
-\mathcal{A}_{\Pi \beta}\J^{ij}_{0,0}+\frac{4}{3}\mathcal{A}_{\Pi w}\J^{ij}_{1,0}
-\frac{m^2}{3}\mathcal{A}_{\Pi w}\J^{ij}_{-1,0}\Bigr)X_i^\alpha X_j^\nu
\nonumber\\
&&\hspace{1.2cm}
+\lambda_{\Pi,\sigma\lambda}\Bigl(\J^{ijk\ell}_{-1,0}X_k^\sigma X_\ell^\lambda+\Delta^{\sigma\lambda}\J^{ij}_{1,1}\Bigr)X_i^\alpha X_j^\nu\Bigr]\;,
\\
\ell^{\mu\nu\alpha\beta}_{V\pi}&\equiv& \Delta^\mu_\lambda\Bigl[\mathcal{A}^{\alpha\beta}_{\pi \alpha}
\J^{ij}_{-1,0} {-}\mathcal{A}^{\alpha\beta}_{\pi \beta}\J^{ij}_{0,0}
+\frac{4}{3}\mathcal{A}^{\alpha\beta}_{\pi w}\J^{ij}_{1,0}
-\frac{m^2}{3}\mathcal{A}^{\alpha\beta}_{\pi w}\J^{ij}_{-1,0} 
\nonumber\\
&&\hspace{1.2cm}
+\lambda_{\pi,\delta\epsilon}^{\alpha\beta}\J^{ijk\ell}_{-1,0}X_k^\delta X_\ell^\epsilon\Bigr]
X_i^\lambda X_j^\nu,
\\
\tau^{\mu\alpha\beta}_{V\pi}&\equiv&\Delta^\mu_\lambda\nabla_\nu
\Bigl[\Bigl(\mathcal{A}^{\alpha\beta}_{\pi \alpha}\J^{ij}_{-1,0} 
-\mathcal{A}^{\alpha\beta}_{\pi \beta}\J^{ij}_{0,0}
+\frac{4}{3}\mathcal{A}^{\alpha\beta}_{\pi w}\J^{ij}_{1,0}
-\frac{m^2}{3}\mathcal{A}^{\alpha\beta}_{\pi w}\J^{ij}_{-1,0}\Bigr)X_i^\lambda X_j^\nu
\nonumber\\
&&\hspace{1.2cm}
+\lambda_{\pi,\delta\epsilon}^{\alpha\beta}\J^{ijk\ell}_{-1,0} 
X_i^\lambda X_j^\nu X_k^\delta X_\ell^\epsilon\Bigr]\;,
\\
\delta^{\mu\nu\alpha\beta}_{VV}&\equiv& \left(2\mathcal{B}_{w,\rho}^\nu\J^{ijk\ell}_{-1,0}
-\mathcal{B}_{v,\rho}^\nu\J^{ijk\ell}_{-2,0}\right)X_i^\mu X_j^\rho X_k^\alpha X_\ell^\beta\;.
\end{eqnarray}
The dissipative forces and transport coefficients for the shear-stress tensor are
\begin{eqnarray}
{\cal K}^{\mu\nu}&\equiv&\dot{\beta}_\rs\left(\J^{ij}_{0,0,1}X^{(\mu}_iX^{\nu )}_j+\Delta^{\mu\nu}\J_{2,1,1}\right)
-\dot{\alpha}_\rs\left(\J^{ij}_{0,0,0}X^{(\mu}_iX^{\nu )}_j+\Delta^{\mu\nu}\J_{2,1,0}\right) \;,\\
{\cal L}^{\mu\nu}&\equiv&\frac{\beta_\rs}{2}\dot{\xi}
\left(\J^{ijzz}_{0,0,-1}X^{(\mu}_iX^{\nu )}_j+\Delta^{\mu\nu}\J^{zz}_{2,1,-1}\right)\;,\\
{\cal H}^{\mu\nu\lambda}&\equiv& -\beta_\rs\xi\left(\J^{ijkz}_{0,0,-1}X^{(\mu}_iX^{\nu )}_jX^\lambda_k+\Delta^{\mu\nu}\J^{zz}_{2,1,-1}z^\lambda\right)\;,\\
{\cal Q}^{\mu\nu\lambda\alpha}&\equiv& \beta_\rs\left(\J^{ijk\ell}_{0,0,-1}X^{(\mu}_iX^{\nu )}_jX^\lambda_kX^\alpha_\ell
+\Delta^{\mu\nu}\J^{k\ell}_{2,1,-1}X^\lambda_kX^\alpha_\ell\right)\;,\\
\ell_{\pi V}^{\mu\nu\alpha\beta} &\equiv&\Delta^{\mu\nu}_{\sigma\rho}\left((2\mathcal{B}_w)_\gamma^\beta\J^{ijkl}_{0,0}-(2\mathcal{B}_v)_\gamma^\beta\J^{ijkl}_{-1,0}\right)X_i^\sigma X_j^\rho X_k^\alpha X_\ell^\gamma\;,\\
\tau_{\pi V}^{\mu\nu\lambda} &\equiv&\Delta^{\mu\nu}_{\alpha\beta}\nabla_\rho\left(2(\mathcal{B}_w)_\sigma^\lambda\J^{ijkl}_{0,0}X_i^\alpha X_j^\beta X_k^\rho X_\ell^\sigma-(\mathcal{B}_v)_\sigma^\lambda\J^{ijkl}_{-1,0}X_i^\alpha X_j^\beta X_k^\rho X_\ell^\sigma\right)\;,\\
\delta_{\pi\Pi}^{\mu\nu\alpha\beta}&\equiv & 
\left[\left(\mathcal{A}_{\Pi \alpha}-\frac{m^2}{3}\mathcal{A}_{\Pi w}\right)\left(\J^{ijk\ell}_{-2,0}X_i^\mu X_j^\nu+\Delta^{\mu\nu}\J^{k\ell}_{0,1}\right) -\mathcal{A}_{\Pi \beta}\left(\J^{ijk\ell}_{-1,0}X_i^\mu X_j^\nu+\Delta^{\mu\nu}\J^{k\ell}_{1,1}\right)\right.\nonumber\\
&&+\left.\frac{4}{3}\mathcal{A}_{\Pi w}\left(\J^{ijk\ell}_{0,0}X_i^\mu X_j^\nu+\Delta^{\mu\nu}\J^{k\ell}_{2,1}\right)\right]X_k^\alpha X_\ell^\beta \nonumber\\
&&+(\lambda_\Pi)_{\sigma\rho}\left(\J^{ijk\ell mn}_{-2,0}X_i^\mu X_j^\nu+\Delta^{\mu\nu}\J^{k\ell mn}_{0,1}\right)X_k^\alpha X_\ell^\beta X_m^\sigma X_n^\rho
\;,\\
\delta_{\pi\pi}^{\mu\nu\alpha\beta\sigma\rho}&\equiv &
\left[\left(\mathcal{A}^{\sigma\rho}_{\pi \alpha}-\frac{m^2}{3}\mathcal{A}^{\sigma\rho}_{\pi w}\right)\left(\J^{ijk\ell}_{-2,0}X_i^\mu X_j^\nu+\Delta^{\mu\nu}\J^{k\ell}_{0,1}\right) -\mathcal{A}^{\sigma\rho}_{\pi \beta}\left(\J^{ijk\ell}_{-1,0}X_i^\mu X_j^\nu+\Delta^{\mu\nu}\J^{k\ell}_{1,1}\right)\right.
\nonumber\\
&&+\left.\frac{4}{3}\mathcal{A}^{\sigma\rho}_{\pi w}\left(\J^{ijk\ell}_{0,0}X_i^\mu X_j^\nu+\Delta^{\mu\nu}\J^{k\ell}_{2,1}\right)\right]X_k^\alpha X_\ell^\beta 
\nonumber\\
&&+\left(\lambda_{\pi}\right)^{\sigma\rho}_{\delta\epsilon}\left(\J^{ijk\ell mn}_{-2,0}X_i^\mu X_j^\nu+\Delta^{\mu\nu}\J^{k\ell mn}_{0,1}\right)X_k^\alpha X_\ell^\beta X_m^\delta X_n^\epsilon\;.
\end{eqnarray}
%

\section{(0+1)-\lowercase{d} transport coefficients}
\label{sec:lambda0p1}
In this appendix we compute the necessary transport coefficients ($\lambda(\xi)$ appearing in Eq.~(\ref{pi_0+1d})) which govern the evolution of a conformal system with no additional conserved currents undergoing one-dimensional boost-invariant expansion. The residual distribution function in this simplified case is given in the 14-moment approximation by
\begin{equation}
\label{eq:dft}
\dft\equiv w_{\mu\nu}(x)p^{\langle\mu}p^{\nu\rangle}f_\rs\tilde{f}_\rs\;,
\end{equation}
Here $w_{\mu\nu}(x)$ solves the system of linear equations
\begin{equation}
\lambda^{\alpha\beta\mu\nu}w_{\mu\nu}=\pit^{\alpha\beta}\;.
\end{equation}
We decompose $\lambda^{\alpha\beta\mu\nu}$ as
\begin{equation}
\label{lambda_dec}
\lambda^{\alpha\beta\mu\nu}=\tilde{\lambda}^{-1}_\pi\Delta^{\alpha\beta\mu\nu}+\Omega^{\alpha\beta\mu\nu}\;,
\end{equation}
where
\begin{eqnarray}
\tilde{\lambda}^{-1}_\pi &\equiv& 2\J^{xxyy}_{0,0}\;,\\
\Omega^{\alpha\beta\mu\nu} &\equiv& 4\Omega_{(1)}\left[z^{(\alpha}z^{(\mu}\Delta^{\beta)\nu)}-\frac{1}{3}\left(z^\alpha z^\beta\Delta^{\mu\nu}+\Delta^{\alpha\beta}z^\mu z^\nu+\frac{1}{3}\Delta^{\alpha\beta}\Delta^{\mu\nu}\right)\right]\nonumber\\
&&\hspace{2cm}+\Omega_{(2)}\left(z^\alpha z^\beta z^\mu z^\nu
+\frac{1}{3}z^\alpha z^\beta\Delta^{\mu\nu}+\frac{1}{3}\Delta^{\alpha\beta}z^\mu z^\nu+\frac{1}{9}\Delta^{\alpha\beta}\Delta^{\mu\nu}\right)\label{omega_tensor}\;.
\end{eqnarray}
The coefficients $\Omega_{(1)}\equiv\J^{xxyy}_{0,0}-\J^{xxzz}_{0,0}$ and $\Omega_{(2)}\equiv\J^{zzzz}_{0,0}+3\J^{xxyy}_{0,0}-6\J^{xxzz}_{0,0}$ arise due to the breaking of momentum-space isotropy along the $z$-direction (in the local rest frame) and are zero in the isotropic, $\xi\to 0$, limit. In this limit, it is clear that $\epsilon^{\alpha\beta}=2{\cal J}_{42}\pi^{\alpha\beta}$, as given in standard IS theory.
The only components that need to be considered are $\alpha\beta=(11,12,13,22,23,33)$. We can then write Eq.~(\ref{eq:mateq}) in matrix form
\begin{equation}
\label{matrix_eqn}
\Lambda \bm{w}=\bm{\pit}\;,
\end{equation}
where
\begin{eqnarray}
&&\Lambda\equiv \begin{pmatrix}
  \lambda^{1111} & 0 & 0 & \lambda^{1122} & 0 & \lambda^{1133} \\
  \cdot & 2\lambda^{1212} & 0 & 0 & 0 & 0 \\
  \cdot  & \cdot  & 2\lambda^{1313} & 0 & 0 & 0  \\
  \lambda^{1122} & \cdot  & \cdot & \lambda^{1111} & 0 & \lambda^{1133} \\
  \cdot  & \cdot  & \cdot  & \cdot  & 2\lambda^{1313} & 0 \\
  \lambda^{1133} & \cdot  & \cdot & \lambda^{1133} & \cdot & \lambda^{3333}
 \end{pmatrix}\;,\label{eq:mateq}
\\
&&\bm{w}\equiv \begin{pmatrix}
  w_{11} & w_{12} & w_{13} & w_{22} & w_{23} & w_{33} 
 \end{pmatrix}^T\;,
\\
&&\bm{\pit}\equiv \begin{pmatrix}
  \pit_{11} & \pit_{12} & \pit_{13} & \pit_{22} & \pit_{23} & \pit_{33} 
 \end{pmatrix}^T\;.
\end{eqnarray}
In the moment approximation (\ref{eq:dft}) for $\dft$, $w_{\mu\nu}$ is determined by solving the matrix equation (\ref{eq:mateq}) $\bm{w}=\Lambda^{-1}\bm{\pit}$. The term
$\delta_{\pi\pi}^{\mu\nu\alpha\beta\sigma\lambda}\pit_{\sigma\lambda}\nabla_\alpha u_\beta$
in Eq.~(\ref{eq:pimunu}) can be rewritten as
\begin{equation}
\begin{split}
\delta_{\pi\pi}^{\mu\nu\alpha\beta\sigma\lambda}\pit_{\sigma\lambda}\nabla_\alpha u_\beta &=
\left(\int dP\,E^{-2}p^{\langle\mu}p^{\nu\rangle}p^{\langle\alpha\rangle}p^{\langle\beta\rangle}p^{\langle\sigma\rangle}p^{\langle\rho\rangle}f_\rs\tilde{f}_\rs\right)w_{\sigma\rho}\nabla_\alpha u_\beta \\
&=-\frac{\pit}{\tau}\sum_i\left(\frac{1}{3}{\cal R}^{iizz}_{-1}(\xi)-{\cal R}^{iizzzz}_{-3}(\xi)\right)\bar{w}_{ii}(x)\;,
\end{split}
\end{equation}
where we defined $\bar{w}_{ii}(x)\equiv w_{ii}(x)\J_{4,0}(\Lambda)/\pit(x)$ and the ${\cal R}$-functions in Appendix~\ref{sec:themoInt}.
In the last line we used the fact that, for transversely homogeneous systems undergoing boost-invariant longitudinal expansion, the only nonzero contribution to $\nabla_\alpha u_\beta$ is $\nabla_z u_z=-1/\tau$. Additionally, for 0+1d systems, $\pit^{\mu\nu}\theta\to \pit/\tau$, $\pit_\lambda^{\langle\mu}\sigma^{\nu\rangle\lambda}\to\pit/3\tau$, and $\pit_\lambda^{\langle\mu}\omega^{\nu\rangle\lambda}\to 0$. Therefore, there is only one transport coefficient which controls the evolution of $\pit$ for conformal systems with vanishing chemical potential:
\begin{equation}
\label{lambda}
\lambda(\xi)\equiv\frac{7}{3}-2\left[
\left(\frac{1}{3}{\cal R}^{xxzz}_{-1}(\xi)-{\cal R}^{xxzzzz}_{-3}(\xi)\right)
-\left(\frac{1}{3}{\cal R}^{zzzz}_{-1}(\xi)-{\cal R}^{zzzzzz}_{-3}(\xi)\right)
\right]\bar{w}_{xx}(x)\;.
\end{equation}
The appearance of only $\bar{w}_{xx}(x)$ is due to the traceless (and transverse to $u_\mu$) condition $\Delta_{\mu\nu}w^{\mu\nu}=0$. Solving the matrix equation (\ref{matrix_eqn}) together with the transverse and traceless constraint $w_{xx}+w_{yy}+w_{zz}=0$ gives
\begin{eqnarray}
w_{xx}(x)&=&\frac{\pit}{2}\frac{1}{\lambda^{1111}(\xi){+}\lambda^{1122}(\xi){-}2\lambda^{1133}(\xi)}=w_{yy}(x)\;,\\
w_{zz}(x)&=&-\frac{\pit}{\lambda^{1111}(\xi){+}\lambda^{1122}(\xi){-}2\lambda^{1133}(\xi)}\;,
\end{eqnarray}
where the $\lambda^{ijk\ell}(\xi)$ functions are defined in Eq.~(\ref{lambda_dec}). Explicitly working out ${\bar w}_{xx}$ results in 
\begin{equation}
\begin{split}
\bar{w}_{xx}(x)&=
\frac{3\J_{4,0}}{6\tilde{\lambda}^{-1}_\pi{-}4(4\Omega_{(1)}{-}\Omega_{(2)})}
=
\frac{3}{8{\cal R}^{xxyy}_{-1}(\xi){-}8{\cal R}^{xxzz}_{-1}(\xi){+}4 {\cal R}^{zzzz}_{-1}(\xi)}
\\
&=
\frac{24 \xi^{5/2} (1 + \xi)^2}{
\sqrt{\xi} (3 + \xi) [-9 + \xi (-10 + 3 \xi)] + 
 3 (1 + \xi)^2 [9 + (-2 + \xi) \xi] \mathrm{tan}^{-1}\sqrt{\xi}}\;.
 \end{split}
\end{equation}
In the isotropic limit, $\lambda(\xi)$ reduces to the same numerical value\footnote{We point out that there is a typo for the numerical value quoted in Ref.~\cite{Denicol:2010xn}. In follow up work they list all of the transport coeffcients for the dissipative currents. Using Eqs.~(153), (162), and (163) from Ref.~\cite{Denicol:2012es}, then
\begin{equation*}
-2\lambda^r_{\pi\pi}\pi_\alpha^{\langle\mu}\sigma^{\nu\rangle\alpha}-2\delta^r_{\pi\pi}\pi^{\mu\nu}\theta\to -2\left(\frac{5}{7}\right)\frac{\pi}{3\tau}-2\left(\frac{2}{3}\right)\frac{\pi}{\tau}=-\frac{38}{21}\frac{\pi}{\tau}\;.
\end{equation*}
} obtained in Ref.~\cite{Denicol:2010xn},
\begin{equation}
\lambda(\xi\to 0)=38/21\;.
\end{equation} 
%
\section{Evaluation of thermodynamic integrals}
\label{sec:themoInt}

In this appendix we compute the thermodynamic function $\I^{i_1\cdots i_\ell}_{nqr}$ and $\J^{i_1\cdots i_\ell}_{nqr}$ for the case of massless particles, which allows us to factor out the anisotropic degree of freedom. We first notice that $\Delta^{\alpha\beta}p_\alpha p_\beta=-E^2$ and then write
\begin{equation} 
\label{eq:alpha2}
\I^{i_1\cdots i_\ell}_{nqr}\left(\Lambda,\xi\right)=
\frac{1}{(2q+1)!!}\int dP E^{n}E^r_\rs \, 
p_{i_1}\cdots p_{i_\ell} f_\rs \; , \quad \forall \, q\;.
\end{equation}
We can separate $\I^{i_1\cdots i_\ell}_{nqr}$ into a function that depends only on the anisotropic deformation parameter and the isotropic thermodynamical integral ${\cal I}_{km}(\Lambda)$ by using scaled spherical coordinates which characterize the quadric surface (\ref{quadric_surface}) along with Eq.~(\ref{jocabian}). This results in  
\begin{eqnarray} 
\I^{i_1\cdots i_\ell}_{nqr}\left(\Lambda,\xi\right) &=&
\lambda_\parallel\int\frac{d\Omega}{4\pi} \left(\mathrm{sin}^2\theta+\lambda^2_\parallel\mathrm{cos}^2\theta\right)^
{\frac{n-1}{2}}\bar{p}_{i_1}\cdots \bar{p}_{i_\ell}\frac{4\pi}{(2q+1)!!} \int_0^\infty \frac{d\lambda}{(2\pi)^3}\cdot \lambda^{n+r+\ell+1}f_0(\betat\lambda) \nonumber \\
&=& \frac{1}{(2q+1)!!}{\cal R}^{i_1\cdots i_\ell}_{n-1}(\xi){\cal I}_{n+r+\ell,0}\;,
\label{eq:alpha2-2}
\end{eqnarray}
where we have defined the scaled momentum-space Cartesian coordinates $\bar{p}_i\equiv p_i/\lambda$. The same decomposition follows for $\J^{i_1\cdots i_\ell}_{nqr}$:
\begin{equation} 
\label{eq:alpha2-2a}
\J^{i_1\cdots i_\ell}_{nqr}\left(\Lambda,\xi\right) = \frac{1}{(2q+1)!!}{\cal R}^{i_1\cdots i_\ell}_{n-1}(\xi)
{\cal J}_{n+r+\ell,0} .
\end{equation}
We note that in the classical limit ($a=0$), the two functions are identical, $\I^{i_1\cdots i_\ell}_{nqr}(a=0)=\J^{i_1\cdots i_\ell}_{nqr}(a=0)$. The function ${\cal I}_{km}(\Lambda)$ is defined  from $\I_{km}$ as formally identical moments of $f_0$ instead of $f_\rs$. This function can be expressed in terms of the pressure by
\begin{equation}
{\cal I}_{nq}(\Lambda)=\frac{{\cal P}_0(\Lambda)\left(n+1\right)!}{2\betat^{n-2}\left(2q+1\right)!!}\;.
\end{equation}
The integrals defining ${\cal R}_n(\xi)$ can be evaluated in closed form in terms of the hypergeometric function ${}_2F_1(a,b;c;z)$ as
\beq
 {\cal R}_{n}(\xi)\equiv\frac{\lambda_\parallel}{2}\int_{-1}^1 dz\left(1-z^2+\lambda^2_\parallel z^2\right)^{\frac{n}{2}}=\lambda_\parallel\cdot {}_2F_1\!\left(\frac{1}{2},-\frac{n}{2};\frac{3}{2};1-\lambda^2_\parallel\right) \label{eq:rn} \; ,
\eeq
where we have defined the parameter $\lambda^2_\parallel=(1+\xi)^{-1}$. The ${\cal R}_n$ functions for the first few values of $n$ are
\begin{align}
{\cal R}_0(\xi)&=\frac{1}{\sqrt{1{+}\xi}} \, , \nonumber\\
{\cal R}_1(\xi)&\equiv{\cal R}(\xi)=\frac{1}{2}\left(\frac{1}{1{+}\xi}+\frac{{\rm arctan}\sqrt{\xi}}{\sqrt{\xi}}\right)  \, , \nonumber\\
{\cal R}_2(\xi)&=\frac{1}{3\sqrt{1{+}\xi}}\left(2+\frac{1}{1{+}\xi}\right)  \, , \nonumber\\
{\cal R}_3(\xi)&=\frac{1}{8}\left[\frac{5+3\xi}{\left(1+\xi\right)^2}+3\frac{{\rm arctan}\sqrt{\xi}}{\sqrt{\xi}}\right] \; .
\end{align}
In addition, we also tabulate here the remaining ${\cal R}$ functions needed in this paper:
\begin{align}
\label{C6}
{\cal R}^{zz}_{-1}(\xi)&=\lambda^3_\parallel\int_{-1}^1\frac{dz}{2}\left(1-z^2+\lambda^2_\parallel z^2\right)^{-\frac{1}{2}}z^2\equiv\frac{{\cal R}_{\rm L}(\xi)}{3} \;,\nonumber\\
{\cal R}^{zz}_{1}(\xi)&=\frac{(\xi{-}1)+(1{+}\xi)^2{\rm arctan}\sqrt{\xi}/\sqrt{\xi}}
                                           {8\,\xi\,(1{+}\xi)^2}\;,\nonumber\\                                           
{\cal R}^{xxyy}_{-1}(\xi)&=\frac{1}{64 \xi^{5/2}}\left(3 (-1 + \xi) \sqrt{\xi} + (3 + \xi (-2 + 3\xi)) \mathrm{arctan}\sqrt{\xi}\right)                                           \;,
\nonumber\\    
{\cal R}^{xxzz}_{-1}(\xi)&=\frac{\sqrt{\xi}(3 + \xi) + (-3 + \xi) (1 + \xi) \mathrm{arctan}\sqrt{\xi}}{16\xi^{5/2}(1{+}\xi)}                                      \;,
\nonumber\\ 
{\cal R}^{zzzz}_{-1}(\xi)&=\frac{-(3{+}5\xi)+3(1{+}\xi)^2{\rm arctan}\sqrt{\xi}/\sqrt{\xi}}
                                                {8\,\xi^2(1{+}\xi)^2}\;,
\nonumber\\
{\cal R}^{xxzzzz}_{-3}(\xi)&= -\frac{15 + 13 \xi}{16 \xi^3 (1 + \xi)} + \frac{3(5 + \xi) \mathrm{arctan}(\sqrt{\xi})}{16 \xi^{7/2}}
\;,
\nonumber\\
{\cal R}^{zzzzzz}_{-3}(\xi)&=\frac{\sqrt{\xi}(15 + \xi (25 + 8 \xi)) - 
 15 (1 + \xi)^2 \mathrm{arctan}(\sqrt{\xi})}{8 \xi^{7/2} (1{+}\xi)^2}
 \;,
\end{align}
and note that ${\cal R}^{xx}_{-1}={\cal R}^{yy}_{-1}\equiv{\cal R}_\perp/3$. We will now write the asymptotic expansion of ${\cal R}_n$ for small and large $\xi$. For $\xi\ll 0$ we use ({\ref{eq:rn}) with $\lambda^2_\parallel=(1+\xi)^{-1}$ and then write
\begin{eqnarray}
{\cal R}_n(\xi)&=&\frac{1}{2}\left(1-\frac{\xi}{2}+{\cal O}\left(\xi^2\right)\right) \! \int _{-1}^1 dz\!\left(1-\frac{1}{2}\left(n z^2\right)\xi+{\cal O}\left(\xi^2\right)\right)
\nonumber\\
&=& 1-\frac{1}{2}\left(\frac{n}{3}+1\right)\xi+{\cal O}\left(\xi^2\right) .
\end{eqnarray}
The large $\xi$ limit corresponds to $\lambda_\parallel\to 0$. Therefore, ({\ref{eq:rn}) becomes
\begin{equation}
{\cal R}_n(\xi)=\frac{\lambda_\parallel}{2}\int_{-1}^1 dz\left(\left(1{-}z^2\right)^{n/2}+{\cal O}\left(\lambda^2_\parallel\right)\right)=c(n)\lambda_\parallel+{\cal O}\left(\lambda^2_\parallel\right) \; ,
\end{equation}
where $c(n)$ is a constant. Thus, to first order in the expansion parameter, the anisotropy dependence is given asymptotically by ${\cal R}_n(\xi{\to}0)=1-a(n)\xi$ and ${\cal R}_n(\xi{\to}\infty)\sim (1{+}\xi)^{-1/2}$. 

\section{Exact solution of the (0+1)-d Boltzmann equation}
\label{sec:exact_solution}

In this appendix we briefly review the exact solution \cite{Florkowski:2013lza,Florkowski:2013lya} of the Boltzmann equation in relaxation time approximation \cite{1954PhRv...94..511B} for a (0+1)-dimensional boost-invariant system:
\begin{equation}
\label{eq:col-term}
C[f] = -\frac{p\cdot u}{\tau_{\rm eq}} \Bigl[f(\tau,p)- f_{\rm eq}\bigl(p\cdot u,T(\tau)\bigr)\Bigr].
\end{equation}
Here $\tau_{\rm eq}\equiv1/\Gamma$ is the relaxation time (which may depend on proper time $\tau$), and $f_{\rm eq}$ is an equilibrium distribution function. In a boost invariant system which is homogenous in the transverse direction, the dynamical variables only depend on the proper time. The effective temperature $T(\tau)$ appearing in the argument of the equilibrium distribution function is fixed by dynamical Landau matching to the evolving energy density \cite{Baym:1984np}.

We will only consider gases of massless particles. The left hand side of the Boltzmann equation can then be written as $p^\mu \partial_\mu f(\tau,w,p_\perp) = (v/\tau) \partial_\tau f(\tau,w,p_\perp)$, where $w\equiv t p_L - z E$ and $v\equiv Et-p_L z$. The kinetic equation in RTA can then be solved exactly,
\begin{equation}
f(\tau,w,p_\perp) = D(\tau,\tau_0) f_0(w,p_\perp)  
+ \int_{\tau_0}^\tau \frac{d\tau^\prime}{\tau_{\rm eq}(\tau^\prime)} \, D(\tau,\tau^\prime) \, 
f_{\rm eq}(\tau^\prime,w,p_\perp) \, ,  
\label{eq:solf}
\end{equation}
where $\tau_0$ is the initial proper time, $f_0$ is the initial non-equilibrium distribution function, and ${D(\tau_2,\tau_1) = \exp\!\left[-\int_{\tau_1}^{\tau_2} d\tau^{\prime\prime} \, \tau^{-1}_{\rm eq}(\tau^{\prime\prime})\right]}$ is the so-called damping function. Using Eq.~(\ref{eq:solf}), the energy density is readily obtained from
\begin{equation}
{\cal E}(\tau) = \frac{g}{\tau^2} \int dP \, v^2\,  f(\tau,w,p_\perp) \, ,
\end{equation}
where $g$ is the degeneracy factor and $dP = 2 \, d^4p \, \delta(p^2) \theta(p^0) = v^{-1} \, dw \, d^2p_T$. Integrating Eq.~(\ref{eq:solf}) one  obtains an integral equation for the energy density
\beq
\bar{\cal E}(\tau) = D(\tau,\tau_0) \frac{{\cal R}\big(\xi_{\rm FS}(\tau)\big)}
                                                            {{\cal R}(\xi_0)}
+ \int_{\tau_0}^{\tau} \! \frac{d\tau^\prime}{\tau_{\rm eq}(\tau^\prime)} \, D(\tau,\tau^\prime) \, 
\bar{\cal E}(\tau^\prime) \, {\cal R}\!\left( \! \left(\frac{\tau}{\tau^\prime}\right)^2{-}1 \right) ,
\label{eq:inteq}
\eeq
where ${\bar{\cal E} = {\cal E}/{\cal E}_0}$ is the energy density scaled by the initial energy density and ${\xi_{\rm FS}(\tau)\equiv(1{+}\xi_0)(\tau/\tau_0)^2-1}$.

Equation~(\ref{eq:inteq}) can be solved numerically by iteration until a given numerical precision threshold is achieved. The result is a stable energy density profile which is invariant (for a given accuracy threshold) under further iterations. From the resulting energy density, one can solve for the effective temperature via ${\cal E}(\tau) = \gamma \, T^4(\tau)$ where $\gamma$ is a constant which depends on the particular equilibrium distribution function assumed and the number of degrees of freedom. The resulting effective temperature allows one to determine the distribution function $f_{\rm eq}$ at all proper times and, with this, the full particle distribution function can be obtained using Eq.~(\ref{eq:solf}). From this one can determine the number density, longitudinal pressure, and transverse pressure, by integrating the distribution function multiplied by $v/\tau$, $w^2/\tau^2$, and $p_T^2/2$, respectively. From Eq.~(\ref{eq:inteq}) one can also determine the exact late-time behavior of the energy density~\cite{Florkowski:2013lya,Baym:1984np}. For $\tau_\mathrm{eq}(\tau)\xrightarrow[\tau\to\infty]{}\tau^\alpha$ with $\alpha<1$ one finds~\footnote{For $\alpha=1$, the late time fixed point has constant 
         anisotropy $\xi$ and $\Lambda \sim \tau^{-1/3}$. For $\alpha > 1$, the late time 
         fixed point is longitudinal free streaming with $\xi \sim \tau^2$ and constant $\Lambda$.}
\begin{equation}
\lim_{\tau \rightarrow \infty} {\cal E}(\tau) = A\left(\frac{\tau_{\rm eq}}{\tau}\right)^{4/3} \left( 1 - \frac{16}{45} \frac{\tau_{\rm eq}}{\tau} + {\cal O}\left(\tau^{-2}\right) \right).
\label{epsKIN}
\end{equation}
We use this result in the main body of the text to fix the relaxation rate for {\sc vaHydro}.

\bibliography{vahydro}

\end{document}